\documentclass{nature_hack}

\usepackage{deluxetable}
\usepackage{graphicx}   
\usepackage{subcaption}
\usepackage{amsmath}    
\usepackage{amssymb}
\usepackage{times}
\usepackage{latexsym}
\usepackage{deluxetable}
\usepackage{fnpos}
\usepackage{scrextend}
\usepackage{scalerel}
\usepackage{hyperref}
\usepackage{tabularx}
\usepackage{xspace}
\usepackage{enumerate}
\usepackage{longtable}
\usepackage{lscape}
\usepackage{xurl}
\usepackage{lineno}
\usepackage{caption}
\usepackage{comment}
\usepackage{ragged2e}
\usepackage{pdflscape}
\usepackage[inner=2.5cm,outer=2.5cm,top=2.5cm,bottom=2.5cm]{geometry}

\usepackage{xcolor}
\definecolor{tabblue}{HTML}{1F77B4}
\definecolor{taborange}{HTML}{FF7F0E}
\definecolor{tabgreen}{HTML}{2CA02C}
\definecolor{tabred}{HTML}{D62728}
\definecolor{tabpurple}{HTML}{9467BD}
\definecolor{tabbrown}{HTML}{8C564B}
\definecolor{tabpink}{HTML}{E377C2}
\definecolor{tabgray}{HTML}{7F7F7F}
\definecolor{tabolive}{HTML}{BCBD22}
\definecolor{tabcyan}{HTML}{17BECF}

\newcommand*\aap{Astron. Astrophys.}

\newcommand*\aj{Astron. J.}

\newcommand*\apj{Astrophys. J.}
\newcommand*\apjl{Astrophys. J. l.}

\newcommand*\apjs{Astrophys. J. S.}

\newcommand*\apss{Astrophys.~Space~Science}
\newcommand*\araa{An. Rev. Astron. Astrophys.}

\newcommand*\mnras{Mon. Not. R. Astron. Soc.}

\newcommand*\nar{New A Rev.}
\newcommand*\nat{Nature}

\newcommand*\pasj{PASJ}
\newcommand*\pasp{Publications of the Astronomical Society of the Pacific}

\date{}

\newcommand{\eal}[2]{\ifmmode{\mathrm{#1\,#2}}\else{#1\textsc{$\,$\lowercase{#2}}}\fi\xspace}
\newcommand{\feal}[2]{\ifmmode{\mathrm{#1\,#2}}\else{[#1\textsc{$\,$\lowercase{#2}}]}\fi\xspace}
\newcommand{\hfeal}[2]{\ifmmode{\mathrm{#1\,#2}}\else{#1\textsc{$\,$\lowercase{#2}}]}\fi\xspace}

\newcommand{\beginmain}{%
        \setcounter{figure}{0}
        \renewcommand{\figurename}{Figure}
        \renewcommand{\thefigure}{\arabic{figure}}%
     }

\newcommand{\beginsupplement}{%
        \setcounter{table}{0}
        \renewcommand{\tablename}{Supplementary\,\,Table}
        \renewcommand{\thetable}{\arabic{table}}%
        \setcounter{figure}{0}
        \renewcommand{\figurename}{Supplementary\,\,Figure}
        \renewcommand{\thefigure}{\arabic{figure}}%
     }

\title{Multiple outflows and delayed ejections revealed by early imaging of novae}

\author{Elias Aydi$^{1,2*}$, John D. Monnier$^{3*}$, Antoine M\'erand${^4}$, Gail H. Schaefer$^{5}$, Laura Chomiuk$^{2}$, Magdalena Otulakowska-Hypka$^{6}$, Jhih-Ling Fan$^{7}$, Kwan Lok Li$^{7}$, Kirill V.\ Sokolovsky$^{8}$, Ricardo Salinas$^{9}$, Michael Tucker$^{10,11,12}$, Benjamin Shappee$^{12}$, Richard Rudy$^{13}$, Kim L. Page$^{14}$, N. Paul M. Kuin$^{15}$, David~A.~H.~Buckley$^{16, 17}$, Peter Craig$^{2}$, Luca Izzo$^{18,19}$, Justin Linford$^{20}$, Brian D.\ Metzger$^{21,22}$, Koji Mukai$^{23,24}$, Marina Orio$^{25,26}$, Ken J.\ Shen$^{27}$, Jay Strader$^{2}$, Jennifer L.\ Sokoloski$^{21}$, Robert E. Williams$^{28,29}$, Montana N. Williams$^{30}$, Gesesew R. Habtie$^{31}$, Stefan Kraus$^{32}$,  Narsireddy Anugu$^{5}$, Jean-Baptiste Le Bouquin$^{33}$, Sorabh Chhabra$^{32}$, Isabelle Codron$^{32}$, Tyler Gardner$^{34}$, Mayra Gutierrez$^{3}$, Noura Ibrahim$^{3}$, Cyprien Lanthermann$^{5}$, Benjamin R. Setterholm$^{35}$, Christopher Ashall$^{12}$, Jason T. Hinkle$^{12}$, Thomas de Jaeger$^{36}$, and Anna V. Payne$^{29}$}

\begin{document} 
\spacing{1}
\maketitle
\begin{affiliations}

\item Department of Physics \& Astronomy, Texas Tech University, Box 41051, Lubbock, TX, 79409-1051, USA

\item Center for Data Intensive and Time Domain Astronomy, Department of Physics and Astronomy, Michigan State University, East Lansing, MI 48824, USA
\item Astronomy Department, University of Michigan, Ann Arbor, MI 48109, USA

\item European Southern Observatory, Karl-Schwarzschild-Str. 2, 85748 Garching, Germany

\item The CHARA Array of Georgia State University, Mount Wilson Observatory, Mount Wilson, CA 91203, USA

\item Astronomical Observatory Institute, Faculty of Physics and Astronomy, Adam Mickiewicz University, S{\l}oneczna 36, 60-286 Pozna{\'n}, Poland

\item Department of Physics, National Cheng Kung University,
No.\ 1 University Road, Tainan City 70101, Taiwan

\item Department of Astronomy, University of Illinois at Urbana-Champaign, 1002 W. Green Street, Urbana, IL 61801 USA

\item Nicolaus Copernicus Astronomical Center, Polish Academy of Sciences, Bartycka 18, 00-716 Warszawa, Poland

\item Center for Cosmology and AstroParticle Physics, 191 W Woodruff Avenue, Columbus, OH 43210, USA;

\item Department of Astronomy, The Ohio State University, 140 W 18th Avenue, Columbus, OH 43210, USA

\item Institute for Astronomy, University of Hawai'i, 2680 Woodlawn Drive, Honolulu, HI 96822, USA

\item Kookoosint Scientific,
1530 Calle Portada, Camarillo, CA 93010, USA

\item School of Physics and Astronomy, University of Leicester, Leicester LE1 7RH, UK

\item University College London, Mullard Space Science Laboratory, Holmbury House, Dorking, Surrey, UK

\item South African Astronomical Observatory, P.O.\ Box 9, 7935 Observatory, South Africa

\item Department of Astronomy, University of Cape Town, Private Bag X3, Rondebosch 7701, South Africa

\item Osservatorio Astronomico di Capodimonte, INAF, Salita Moiariello 16, Napoli, 80131, Italy

\item Niels Bohr Institute, University of Copenhagen, DARK, Jagtvej 128, Copenhagen, 2200, Denmark

\item National Radio Astronomy Observatory, P.O. Box O, Socorro, NM 87801, USA

\item Columbia Astrophysics Laboratory and Department of Physics, Columbia University, New York, NY 10027, US

\item Center for Computational Astrophysics, Flatiron Institute, 162 5th Avenue, New York, NY 10010, US

\item CRESST and X-ray Astrophysics Laboratory, NASA/GSFC, Greenbelt, MD 20771, USA

\item Department of Physics, University of Maryland, Baltimore County, 1000 Hilltop Circle, Baltimore, MD 21250, USA

\item INAF--Osservatorio di Padova, vicolo dell'Osservatorio 5, I-35122 Padova, Italy

\item Department of Astronomy, University of Wisconsin, 475 N.\ Charter St., Madison, WI 53704, USA

\item Department of Astronomy and Theoretical Astrophysics Center, University of California, Berkeley, CA 94720, US

\item Department of Astronomy \& Astrophysics, University of California, Santa Cruz, 1156 High Street, Santa Cruz, CA 95064, USA

\item Space Telescope Science Institute, 3700 San Martin Drive, Baltimore, MD 21218, USA

\item Physics Department, New Mexico Tech, 801 Leroy Pl., Socorro, NM 87801, USA

\item Department of Physics, Debre Berhan University, P.O. Box 445, Debre Berhan, Ethiopia

\item Astrophysics Group, Department of Physics \& Astronomy, University of Exeter, Stocker Road, Exeter, EX4 4QL, UK

\item Institut de Planetologie et d'Astrophysique de Grenoble, Grenoble 38058, France

\item Cooperative Institute for Research in Environmental Sciences, 216 UCB, University of Colorado Boulder,  Boulder, CO, 80309, USA

\item Max-Planck-Institut f$\ddot{u}$r Astronomie; Heidelberg, Germany

\item L'École Pour l’Informatique et les Techniques Avancées, EPITA, 94270 Le Kremlin-Bicêtre, France. 
\end{affiliations}
\newpage



\begin{abstract}

Novae are thermonuclear eruptions on accreting white dwarfs in interacting binaries. Although most of the accreted envelope is expelled, the mechanism---impulsive ejection, multiple outflows or prolonged winds, or a common-envelope interaction---remains uncertain. GeV $\gamma$-ray detections from $>20$ Galactic novae establish these eruptions as nearby laboratories for shock physics and particle acceleration, underscoring the need to determine how novae eject their envelopes. Here we report on near-infrared interferometry, supported with multiwavelength observations, of two $\gamma$-ray detected novae. The images of the very fast 2021 nova V1674~Her, taken just 2--3 days after discovery, reveal the presence of two perpendicular outflows. The interaction between these outflows likely drives the observed $\gamma$-ray emission. Conversely, the images of the very slow 2021 nova V1405~Cas suggest a delay in the ejection of the bulk of the accreted envelope of more than 50 days after the start of eruption, as the nova slowly rises to visible peak and during which the envelope engulfed the system in a common envelope phase. These unprecedented images offer direct observational evidence that the mechanisms driving mass ejection from the surfaces of accreting white dwarfs are not as simple as previously thought, revealing multiple outflows and delayed ejections. 

\end{abstract}

\beginmain

A classical nova is a transient astronomical event characterized by sudden dramatic brightening of an interacting binary star system, 
resulting from a thermonuclear runaway on the surface of a white dwarf that has accreted hydrogen-rich material from its companion\cite{Gallagher_Starrfield_1976_Main,Starrfield_1989_Main}. The energy released by the thermonuclear runaway causes the accreted envelope to expand tremendously, 
though this energy alone may not be sufficient for completely ejecting it from the binary system \cite{Shen_Quataert_2022_Main}. As novae are observed to expel material at velocities ranging from $\approx$ 500 to 10000\,km\,s$^{-1}$ \cite{McLaughlin_1964_Main,Aydi_etal_2020b_Main}, 
some additional mechanism might be at work. The possibilities include multiple outflows and winds powered by continuing nuclear burning on the white dwarf\cite{Friedjung_1966_I_Main,Friedjung_1987_Main,Friedjung_2011_Main,Aydi_etal_2020b_Main}, 
and/or expulsion through a common-envelope interaction drawing energy from the orbital motion of the binary companion\cite{Livio_1992_Main,Chomiuk_etal_2014_Main}. Understanding how novae expel their envelopes has never been more important with the now-routine detection of GeV $\gamma$-ray emission from more than 20 Galactic novae by the Large Area Telescope (LAT) on the \textit{Fermi} Gamma-Ray Space Telescope \cite{Ackermann_etal_2014_Main,Cheung_etal_2016_Main,Franckowiak_etal_2018_Main,Chomiuk_etal_2020_Main}. These detections have established novae as a new class of particle accelerators and confirmed the presence of strong, energetic shocks within the nova ejecta \cite{Metzger_etal_2014_Main,Metzger_etal_2015_Main}. More recent studies showed that these shocks can contribute a substantial fraction of the nova luminosity, highlighting their importance in powering these enigmatic events \cite{Li_etal_2017_nature_Main,Aydi_etal_2020a_Main}. The fact that the luminosity of some novae could be powered by shocks means that novae are particularly valuable laboratories to better understand the shock-powered transients, with substantial implications for other interaction-powered events such as super-luminous supernovae (SNe; e.g., \cite{Chevalier_Irwin_2011_Main}), Type Ia-Circumstellar Medium (CSM) SNe (e.g., \cite{Silverman_etal_2013_Main}), and stellar mergers (e.g., \cite{Metzger_Pejcha_2017_Main}). 

\begin{figure*}[t]
  \centering
\includegraphics[width=0.9\textwidth]{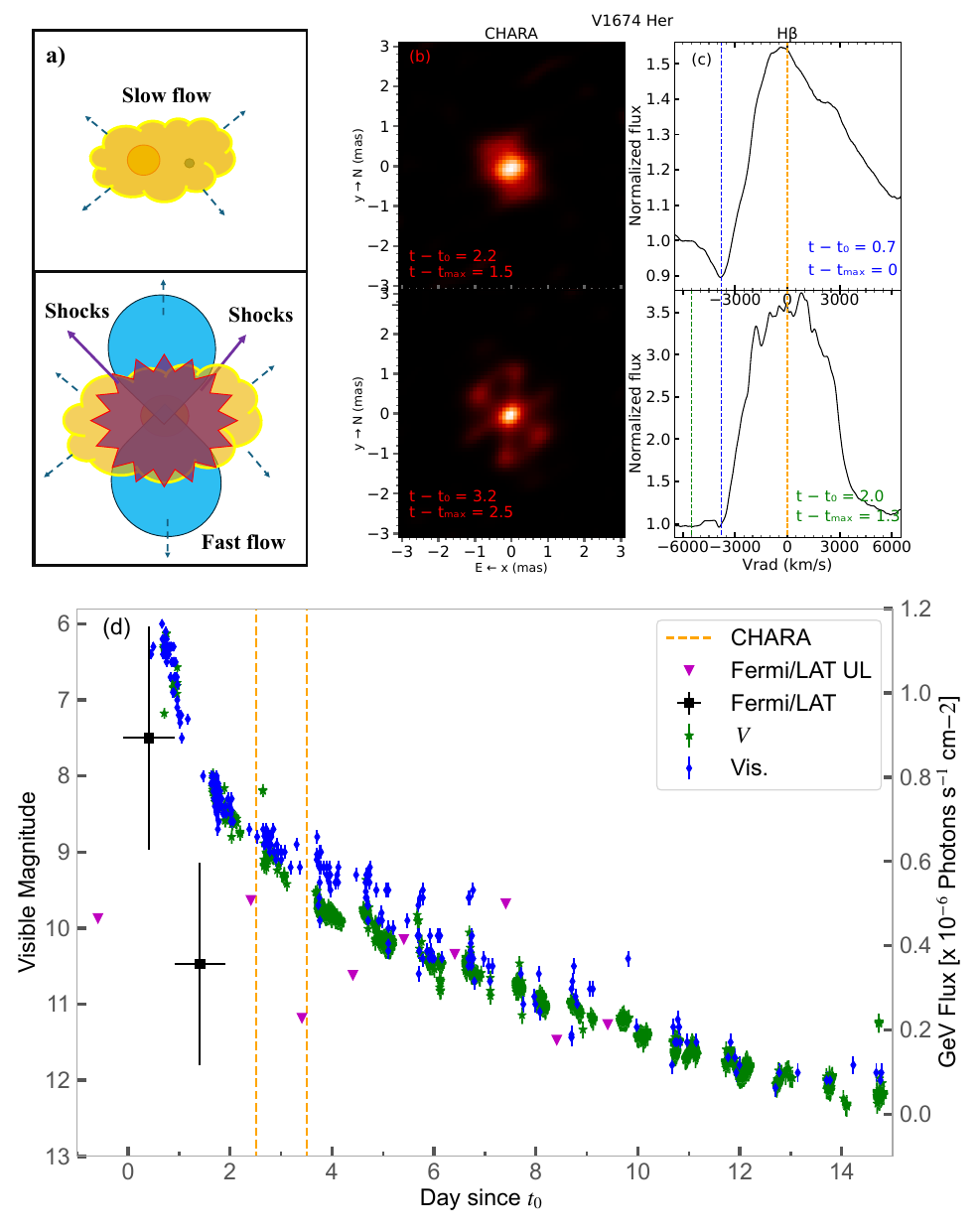}
  \vspace{-0.3cm}  \caption{\footnotesize{\textbf{Early imaging of nova V1674~Her reveals a shift from spherical symmetry with potential multiple outflows.} \textit{Panel (a)}: multiple-ejection schematic illustration---an early slower flow followed near optical peak by a faster outflow; their collision produces GeV $\gamma$-ray–emitting shocks (purple arrows). \textit{Panel (b)}: CHARA images at $t=2.2$ and $3.2$\,d after discovery ($t_0=$ 2021-06-12.19 UT), reconstructed with \textsc{BSMEM} (see Section~Methods). The images show a central elongated component is surrounded by an extended structure elongated roughly perpendicular; we interpret the inner feature as the slow flow and the extended feature as the fast outflow. \textit{Panel (c)}: H$\beta$ spectral line profiles taken on days 0.7 and 2 since $t_0$. The orange, blue, and green dashed lines represent $v_{\mathrm{rad}}$ = 0\,km\,s$^{-1}$, = $-3800$\,km\,s$^{-1}$, and = $-5500$\,km\,s$^{-1}$ relative to rest wavelength, respectively. {Panel (d)}: the AAVSO visible (green stars for $V$-band and blue points for Visual measurements) and GeV $\gamma$-rays (black squares for detections and magenta triangles for 1$\sigma$ upper limits, noted UL in the legend) light curves of V1674~Her. The error bars represent 1-$\sigma$ uncertainties. The vertical dashed lines represent the dates of the CHARA imaging.}}
  \label{Fig:V1674_Her_main}
\end{figure*}

The formation mechanisms of the energetic shocks that lead to the GeV $\gamma$-ray emission from novae are still poorly constrained. Recent models suggest that these shocks are internal to the nova ejecta, more specifically, they occur at the interface of multiple (at least two) ejections: an initial, \textit{slower} ejection/outflow possibly directed towards the orbital plane due to the binary motion, followed by a faster outflow/wind which travels more freely in the polar directions -- possibly driven by radiation powered by the near-Eddington luminosity of the nuclear burning WD \cite{Chomiuk_etal_2014_Main,Metzger_etal_2015_Main,Li_etal_2017_nature_Main,Aydi_etal_2020a_Main,Aydi_etal_2020b_Main,Shen_Quataert_2022_Main,Hachisu_Kato_2022_Main} (see \textit{Panel a} in Figure~\ref{Fig:V1674_Her_main}). As these flows interact, they lead to the shocks responsible for particle acceleration and high-energy emission. This picture was first inspired by the radio imaging of the 2012 \textit{Fermi}-detected nova V959~Mon (see figure~2 in \cite{Chomiuk_etal_2014_Main}), which showed evidence for two outflows expanding in orthogonal directions. This scenario garnered further support based on early spectroscopic follow up of a large sample of novae by \cite{Aydi_etal_2020b_Main}, which showed consistent evidence for multiple spectral components, characterized by distinct velocities, and which were associated with multiple physically distinct ejections (see also \cite{McLaughlin_1947_Main,Payne-Gaposchkin_1957_Main,McLaughlin_1964_Main,Friedjung_1966_I_Main,Friedjung_1966_II_Main,Friedjung_1966_III_Main,Friedjung_1987_Main,Friedman_etal_2011_Main,Hachisu_Kato_2022_Main}). If the binary motion indeed plays a role in expelling the nova envelope during the early stages of the eruption \cite{Lloyd_etal_1997_Main,Chomiuk_etal_2014_Main,Shen_Quataert_2022_Main}, by transferring angular momentum and energy to the ejecta, novae would become \textit{miniature} common envelope systems evolving in real time, i.e., on observable timescales (days/weeks). This would make novae ideal laboratories in our Galactic backyard to constrain the physics of common envelope interaction, which dictates the future evolution of more than 10\% of stars \cite{2011A&A...536A..43N_Main}, yet these processes remain poorly constrained observationally and uncertain theoretically (e.g., \cite{Ivanova_etal_2013_Main,Schneider_etal_2025_Main})

Decades of imaging of old nova shells show a clear departure from spherical symmetry in many novae, with diverse structures, including elliptical morphologies, rings, polar caps/knots, and clumps \cite{Obrien_etal_2006_Main,Obrien_Bode_2008_Main,Lyke_etal_2009_Main,Chesneau_etal_2012_Main}. While some of these structures could have been influenced by the binary companion and its role in shaping the ejecta \cite{Lloyd_etal_1997_Main}, static images of nova remnants observed years to decades after eruption do not allow us to build a complete chronological picture of the ejecta’s shaping and morphological evolution. High-resolution imaging taken during the early days of the eruption is necessary to better understand the ejection scenario in novae -- a challenging task given the small size of the ejecta during these early stages (a few milliarcseconds projected on the sky). However, recent advances in optical/near-IR (NIR) interferometry with facilities like the Center for High Angular Resolution Astronomy (CHARA; \cite{tenBrummelaar_etal_2005_Main}) Array and the Very Large Telescope Interferometer (VLTI; \cite{Petrov_etal_2007_Main,GRAVITY_Collaboration_2017_Main}), which are capable of resolving structures down to a few milliarcseconds (mas), means that resolving nova ejecta during the early days of the eruption is now a possibility. Since some novae peak at brighter than 5--6 magnitudes in the optical/NIR, they are ideal targets for such facilities. Previous efforts were carried out for some novae, such as the recurrent nova T~Pyx (2011) \cite{Chesneau_etal_2011_Main}, which revealed evidence for bipolar ejections, and the \textit{classical} nova V339~Del \cite{Schaefer_etal_2014_Main}, where the size of the early expanding fireball was measured. Here, we report on early (days/weeks) imaging of two Galactic \textit{Fermi}-detected novae, namely V1674~Her and V1405~Cas (both erupted in 2021), obtained with the CHARA Array, revealing unprecedented evidence for multiple and delayed ejections. These novae sample the two extremes of the nova population, with V1674~Her being
one of the fastest novae ever observed and V1405~Cas evolving very slowly, staying
near its peak brightness for around 200 days. 

Nova V1674~Her (Nova Herculis 2021) was detected in eruption on 2021 June 12.19 UT ($t_0$) at visible brightness of around 8.5\,mag. It rose to peak very rapidly, reaching a maximum visible brightness of 6.0\,mag in less than 16 hours. It then declined by 2 magnitudes in around a day, making it one of the fastest evolving novae on record \cite{Woodward_etal_2021_Main,Sokolovsky_etal_2023_Main} (see Supplementary Figure~\ref{Fig:V1674_Her_optical_LC}). The nova was detected as a \textit{Fermi}-source during the first two days of the eruption, implying the presence of energetic shocks leading to high-energy GeV emission\cite{Sokolovsky_etal_2023_Main,Kato_etal_2025_Main,Hachisu_Kato_2025_Main}. We obtained CHARA NIR interferometry for V1674~Her on UT 2021 June 14 and 15, with the Michigan Infra-Red Combiner-eXeter (MIRC-X) instrument \cite{Anugu_etal_2020_Main}, just $\approx$ 2 and 3 days into the eruption (see Figure~\ref{Fig:V1674_Her_main}). The CHARA images show a clear deviation from spherical symmetry, with evidence for two ejecta components: (1) a bipolar outflow moving toward the northeast and southwest directions and expanding at a greater rate relative to (2) a central structure, with an ellipsoidal morphology, extended in the perpendicular direction compared to the other flow. This is in striking agreement with the multiple-ejections scenario suggested to explain the formation of shocks in novae. 
This is also consistent with the early spectroscopic follow up of the nova, which revealed unusually high-velocities for a classical nova -- consistent with the rapid light curve evolution -- and exhibiting two spectral components in the hydrogen Balmer lines: a pre-maximum P Cygni profile with an absorption trough at blueshifted velocity of around 3800\,km\,s$^{-1}$, and a broad emission which emerges around a day after peak, with absorption  at blueshifted velocities of around 5500\,km\,s$^{-1}$ -- co-existing with the pre-maximum absorption (Figure~\ref{Fig:V1674_Her_main} and Supplementary Figures~\ref{Fig:V1674_Her_Halpha_line_profile} and~\ref{Fig:spec_zoom}). This spectral evolution is common in nova spectra as discussed by \cite{Aydi_etal_2020b_Main}, who associated these two components with physically distinct ejections/outflows (the full spectral evolution of V1674~Her is presented in Supplementary Figures~\ref{Fig:V1674_Her_spec_1}, ~\ref{Fig:V1674_Her_spec_2}, and ~\ref{Fig:V1674_Her_spec_3}). These outflows will rapidly interact, leading to bright high-energy GeV emission \cite{Metzger_etal_2015_Main}, which was an indeed defining feature of nova V1674 Her during the first few days of the eruption \cite{Sokolovsky_etal_2023_Main}.
Therefore, it is reasonable to suggest that the central ellipsoidal structure is associated with the pre-maximum P Cygni profile, while the rapidly expanding caps/outflow is associated with the faster spectral absorption component which emerged after peak. 

We use the \textsc{PMOIRED} tool to fit
the V1674~Her visibilities and closure phases measured by the interferometer
with different models approximating the nova image and to measure the projected size of the ejecta (e.g., central component + elongated ring; see Section Supplementary Information for more detail). Based on these models, we measure angular radii of 0.90$\pm$0.07\,mas and 1.23$\pm0.11$\,mas for the faster component, in the first and second epochs, respectively (see Supplementary Figures~\ref{fig:PMOIRED_V1674_Her_1} and~\ref{fig:PMOIRED_V1674_Her_2}, and Supplementary Table~\ref{tab:PMOIRED_V1674_Her}). We assume that the faster flow started expanding around the time the nova was first detected in $\gamma$-rays (the main tracer of the shocks between the slower and faster flows; see \cite{Aydi_etal_2020a_Main,Aydi_etal_2020b_Main}), which started around $t = 0.5\pm0.5$ days.   
Using a velocity of 5500$\pm$500\,km\,s$^{-1}$ (measured from the optical spectra, as a proxy for the total expansion velocity), an expansion duration $\Delta(t) = 1.7\pm0.5$ days, a radius 0.90$\pm$0.07\,mas, and assuming we are observing the system at 67 degrees inclination \cite{Habtie_etal_2024_Main}, we derive a distance $d \approx$ 5.5$\pm$1.7\,kpc towards V1674~Her (based on the expansion parallax). Similarly, using the second epoch (an angular radius of 1.23$\pm$0.11\,mas), an expansion duration $\Delta (t) = 2.7\pm0.5$ days, with the same assumptions, we derive a distance $d \approx 6.4\pm$1.2\,kpc towards V1674~Her.
These distances are consistent with distances derived from other methods (see Section Supplementary Information and Supplementary Figure~\ref{Fig:V1674_Her_DIBs}).
This implies that the association of the rapidly expanding ejecta component in the CHARA images with the faster spectral component in the Balmer lines is reasonable and supports our interpretation of the structures observed in the CHARA images.

\textit{Resolving the evolution and asymmetry of multiple ejecta components just 2--3 days into a nova event is remarkable. These images provide direct observational evidence that nova ejection is more complex than a single ballistic ejection, consistent with the multiple-outflow scenario, and offer valuable insights into the formation of shocks in novae.}

\begin{figure*}
\begin{center}
\includegraphics[width=0.75\textwidth]{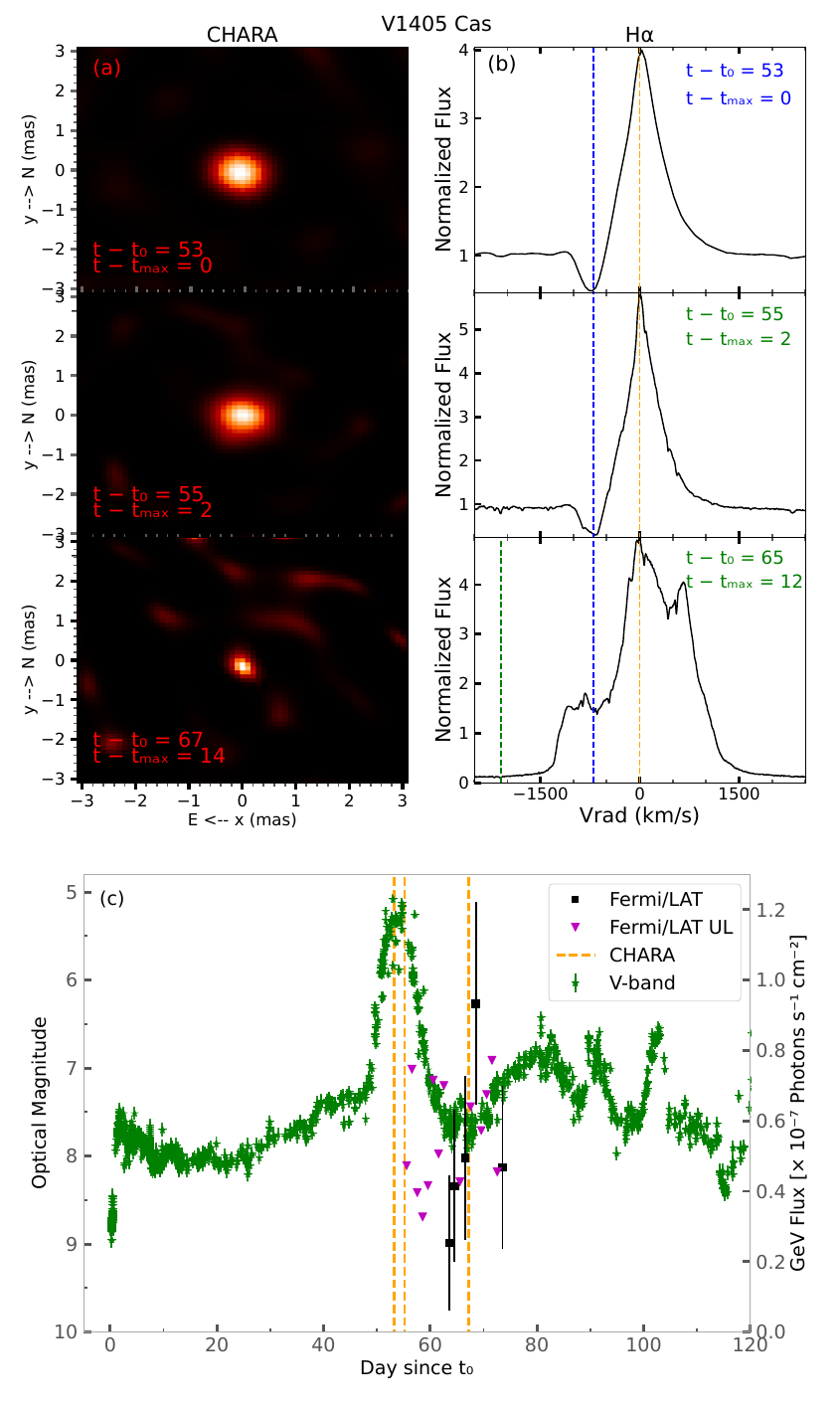}
\vspace{-0.6cm}\caption{\footnotesize{\textbf{Early imaging of nova V1405~Cas reveals a delay in ejection of more than 50 days into eruption.} \textit{Top-left}: three CHARA images obtained on days 53, 55, and 67 since discovery ($t_0$ = 2021-03-18.42 UT), reconstructed with \textsc{BSMEM} (see Section Methods). \textit{Top-right}: H$\alpha$ spectral line profiles taken on days 53, 55, and 65, since $t_0$. The orange, blue, and green dashed lines represent $v_{\mathrm{rad}}$ = 0\,km\,s$^{-1}$, $-700$\,km\,s$^{-1}$, and $-2100$\,km\,s$^{-1}$ relative to rest wavelength, respectively. \textit{Panel (c)}: shows the AAVSO $V$-band (green stars) and GeV $\gamma$-ray (black squares for detections and magenta triangles for 1$\sigma$ upper limits, noted UL in the legend) light curves of V1405~Cas (see Section Methods). The error bars represent 1-$\sigma$ uncertainties. The vertical dashed lines represent the dates of the CHARA epochs.}}
\label{Fig:V1405_Cas_main}
\end{center}
\end{figure*}

The second case presented in this study is nova V1405~Cas, which was discovered on 2021-March 18.42 UT ($t_0$) at a visible brightness of 9.6 mag, and is the extreme opposite of V1674~Her, given its slow evolution. The nova reached a $V-$band magnitude $\approx$ 7.5\,mag a few days after discovery. Thereafter, the brightness-increase halted for a couple of weeks, before rising again to reach a peak in the $V-$band of 5.1\,mag around 53 days after its initial discovery (see Figure~\ref{Fig:V1405_Cas_main}). During this phase (which is often referred to as ``pre-maximum halt''), the optical spectral lines exhibited P Cygni profiles characterized by shallow absorptions and velocities that initially measured around 1500\,km\,s$^{-1}$ but rapidly decreased to about 700\,km\,s$^{-1}$ (see panel (b) in Figure~\ref{Fig:V1405_Cas_main} and Supplementary Figures~\ref{Fig:V1405_Cas_Halpha_line_profile} and~\ref{Fig:V1405_Cas_Dynamic_spectrum}). This apparent deceleration is likely the result of decreasing opacity in the ejecta \cite{Shore_2014_Main,Aydi_etal_2020b_Main}. The presence of shallow absorptions and the rapid drop in velocity point to an origin in a low-density wind emanating from the system during the pre-maximum halt. While most novae show a pre-maximum halt during the rise to visible peak, this halt could last between a few hours in fast novae and up to several weeks in slowly evolving novae \cite{Warner_1985_Main,Strope_etal_2010_Main}. 

After this first peak, V1405~Cas was detected as a GeV source detected by \textit{Fermi} (more than two months into the eruption), implying the presence of strong energetic shocks within the nova ejecta\cite{ATel_14658_Main}. This also implies that the shock onset appears delayed by more than two months after the outburst (in contrast to the case of V1674~Her where GeV emission was detected after less than a day), raising questions about the mass-loss history—whether a substantial part of the envelope was ejected late and/or subsequent winds from the white dwarf were launched only after a substantial delay.

The nova stayed bright for more than 200 days, showing multiple flares/maxima, where the nova colors show substantial changes (Supplementary Figure~\ref{Fig:V1405_Cas_colors}). It then gradually dimmed in the visible over several months (see full light curve in Supplementary Figure~\ref{Fig:V1405_Cas_optica_LC}). Such slow evolution is the antithesis of V1674~Her’s rapid eruption, emphasizing the diversity of nova ejection mechanisms and the complexity of mass-loss processes.

Several slow novae are observed to show a similar long-lasting pre-maximum halt and a gradual rise to visible peak spanning several weeks (REF \cite{Munari_etal_1996_Main,Hachisu_Kato_2004_Main,Kato_Takamizawa_2001_Main,Strope_etal_2010_Main}), the origins of which have not been fully understood yet. Nevertheless, a few ideas have been proposed, some of which suggested that during this early rise to visible peak, the binary motion might play a role in driving the mass loss and eventually expelling the nova envelope (e.g., \cite{Livio_etal_1990_Main,Lloyd_etal_1997_Main}), during a short but important phase of common envelope interaction  (see Section Supplementary Information for more details). With the new insights brought by the presence of strong, energetic shocks during the early days/weeks of a nova, emphasis on the role of the binary motion in driving the mass-loss and expelling the nova envelope has gained traction more recently (e.g., \cite{Chomiuk_etal_2014_Main,Shen_Quataert_2022_Main} ). 
However, direct observational evidence for delayed ejection of the bulk of the nova envelope and the potential role of the binary motion in driving the mass-loss in novae is still lacking. Therefore, resolved imaging of slowly evolving novae like V1405~Cas during their peak brightness can substantially improve our understanding of the role of the binary in driving the mass-loss in novae. 

Motivated by all of this, we obtained two CHARA observations of nova V1405~Cas near visible peak and a third one two weeks after visible peak ($\approx$ 53, 55, and 67 days after discovery; $\approx$ 0, 2, and 14 days from peak, respectively). Surprisingly, the first two observations showed a resolved central source with little to no ejecta around it (Figure~\ref{Fig:V1405_Cas_main}).In these two images, the central source accounts for more than 95\% of the emission, whereas any \textit{potential} extended structure contributes only a few per cent. This extended component is possibly responsible for the pre-maximum P Cygni profiles observed during the rise to peak brightness (panel (b) in Figure~\ref{Fig:V1405_Cas_main} and Supplementary Figure~\ref{Fig:V1405_Cas_Halpha_line_profile}).

The diameter of the resolved central source in the first CHARA epoch is 0.99$\pm$0.02 mas (see Supplementary Figures~\ref{fig:PMOIRED_Nova_cas_1}, ~\ref{fig:PMOIRED_Nova_cas_2}, and~\ref{fig:PMOIRED_Nova_cas_3} and Supplementary Table~\ref{tab:parametersV1405Cas}). Using a distance of 1.73\,kpc measured by \emph{Gaia} (see Section SI), this translates to a source size of around 1.71 AU (i.e., a radius of around 0.85 AU), meaning that the size of the emitting photosphere, in the $H$-band, is indeed similar to that of a red giant star. If the nova envelope was impulsively ejected at $t_0$ and expanded for 53 days with velocities ranging between 700 and 1500\,km\,s$^{-1}$ (as measured from the troughs of the P Cygni profiles in the optical and IR spectra before peak brightness; Supplementary Figures~\ref{Fig:V1405_IR_spec} and~\ref{Fig:V1405_Cas_Halpha_line_profile}), then the line emitting region should have a size between 23 and 46 AU at visible peak. Although the size of the line-emitting regions is not necessarily representative of the continuum-emitting region, at peak brightness it should give us a rough estimate of the size of the photosphere (responsible for the majority of the emission), implying that the majority of the emission should be originating in an extended structure (which would be resolved out by CHARA). 

The large discrepancy between the expected radius of the photosphere ($\sim$23\,--\,46 AU, if the envelope is impulsively ejected at the beginning of the eruption) and the measured radius of the emitting source responsible for the majority of the emission ($\approx$ 0.85 AU), suggests that the bulk of the nova envelope is not fully expelled 53/55 days into the eruption -- indicating delayed ejection of the envelope.  

In the third epoch, there is a remarkable change in the structure of the system and the origin of the emission: the central source is now responsible for only $\approx$ 45--50\% of the emission, while the rest of the emission is originating in an extended structure, which is resolved out by CHARA. At this stage, the nova also shows: (1) a broad emission component characterized by a half width at zero intensity (HWZI) of around 2100\,km\,s$^{-1}$ in the optical spectral lines, (2) GeV $\gamma$-ray emission detected by \textit{Fermi}-LAT (see \cite{ATel_14658_Main} and Figure~\ref{Fig:V1405_Cas_main}), and (3) hard X-ray emission originating in shock-heated plasma detected by the \textit{Neil Gehrels Swift Observatory} 5 days after the third CHARA epoch (see Supplementary Figure~\ref{Fig:V1405_Cas_BVRI_Swift}). The expansion or expulsion of the nova envelope may be driven by multiple mechanisms. The faster outflows observed in the optical spectral can plow into the puffed up envelope helping it to escape the system. Such interaction could lead to energetic shocks, which  power high-energy emission (detected by \textit{Swift} and \textit{Fermi}) --- again, in agreement with the multiple-outflow scenario suggested to explain shocks in novae. In addition, the binary orbital motion can also play an important role in expelling the bulk of the puffed up envelope. \cite{Livio_1992_Main,Lloyd_etal_1997_Main} argued that the binary motion could transfer angular momentum and energy to the ejecta, which might help expelling the nova envelope.

Moreover, the imaged central source shrinks in size between the second and third epoch, which is consistent with the density of the material near the binary dropping. Based on all the above, we conclude that the bulk of the nova envelope was likely still engulfing the binary system for more than 50 days, in a common envelope phase during the rise to peak visible brightness, and it was only expelled after more than 55 days, eventually leading to shock interaction evidenced by the delayed detection of high-energy emission.

After the \textit{first} visible peak, the light curve of V1405~Cas exhibits multiple flares (at least 9) over a period exceeding 200 days, with some reaching amplitudes greater than 2 magnitudes (Supplementary Figure~\ref{Fig:V1405_Cas_optica_LC}). During the flaring activity, new absorption features in the spectra appear at progressively higher velocities (Supplementary Figures~\ref{Fig:V1405_Cas_spec_1}, \ref{Fig:V1405_Cas_spec_2}, \ref{Fig:V1405_Cas_Halpha_line_profile}, \ref{Fig:V1405_Cas_Halpha_line_profile_2}, \ref{Fig:V1405_Cas_Halpha_line_profile_3}, \ref{Fig:V1405_Cas_Halpha_line_profile_4}, and~\ref{Fig:V1405_Cas_Dynamic_spectrum}). This suggests that the multiple flares may result from renewed mass-ejection episodes, with new material launched at increasingly faster velocities. As faster material catch up with previously ejected slower material, further shock interaction manifests, contributing to the nova's multi-wavelength emission \cite{Pejcha_2009_Main,Aydi_etal_2019_I_Main,Aydi_etal_2020a_Main}. This underscore the complexity of mass-loss processes during nova eruptions.

In summary, our results highlight the potential role of the binary motion in driving mass-loss in novae \cite{Livio_etal_1990_Main,Chomiuk_etal_2014_Main,Shen_Quataert_2022_Main} and encourage future theoretical studies/modeling to explore this role. \textit{They also provide us with unmatched observational evidence for a delayed ejection of the bulk of the envelope during a nova, confirming that the ejection scenario in novae is more complex than a single, impulsive ejection at the beginning of the eruption.} By increasing the sample of novae observed with CHARA and other optical/NIR interferometers in the future, we can confirm if this delayed ejection is common in other novae, which would establish novae as ideal laboratories in our Galactic backyard to constrain the physics of common envelope interaction.


\clearpage
\begin{center}
{\bf {\Large Methods}}
\end{center}
\section{CHARA NIR interferometry.}
We utilized the Georgia State University Center for High Angular Resolution Astronomy (CHARA) Array \cite{ten_Brummelaar_etal_2005_Main} to observe nova V1674~Her on UT 2021 June 14 and 15 (days 2 and 3) and nova V1405~Cas on UT 2021 May 10, 12, and 24 (days 53, 55, and 67), in order to probe angular scales with sub-milliarcsecond resolution using the method of NIR interferometry. The Michigan InfraRed Combiner - eXeter (MIRC-X; \cite{Anugu_etal_2020_Main}) combines the light from all 6 CHARA telescopes simultaneously in the near-infrared $H$-band (typical $\lambda_0=1.62\mu$m, $\Delta\lambda=0.25\mu$m), providing sufficient Fourier space (u,v) coverage for simple interferometric imaging. The observations of nova V1674~Her combined all 6 telescopes using the PRISM102 mode with a spectral resolving power of $R \sim 102$. The observations of V1405~Cas combined only 5 telescopes because the S1 telescope was not in delay and used the GRISM190 mode on May 10 and 24 and PRISM50 mode on May 12, providing spectral resolving powers of $R \sim 190$ and 50, respectively. In Supplementary Figures~\ref{Fig:UV_V1674_Her} and~\ref{Fig:UV_V1405_Cas} we present the CHARA (u,v) coverage for the different epochs. The raw data were reduced using the publicly-available data pipeline (available at \url{https://gitlab.chara.gsu.edu/lebouquj/mircx_pipeline}) to produce visibility amplitudes on up to 15 baselines and closure phases on up to 20 closed triangles. We used an integration time of 30 seconds \cite{Monnier_2007_Main}. The calibration of the instrumental transfer function required observations of single stars with known diameters estimated by \textsc{SearchCal} \cite{Bonneau_etal_2011_Main,Bourges_etal_2014_Main,Chelli_etal_2016_Main}. We used a calibration script written in IDL by J. D. Monnier to average the measurements over 2.5 minute intervals with the “deep cleaning” option to remove outliers. For nova V1405~Cas, we used HD 211982 (0.570$\pm$0.015 mas), HD 135969 (0.215$\pm$0.006 mas), HD 145965 (0.211$\pm$0.005 mas), HD 151259 (0.425$\pm$0.011 mas), and HD 219080 (0.69$\pm$0.069mas). The calibrators on UT 2021May12 were located at a declination -23$^\circ$ at a very large angular separation on sky from V1405 Cas (declination +61$^\circ$). For nova V1674~Her, we used HD 177433 (0.537$\pm$0.012mas) and HD 184607 (0.540$\pm$0.012 mas). The calibrated OIFITS files are available through the JMMC Optical Interferometry Database (OIDB; \url{https://www.jmmc.fr/english/tools/data-bases/oidb/}). The log of the CHARA observations is listed in Supplementary Tables~\ref{table:CHARA_log_V1674_Her} and~\ref{table:CHARA_log_V1405_Cas}.

At these early times in the evolution of the novae, we see some modest evidence (see Supplementary Figures~\ref{fig:PMOIRED_V1674_Her_1} and~\ref{fig:PMOIRED_V1674_Her_2}) for small visibility changes across the spectrum when observed in medium resolution (R190) for instance due to strong Hydrogen lines.  However, the effect is relatively small (and hardly seen in the closure phase spectra), so we have combined all the spectral channels together for imaging, assuming grey emission. In the future, a more sophisticated treatment could account for the line emission as a separately imaged component, but for now we are interested in the overall geometrical changes during the first few weeks of the main continuum emission (first few weeks of the eruption).

For the CHARA aperture synthesis imaging presented here, we used a maximum entropy imaging algorithm that is directly fitted to the $V^2$ and closure phase quantities, first described by \cite{Buscher_1994_Main} and later implemented in a publicly-available package called the \textsc{Bi-Spectrum Maximum Entropy Method} (\textsc{BSMEM}; \cite{Malbet_etal_2010_Main}). We used \textsc{BSMEM} (v1.5) to fit to OIFITS data that were averaged on 15\,min timescales. To provide a prior for \textsc{BSMEM}, an elliptical Gaussian was fit to the visibility data for each epoch. All imaging presented used the same \textsc{BSMEM} settings: pixel scale 0.1\,mas, width 64x64 pixels, constrained $V^2$ at origin to 1.00$\pm$0.05, Gaussian elliptical prior, maximum 50 iterations, and regularization hyperparameter evaluation using classical Bayesian with known noise scale ($-rt 1$).  For more discussion on our approach to imaging with NIR interferometry data, see the overview by REF. \cite{baron2016_Main}.

Our results are shown in Supplementary Figures~\ref{Fig:methods_images_1} and~\ref{Fig:methods_images_2}. Each column shows a different epoch. The top row for each nova shows the prior image used based on an elliptical Gaussian fit to the visibility data. The middle row shows the \textsc{BSMEM} image using a linear scale. We also include in the middle panel the percent of the flux contained within 1\,mas of the brightest pixel. The bottom row shows the square-root intensity, to highlight low-surface brightness emission within the field-of-view.

We note a few further details here. The bright spots around the ring for nova V1674~Her on day 3 are likely due to the limited CHARA baselines -- the visibility data themselves are consistent with a more continuous, elongated ring (the brightness of the ring varies with azimuth, and is brighter on the diagonal; see Section SI).  For the final epoch of nova V1405~Cas on day 63, about 50\% of the flux is over-resolved, but we lack short baselines to make a reliable image; thus, the spots in the background are likely artifacts from a diffuse low-surface brightness shell with scale $>$5\,mas. This epoch also showed signs of line-emission in the H-band spectrum and we also carried out imaging limiting to 1.5-1.6 microns, avoiding most of the strong emission lines. These results were consistent with the full spectrum results, albeit with lower reduced chi-squared, and so were not presented in Supplementary Figures~\ref{Fig:methods_images_1} and~\ref{Fig:methods_images_2} for clarity.

\section{Optical/IR Spectroscopy and Photometry.} We obtained optical spectra for novae V1674~Her and V1405~Cas using a diversity of facilities and instruments. In Supplementary Tables~\ref{table:spec_log_V1674_Her} and~\ref{table:spec_log_V1405_Cas} we list the log of these observations. Below we summarize the observations and data reduction. 

We have used the Gemini-North Multi-Object Spectrograph (GMOS-N; \cite{Hook_etal_2004_Main}) on the 8.1\,m Gemini-North to observe nova V1674~Her on 2021-06-17, 06-19, 06-27, 07-01, 07-04, 07-07, and 07-09 (days 5, 7, 15, 19, 22, 25, and 27). The observations were taken using the 0.5$^{\prime\prime}$ slit and the R831 grating to provide a range of
4500\,--\,6800\,\AA{} at a resolving power $R \approx$ 4400). GMOS spectra were reduced with the aid of DRAGONS v3.1.0 \cite{labrie19_Main}, applying bias subtraction and flat fielding, and establishing a wavelength solution based on the arc spectra. Since the latter were obtained during daytime, the zero-point of the wavelength solution was checked for instrument flexure by measuring the Oxygen
5577.33\,\AA{} sky emission. Deviations were found to be at most of 0.1\,\AA{} and were corrected when necessary.

Optical spectra were obtained for nova V1674~Her with the SuperNova Integral Field Spectrograph (SNIFS; \cite{Lantz_etal_2004_Main}) on the University of Hawaii UH2.2m telescope through the Spectroscopic Classification of Astronomical Transients (SCAT; \cite{Tucker_etal_2022_Main}) survey. SNIFS is an Integral Field Unit (IFU) covering
$3400-9000$\,\AA{} at $R\approx 1000$ over a $6''\times6''$ field of view with a spatial sampling of 0.4\,arcsec/spaxel. The 2D frames are converted into 3D ($x,y,\lambda$) datacubes, flat-fielded with a continuum lamp exposure, and extracted into a 1D spectrum by applying aperture photometry to each wavelength slice. Spectra are then corrected for the optical system throughput and atmospheric extinction \cite{Buton_etal_2013_Main} using observations of spectrophotometric standard stars \cite{Rubin_etal_2022_Main}. \cite{Tucker_etal_2022_Main} provide a detailed description of the processing pipeline.

We make use of publicly available spectra reported to the  Astronomical Ring for Access to Spectroscopy (ARAS; \cite{Teyssier_2019_Main}). The data consist of a combination of low-resolution ($R\approx 1000)$, medium-, and high-resolution (up to $R\approx 20,000)$ spectra obtained by citizen scientists. 

On 2021 March 23.13 and 28.12 UT (days 5 and 10 since $t_0$) we obtained observations of nova V1405~Cas using the Visible and Near-Infrared Imaging Spectrograph (VNIRIS) mounted on the 1.0 meter telescope of the Aerospace Corporation. A description of the telescope is given by \cite{Rudy_etal_2021_Main}.  The raw spectra were compensated for atmospheric absorption and converted to absolute flux using observations of the G-type standard stars Hip 63333 and Hip 2087.  

We make use of publicly available data reported to the American Association of Variable Stars (AAVSO; \cite{AAVSODATA_Main}) to create the optical light curves for both novae. The data consist of $BVRI$ photometry and visual estimates.

\section{\textit{Fermi}-LAT $\gamma$-ray observations and analysis.}
We downloaded the \textit{Fermi} Pass 8 $\gamma$-ray data (\texttt{P8R3\_V3}) of V1674~Her and V1405~Cas from the LAT Data Server. The obtained data cover an energy range from 100~MeV to 300~GeV within circular regions of radius 10$^{\circ}$ centered on the targets. The standard analysis software, \texttt{Fermitools} (version 2.2.0), was used for the LAT data reduction and analysis. Only type 3 data (front and back conversion types) in the event class 128 (standard source class) were used. As Earth's limb is a bright source of $\gamma$-rays which influences the LAT observations, events with apparent zenith angle larger than 90$^{\circ}$ are removed. We also excluded the time periods with bad data quality flags to get good time intervals.

Daily $\gamma$-ray light curves of the novae were extracted using the binned likelihood method Note that MJD 59339 had a problem, so no data was available on this day. To solve this, we fixed the isotropic background emission and extended the analysis day from 1 day to 1.1 days. In the analysis, emission model files were built for the fields based on the spatial and spectral information of the 4FGL-DR3 cataloged sources located within 20$^{\circ}$ from the targets \cite{Abdollahi_etal_2022_Main}. The $\gamma$-ray spectral models of the two novae were both assumed to be simple power-law model for simplicity. Except for the normalization factors of the novae, all the spectral parameters in the model file including the Galactic diffuse and isotropic emission (i.e., \texttt{gll\_iem\_v07} and \texttt{iso\_P8R3\_SOURCE\_V3\_v1}) were fixed to the 4FGL-DR3 best-fit values to avoid over-fitting due to the low-count statistics. For V1674~Her, the implied (best-fit) photon index is $-2.3$ and the emission was active on 2021 June 12 and 13 (during the first day since $t_0$;  \cite{Lin_etal_2022_Main}). To estimate the photon index of V1405~Cas, we first extracted a preliminary $\gamma$-ray light curve with a fixed photon index. We defined the $\gamma$-ray active phase by the first and last dates with TS $\gtrapprox4$ over a 24\,hr period (where TS is the likelihood ratio test statistic) in the light curve. This interval of data was then re-analyzed using the likelihood method to obtain the best-fit photon index, which is $-2.74\pm0.23$ in the period from 2021 May 21 to June 2 (days 64 to 76 after eruption). Binned likelihood fitting was then performed daily for both novae from 2021 March 18 to June 9. For those daily bins with significant detections (i.e., TS $\gtrapprox4$, over a 24\,hr period), the inferred photon fluxes with 1-$\sigma$ uncertainties are presented (where $\sigma$ represents the standard deviation). Otherwise, 95\% upper limits are reported. 

\section{\textit{Swift} XRT and UVOT observations.}
The \emph{Neil Gehrels Swift Observatory} (\textit{Swift}) observations of V1405~Cas began on 2021 March 24, 5.6 days after the discovery, with exposures initially obtained every 3 days, with the cadence subsequently dropping to weekly and then fortnightly until the end of 2021 June. The data have been downloaded from the UK Swift Science Data Centre (UKSSDC).

The \textit{Swift} X-ray Telescope (XRT) data were processed with the
standard \textsc{HEASoft} analysis software, together with the corresponding
calibration files. Given the brightness of the UV/optical emission, only
grade 0 (single-pixel X-ray events) were considered in this work, in order
to minimize contamination by optical
loading (\url{https://www.swift.ac.uk/analysis/xrt/optical_loading.php}).
Upon this more detailed examination, the XRT detection announced in \cite{ATel_14530_Main} (using grade 0-12 multi-pixel events) was determined to be spurious.
During the first three months following the eruption, the vast majority of
the \textit{Swift} datasets did not show significant X-ray detections, and
so 3$\sigma$ upper limits are reported in
Supplementary Figure~\ref{Fig:V1405_Cas_BVRI_Swift}. However, the observations on 2021 May
31 and June 15 (days 74 and 89 since $t_0$) did reveal an X-ray source at the location of the nova.
Detailed checks show that the spectrum of this source is inconsistent with optical loading: the combined emission from the two detections can be approximated with a single-temperature optically-thin thermal plasma of kT $>$2.6~keV, assuming an absorbing column N$_{\rm H}$ = 3.8~$\times$~10$^{21}$~cm$^{-2}$ \cite{ATel_14530_Main}, providing an observed (unabsorbed) flux estimate of 1.2 (1.5)~$\times$~10$^{-12}$~erg~cm$^{-2}$~s$^{-1}$ over 0.3--10~keV.
Since `false' X-rays from optical loading would appear much softer, the X-ray detections on these two days are therefore believed to be real. Subsequent observations from 2021 June 30 to July 30 (days 104--134) did not show significant X-ray emission, however the upper limits on the count rates are consistent with the two detections.

The \textit{Swift} UltraViolet and Optical Telescope (UVOT) obtained photometry in three ultraviolet filters: $uvw2$, $uvm2$, and $uvw1$ centered around 190, 220, and 260 nm respectively. Due to the brightness of the source the calibration of the readout-streak \cite{page-Mat_Main} was used with the UVOTPY \cite{Kuin_2014_Main} software. 
\clearpage

\begin{addendum}

\item[Data availability]  The data used in this manuscript are available through: \cite{Aydi2025Figshare}

\item This work was completed using support from NASA award 13-Fermi 80NSSC20K01237. This work is based upon observations obtained with the Georgia State University Center for High Angular Resolution Astronomy Array at Mount Wilson Observatory.  Time at the CHARA Array was granted through the NOIRLab community access program (NOIRLab PropID: 2021A-0042; PI: E. Aydi). The CHARA Array is supported by the National Science Foundation under Grant No. AST-2034336 and AST-2407956. Institutional support has been provided from the GSU College of Arts and Sciences and the GSU Office of the Vice President for Research and Economic Development. E.Ay. acknowledges support by NASA through the NASA Hubble Fellowship grant HST-HF2-51501.001-A awarded by the Space Telescope Science Institute, which is operated by the Association of Universities for Research in Astronomy, Inc., for NASA, under contract NAS5-26555.
L.Ch. acknowledges NSF awards AST-1751874 and AST-2107070, NASA awards 80NSSC23K0497, 80NSSC23K1247
, and 80NSSC25K7334,  and a Cottrell fellowship of the Research Corporation. J.St. was supported by the Packard Foundation. J.So. is supported by NASA grants 80NSSC25K7082 and 80NSSC25K7068.  M. Or acknowledges support from Chandra award GO2-23013X.  M.O.H. was supported by the Polish National Science Center grant 2019/32/C/ST9/00577 and the ”Initiative of Excellence - Research University” (ID-UB) programme
at A. Mickiewicz University in Poznań. KLP and NPMK acknowledge funding from the UK Space Agency.  B.D.M. acknowledges support by NASA through the ATP program (80NSSC22K0807), the Fermi Guest Investigator Program (grant number 80NSSC24K0408) and the Simons Foundation (grant number 727700). The Flatiron Institute is supported by the Simons Foundation. KJS acknowledges support by NASA through the ATP program (80NSSC20K0544). SK acknowledges support from an UK Science and Technology Facilities Council Small Award (ST/Y002695/1). We thank the ARAS and AAVSO observers from around the world who contributed their spectra to the ARAS database and their magnitude measurements to the AAVSO International Database, used in this work. SK acknowledges funding for MIRC-X from the European Research Council (ERC) under the European Union's Horizon 2020 research and innovation programme (Starting Grant No. 639889 and Consolidated Grant No. 101003096). JDM acknowledges funding for the development of MIRC-X (NASA-XRP NNX16AD43G, NSF-AST 2009489) and MYSTIC (NSF-ATI 1506540, NSF-AST 1909165). This research has made use of the Jean-Marie Mariotti Center Aspro and SearchCal services. JTH was supported by NASA grant 80NSSC23K1431.

\item[Author Contributions] E.Ay.\ wrote the text. E.Ay., J.Mo., A.Me., G.Sc., M.Ot., R.Sa., M.Tu., K.L.Li., J-L.Fa., K. Pa., P Ku., and R.Ru obtained and reduced the data, and contributed analysis to the manuscript. All authors contributed to the interpretation of the data and commented on the final manuscript.

\item[Competing Interests] The authors declare no competing interests.

\item[Correspondence] Correspondence and requests for materials should be addressed to E.A. (email: eaydi@ttu.edu), and J.M. (email: monnier@umich.edu).

\end{addendum}



\clearpage

\clearpage

\clearpage

\beginsupplement
\spacing{1}
\noindent {\bf {\Huge Supplementary Information}}
\label{SI_1}

\noindent {\bf {\large SI 1. Nova V1674~Her}}

\noindent Nova V1674~Her (Nova Herculis 2021) was discovered by Seiji Ueda on 2021-06-12.5484\,UTC at a magnitude 8.4 in the visible
\cite{2021CBET.4976....1U,2021CBET.4977....1K}. Pre-discovery observations from the All-Sky Automated Survey for Supernovae (ASAS-SN; \cite{Shappee_etal_2014,Kochanek_etal_2017,Hart_etal_2023},) detected V1674~Her on 2021 June 12.19 (HJD 2459377.69) at $g=16.62$, around 8 hours before discovery. We will use this date as $t_0$ for V1674~Her throughout the manuscript. Spectroscopic follow-up observations by \cite{ATel_14704} and \cite{ATel_14710} confirmed that the transient is a Galactic classical nova. 

\noindent \textbf{Light curve evolution.} 
The optical light curve of V1674~Her is presented in Supplementary figure~\ref{Fig:V1674_Her_optical_LC}. The light curve shows a very rapid evolution, with the time for the nova to decline by 2 magnitudes from optical peak, $t_2 \approx 1.2$\,days \cite{Quimby_etal_2021}, making it one of the fastest novae on record \cite{Woodward_etal_2021,Sokolovsky_etal_2023}.  Photometric follow up of the system during the rapid decline from visible peak, revealed periodic signals at 0.152921(3)\,days and 501.486(5)\,s, which were suggested to be the orbital period and white dwarf spin period, respectively\cite{Patterson_etal_2022}. This implies that the underlying binary is an intermediate polar system \cite{Warner_1995}.

\noindent \textbf{Spectral evolution.} In Supplementary figures~\ref{Fig:V1674_Her_spec_1}, \ref{Fig:V1674_Her_spec_2}, and~\ref{Fig:V1674_Her_spec_3} we show the spectral evolution of nova V1674~Her during the first 50 days of the eruption. The first two spectral epochs, taken on the first day of the eruption show P Cygni profiles of Balmer, He I, and N III, typical for a nova during the rise to peak -- the early He/N phase (see \cite{Williams_1992,Williams_2012,Aydi_etal_2023b}). A couple of days later, as the nova starts declining from peak, the P Cygni profiles turn into broad emission lines, evolving into the second He/N phase. During the first few days, we also detect weak Fe II lines of the (42) and (48) multiplets, meaning that the nova went through a very rapid Fe II phase (and iron curtain phase) as it transitioned from the first to the second He/N phases, typical for very fast novae \cite{Aydi_etal_2023b}.

In Supplementary figure~\ref{Fig:V1674_Her_Halpha_line_profile} we present the line profile evolution of the H$\alpha$ Balmer line during the first two weeks of the eruption. In the first two spectral epochs (during the quick rise to peak), H$\alpha$ and other Balmer lines were characterized by P Cygni profiles with absorption troughs at around $-3800$\,km\,s$^{-1}$. A day after, a broad emission component emerges with velocities extending to $-5500$\,km\,s$^{-1}$, co-existing with the lower velocity components for a short time, before the absorption fades relative to the strong emission (see Supplementary figure~\ref{Fig:spec_zoom} for a zoom-in plot on the base of the emission line profiles of both H$\alpha$ and H$\beta$ to see the velocities of the fast components). These are high velocities in comparison to other classical novae, which is consistent with the rapid evolution in the light curve. These velocities and the rapid evolution of the nova made it an ideal target to be resolved by CHARA during the first few days of the eruption. Thereafter, the spectra were dominated by broad emission lines, showing some decelaration --- likely apparent due to decreased emissivity of the ejecta at the base of the lines (i.e., the highest-velocity material becoming too dispersed).

We use the high-resolution spectra of V1674~Her to estimate the reddening to the nova by using absorption features from diffuse interstellar bands (DIBs), which are a good proxy for the interstellar reddening. We use absorption features at 5487.7, 5780.5, 5797.1, 6196.0, 6204.5, 6283.8, and 6613.6\,\AA{} (see Supplementary figure~\ref{Fig:V1674_Her_DIBs}) and the relations from \cite{Friedman_etal_2011} to derive an average $E(B-V) \simeq 0.63\pm0.1$ and $A_V = 1.9\pm0.3$ (assuming $R_V = 3.1$). This value is comparable to the one derived by \cite{ATel_14723}. We use the reddening values to derive a distance estimate to the nova by placing it in the Galactic reddening maps of \cite{Chen_etal_2019}. To do so, we transform $E(B-V) \simeq 0.63\pm0.1$ into $E(G-Ks) \simeq 1.35\pm0.1$, $E(B_p-G_r) \simeq 0.78\pm0.1$, and $E(H-K_s) \simeq 0.10\pm0.05$, using the extinction law from \cite{Wang_etal_2019}. The maps of \cite{Chen_etal_2019} only extend to around 4.5 to 5\,kpc in the direction of V1674~Her. The reddening values we derive all place the nova at a distance $d \gtrsim 5\,kpc$. This value is consistent with other values from the literature. \cite{Sokolovsky_etal_2023} adopted a distance to the nova of 6.3\,kpc based on the quiescent brightness of the system and its intermediate polar nature. \cite{Craig_etal_2025b} estimated a distance of $6.3^{+2.1}_{-3.5}$ kpc based on Galactic reddening maps combined with the Galactic stellar distribution. \cite{Habtie_etal_2024} adopted a distance of around 5.24$\pm0.35$\,kpc based on the maximum magnitude--rate of decline relation (MMRD).
\cite{Woodward_etal_2021} estimated a distance of around 4.8$\pm1.3$\,kpc also based on a combination of MMRD relations and reddening values. Note that the MMRD relations are now established to be unreliable in deriving accurate distances to novae \cite{Shara_etal_2017_apr,Schaefer_2018}. The \emph{Gaia} parallax of the system is negative and suffers from large uncertainty. Therefore, the \emph{Gaia} distance is not reliable, however, the \emph{Gaia} EDR3 constraint with the priors of \cite{Bailer-Jones_etal_2021} places the nova at a distance of $6.0^{+3.8}_{-2.8}$\,kpc. All of these suggest that the distance to the nova is $\sim$ 5.0--7.0\,kpc.

\noindent Does the observed expansion in the CHARA images agree with the distance estimates? We performed fitting to the CHARA data using different models (see below). The first model, which fits the extended structure with  ring-like structure (allowed to vary azimuthally in brightness), results in outer radii of the resolved ejecta of around 0.9 and 1.23\,mas in the first and second CHARA epochs, respectively. Using these values, and the radial velocity measured from the optical spectra for the fastest component (5500$\pm$ 500\,km\,s$^{-1}$), an inclination of 67 degrees\cite{Habtie_etal_2024}, and expansion duration of around 1.7$\pm$0.5 days (for the first epoch), we derive a distance based on the expansion parallax of around 5.5$\pm$1.7\,kpc (we assume that the fast flow started expanding around the time the nova was detected emitting $\gamma$-rays, which started around $t = 0.5\pm0.5$ days). If we do the same for the second epoch (expansion duration of 2.7$\pm$0.5 days and an angular radius of 1.49\,mas), we derive a distance of around 6.4 $\pm$ 1.2\,kpc, using the expansion parallax. These distances are in agreement with the distances discussed above, arguing that the measured size of the ejecta, their morphology, and their velocity are a good representation of reality.

\cite{Habtie_etal_2024} used optical spectroscopy taken during the first few days/weeks of the eruption to perform morpho-kinematical modeling of the ejecta of nova V1674~Her. The models they obtained show striking resemblance to the CHARA images, with equatorial and bipolar outflows moving at perpendicular directions (see figure~14 in \cite{Habtie_etal_2024}), offering additional evidence for the validity of the structures resolved in the CHARA data.  Based on their models, \cite{Habtie_etal_2024} derived an inclination of 67 degrees (which we use above) and have calculated the angular radii of the polar and equatorial outflows during different epochs. Assuming a linear expansion, they derive a radius of 11.24 and 17.44 AU, at 1.7 and 2.7 days, respectively. Using the angular radii from \cite{Habtie_etal_2024}, a 67 degrees inclination, and the angular radius we measure from the CHARA images, we derive distances of around 10.7 and 10.8\,kpc, from the first and second epoch, respectively. This distance is larger than the distances derived above and the ones in the literature. There are multiple reasons to explain the larger distance estimates: (1) the expansion velocity law adopted in \cite{Habtie_etal_2024} is assumed to be the same as that of RS~Oph (from \cite{Ribeiro_etal_2009}). While RS~Oph and V1674~Her are comparably fast novae, they are still distinct novae with different expansion velocities. Moreover, RS~Oph is a nova taking place in a symbiotic system while V1674~Her is a nova in an intermediate polar system; (2) the size of the NIR continuum photosphere is different than the size of the region responsible for the optical emission lines; (3) the size of the ejecta measured in the CHARA images is underestimated, mostly due to a drop in the emissivity from the outer regions of the ejecta where the density is lower compared to the inner regions and/or due to the sensitivity of the instrument.

\noindent \subsection{Fitting the CHARA data of V1674~Her.} The fitting of the CHARA data in the uv-plane was performed using the \textsc{PMOIRED} Python3 tool (which can be found at \url{https://github.com/amerand/PMOIRED}) developed by Antoine M\'erand \cite{2022SPIE12183E..1NM,2022ascl.soft05001M}. 
\textsc{PMOIRED} can perform a parametric modelling of spectro-interferometric OIFITS data.
This is done by adopting blocks of possible components (disks, rings, Gaussians, with their highly developed parameters) which can be combined together to describe even a complex object geometry. 

Qualitative assessment of the interferometric data guides the modeling: the object is partially resolved as it shows a first null in visibility amplitude (Supplementary figures~\ref{fig:PMOIRED_V1674_Her_1} and ~\ref{fig:PMOIRED_V1674_Her_1}). 
The rebound at longer baselines is higher than expected for a uniform disk, indicating the geometry is closer to a ring. 
Moreover, there is small azimuthal variations in visibility amplitude, indicating that the object is slightly oblate. 
Finally, the closure phase is consistent with zero within the measurement uncertainties, indicating no significant departure from central symmetry. 
We tested two different models. The first is a central unresolved object surrounded by a thin elongated ring with first and second order azimuthal variations. 
The whole image is smoothed by a 0.33 mas Gaussian (corresponding to $\sim\lambda/3B_\mathrm{max}$). 
The second model we considered is a collection of three elongated Gaussian ``blobs'', two of which can be anywhere in the field of view (Supplementary figures~\ref{fig:PMOIRED_V1674_Her_1} and~\ref{fig:PMOIRED_V1674_Her_2}). The ring model has 6 free parameters, whereas the three-Gaussians model has 15. Both models lead to reasonable fits ($\chi^2_r\sim1.7$ for the June 14 epoch, $\chi^2_r\sim1.0$ for the June 15 one; see Table~\ref{tab:PMOIRED_V1674_Her}) and qualitatively similar images. There is a clear NE-SW elongation in both epochs, as well as a clear increase in size between June 14 and June 15. To quantify the increase in size, we used different approaches: for the smooth ring model, we use the outer radius in the elongation direction, and for the 3-Gaussians models we use the average separation of the 2 outer Gaussians. These two radii definitions will lead to substantially different values, since the outer radius of the ring is roughly twice the size of the position of the Gaussian blobs which trace where most of the flux is emitted. To compare both models, we also compute the half-light radius, as \cite{Chrenko_etal_2024} have shown this is a robust interferometric observable. Uncertainties are computed using data resampling (bootstrapping). We also compute the position angle of the major axis of the synthetic images, using the algorithm from \cite{Biernaux_etal_2016}. We will use the outer radii derived from the ring-like fit to the extended outflow to estimate the distance to the nova using expansion parallax, since the outer radius is likely more representative of the size of the ejecta, compared to the half radii derived from the Gaussian fit. The quantitative parameters derived from the synthetic images fit to the two epochs (Supplementary table~\ref{tab:PMOIRED_V1674_Her}) are consistent between the 2 models: the position angles are consistent, and the growth in size is fairly linear since the eruption of June 12.55.

\noindent \subsection{Does the strong magnetic field of the white dwarf play a role in shaping the nova ejecta?}
In an intermediate polar system, the white dwarf possesses a moderately strong magnetic field that can disrupt the inner accretion flow, channeling material along magnetic field lines toward the magnetic poles (see \cite{Warner_1985a} for a review). While many novae have been observed to take place in intermediate polar systems (e.g., \cite{Woudt_Warner_2003,Woudt_Warner_2004,Aydi_etal_2018}), the impact of the magnetic field of the white dwarf on the ejecta morphology and shaping is not yet constrained. While during a nova eruption, the magnetic influence may extend to the ejected envelope, introducing anisotropies in the outflow (e.g., the magnetic field can inhibit expansion along the magnetic equator while facilitating polar ejection, leading to a bipolar morphology), this cannot be confirmed with the data presented here and would be considered an open question for future studies to tackle.

\begin{table}[h!]

\def\arraystretch{1.0}
\begin{tabular}{lcccc}
\hline
\hline

\rule{0pt}{2ex} UT Date & Telescopes & Spectral Mode & Target & Calibrators\\
\hline
2021 Jun 14 & E1-W2-W1-S2-S1-E2 & PRISM102 & V1674~Her & HD 184607\\
2021 Jun 15 & E1-W2-W1-S2-S1-E2 & PRISM102 & V1674~Her & HD 177433\\
\hline
\end{tabular}
\caption{Log of CHARA observations of nova V1674~Her. The first column is the date of the observations. The second column represents the CHARA units used in the observation. The third column is the spectral mode. The last two columns are the science target and the calibrator sources.}
\label{table:CHARA_log_V1674_Her}
\end{table}
\vspace{1cm}

\begin{table}[h!]
    \centering
    \begin{tabular}{lcc}
    \hline
    \hline
          & June 14 & June 15 \\
         \hline
         \multicolumn{3}{c}{3-Gaussians} \\
         \hline
         $\chi^2_r$ & 1.71 & 1.04\\
         separation (mas) & $0.32\pm0.06$  & $0.79\pm0.03$ \\
         HLR  (mas) & $0.48\pm0.03$ & $0.71\pm0.03$\\
         PA$^{(1)}$ (deg) & $41\pm9$ & --- $^{(2)}$ \\ 
         \hline

         \multicolumn{3}{c}{smooth ring} \\
         \hline
         $\chi^2_r$ &  1.72  & 0.97 \\    
         outer radius (mas) & $0.90\pm0.07$ & $1.23\pm0.11$ \\
         HLR (mas) & $0.46\pm0.03$ & $0.71\pm0.06$ \\
         PA (deg) & $47\pm6$ & $38\pm12$ \\ 
         \hline
         
    \end{tabular}
    \caption{Parameters of the analytic models fit using \textsc{PMOIRED}. "separation" is the average radial position of the 2 outer Gaussians, "HLR" is the half-light radius. Based on the respective definitions, "outer radius" is expected to be about twice "separation" and "HLR". Notes: $^{(1)}$ defined as 0 for North and 90 for East. $^{(2)}$ algorithm does not converge.}
    \label{tab:PMOIRED_V1674_Her}
\end{table}

\clearpage

\begin{figure*}
\begin{center}
\includegraphics[width=\textwidth]{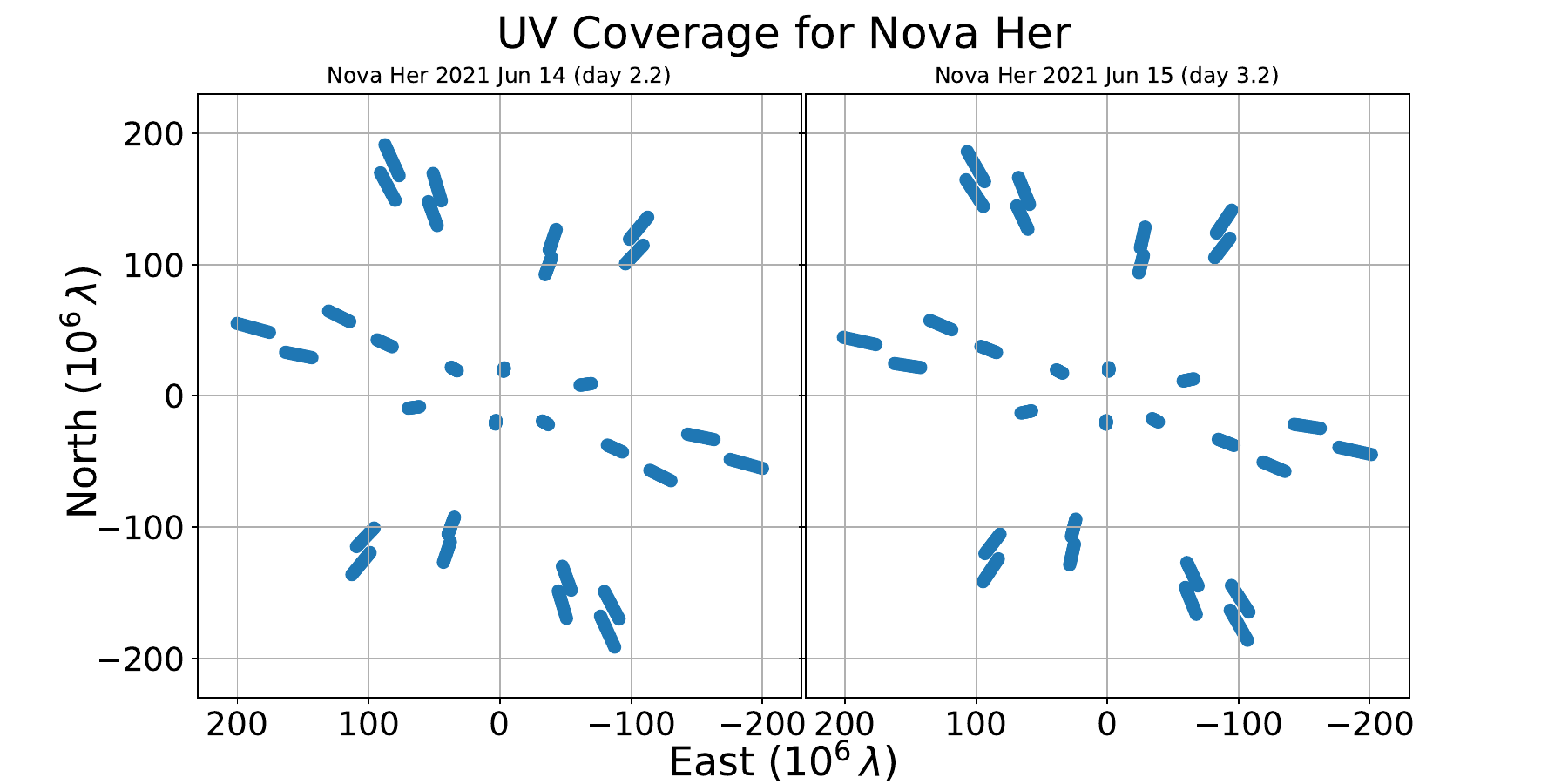}
\caption{The CHARA (u,v) coverage of interferometric baselines projected on the  plane of the sky in right ascension (RA) and declination (Dec.), during the observations of nova V1674~Her. \textit{Left} panel is the 2021 June 14 epoch (day 2.2) and \textit{right} panel is the 2021 June 15 epoch (day 3.2).}
\label{Fig:UV_V1674_Her}
\end{center}
\end{figure*}

\begin{figure*}
\begin{center}
\includegraphics[width=0.6\textwidth]{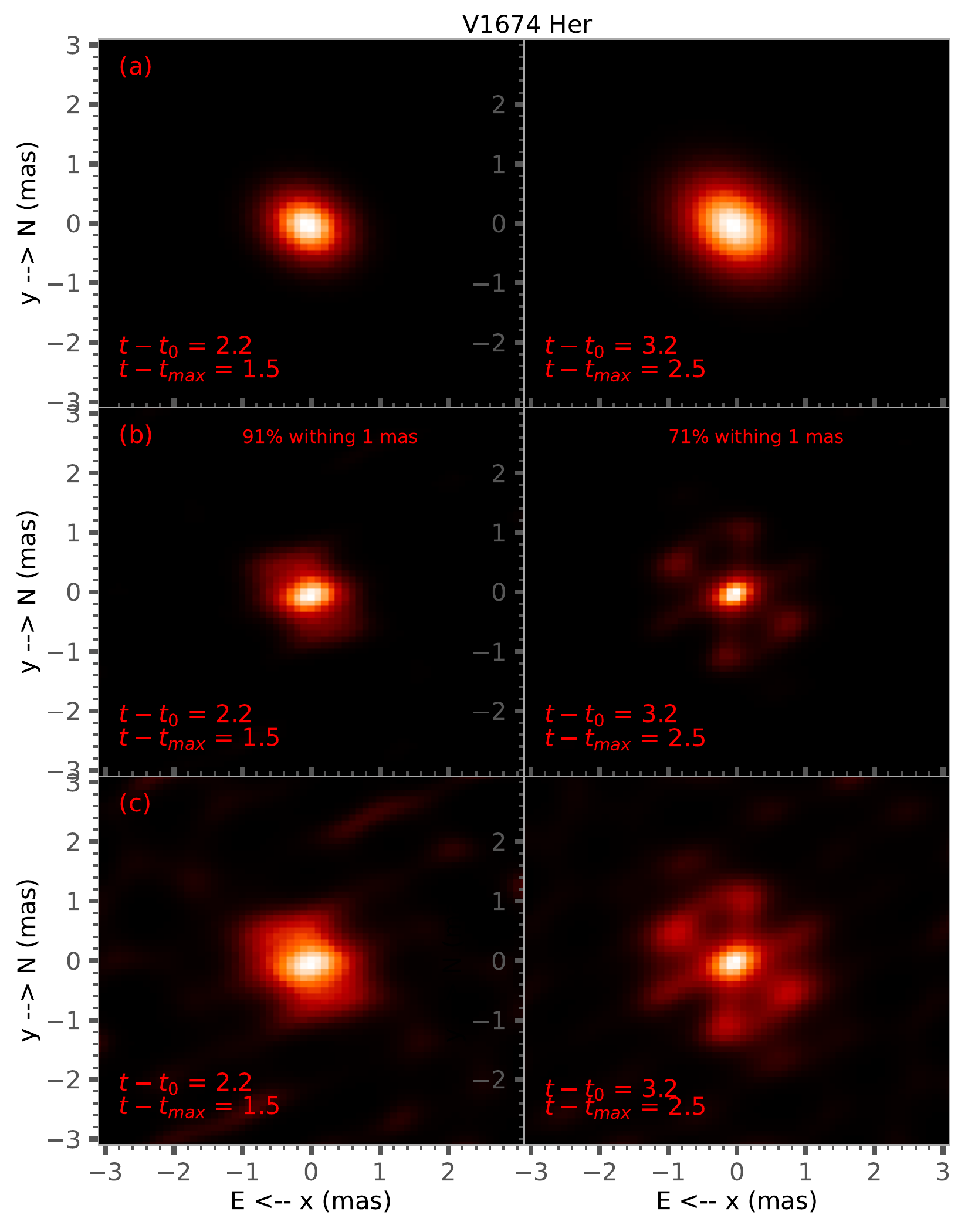}
\vspace{-0.5cm}
\caption{\textbf{The prior (top) and \textsc{BSMEM} images of nova V1674~Her.} \textit{Top panels}: the prior images used based on an elliptical Gaussian fit to the visibility data. \textit{Middle panels}: the \textsc{BSMEM} images using a linear scale. \textit{Bottom panel}: the \textsc{BSMEM} images using square-root intensity, to highlight low-surface brightness emission within the field-of-view.}
\label{Fig:methods_images_1}
\end{center}
\end{figure*}

\begin{figure*}[h!]
\begin{center}
\includegraphics[width=0.8\textwidth]{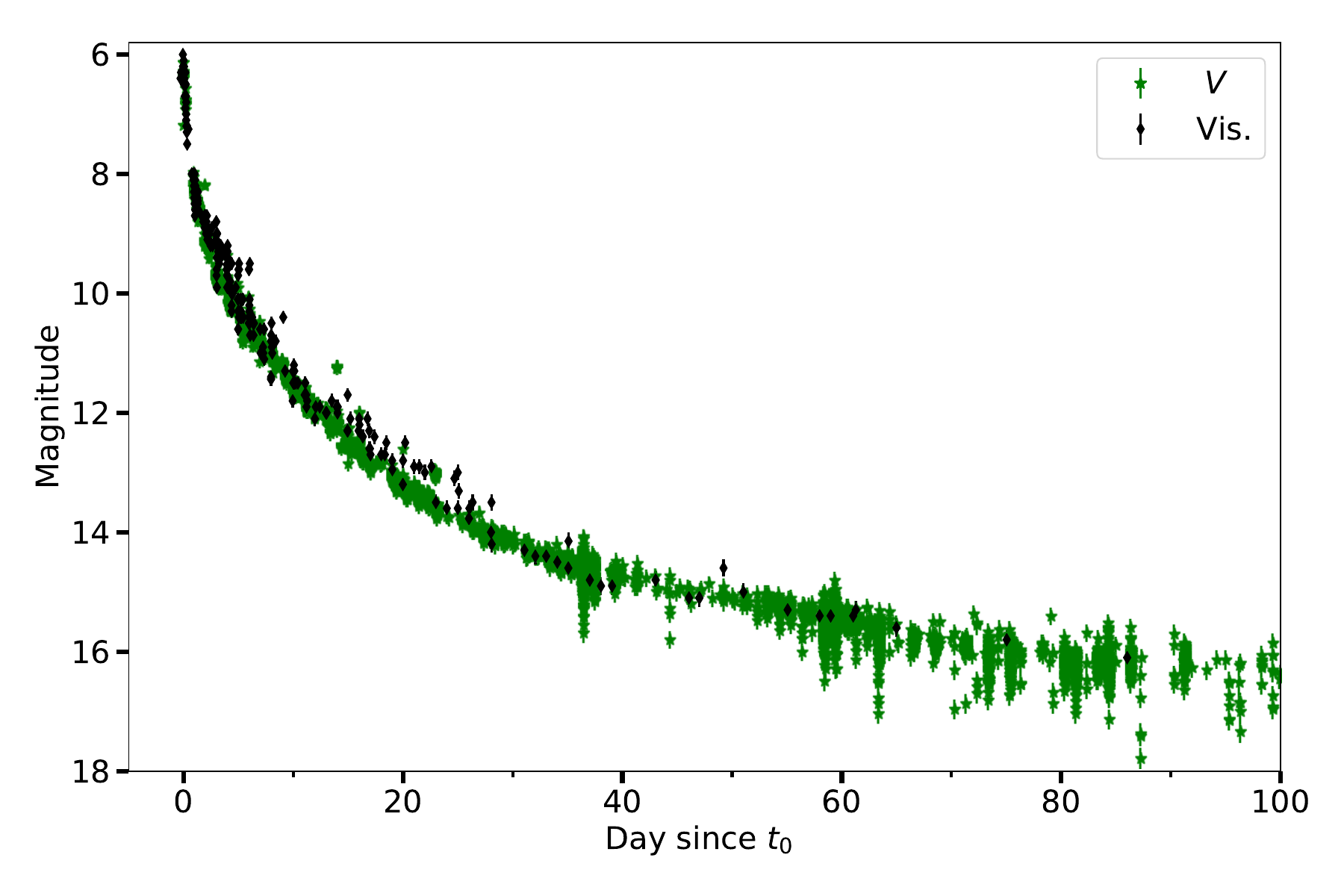}
\caption{The AAVSO optical light curve of nova V1674~Her during the first 100 days of the eruption \cite{AAVSODATA}. The green stars are $V$-band measurements while the black dots are visutal estimates. The error bars represent 1-$\sigma$ uncertainties.}
\label{Fig:V1674_Her_optical_LC}
\end{center}
\end{figure*}


\begin{figure*}
\begin{center}

\includegraphics[width=\textwidth]{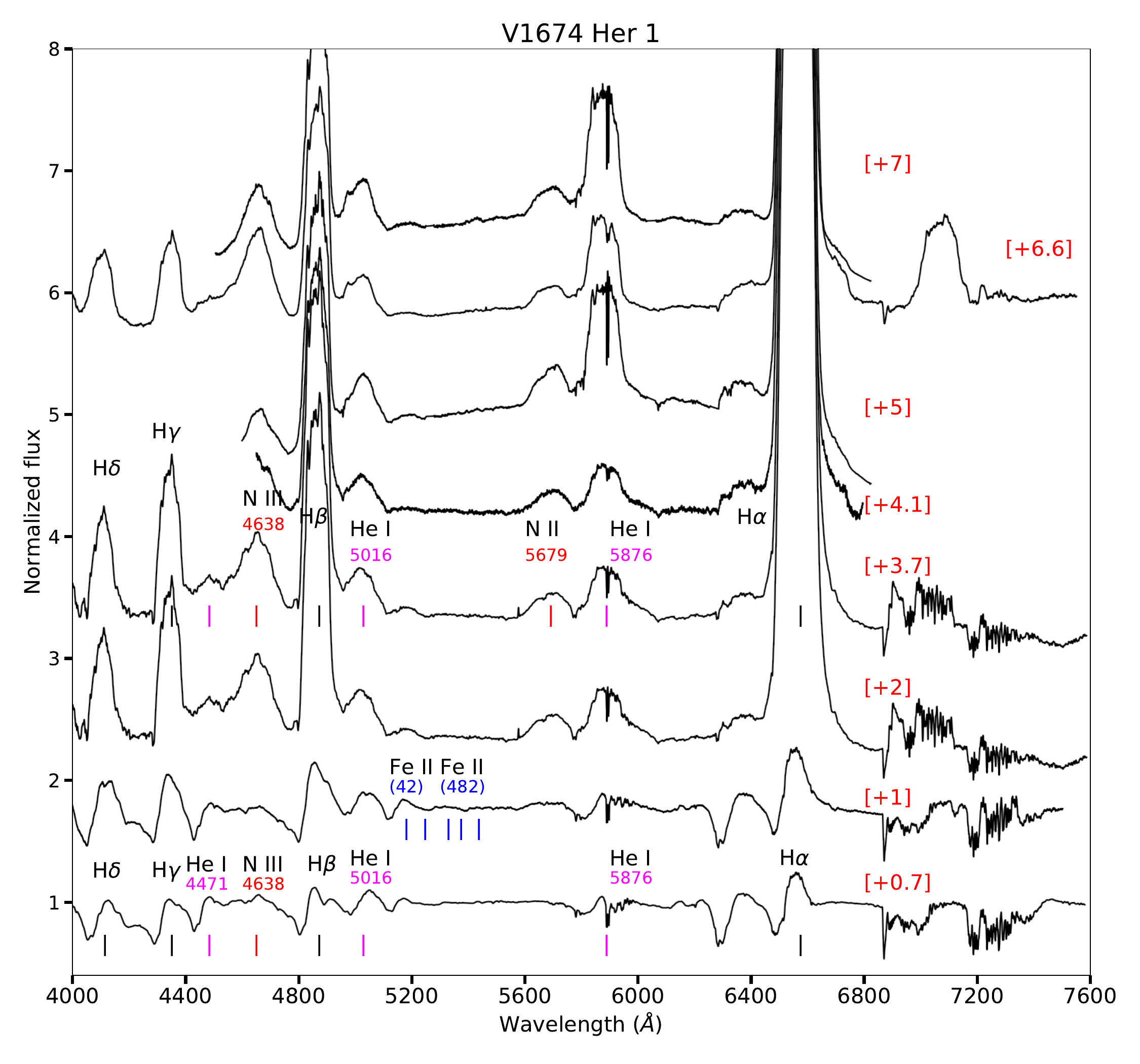}
\caption{\textbf{The spectral evolution of nova V1674~Her (part 1)}. The numbers between brackets are days since $t_0$. Line identifications are presented with tick marks under the lines for easier identification and they are color
coded based on the line species.}
\label{Fig:V1674_Her_spec_1}
\end{center}
\end{figure*}

\begin{figure*}
\begin{center}
\includegraphics[width=\textwidth]{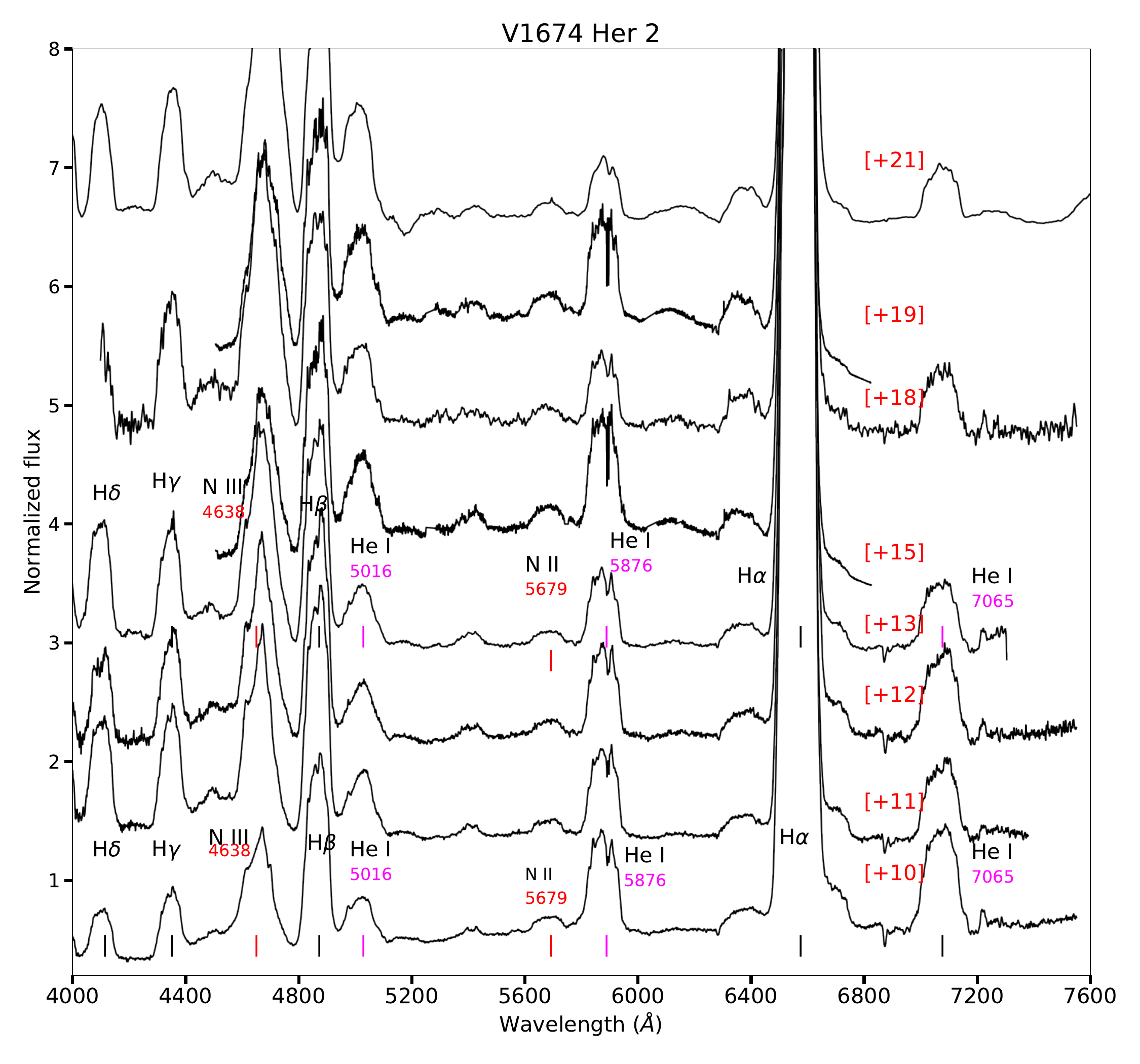}
\caption{\textbf{The spectral evolution of nova V1674~Her (part 2)}. The numbers between brackets are days since $t_0$. Line identifications are presented with tick marks under the lines for easier identification and they are color
coded based on the line species.}
\label{Fig:V1674_Her_spec_2}
\end{center}
\end{figure*}

\begin{figure*}
\begin{center}
\includegraphics[width=\textwidth]{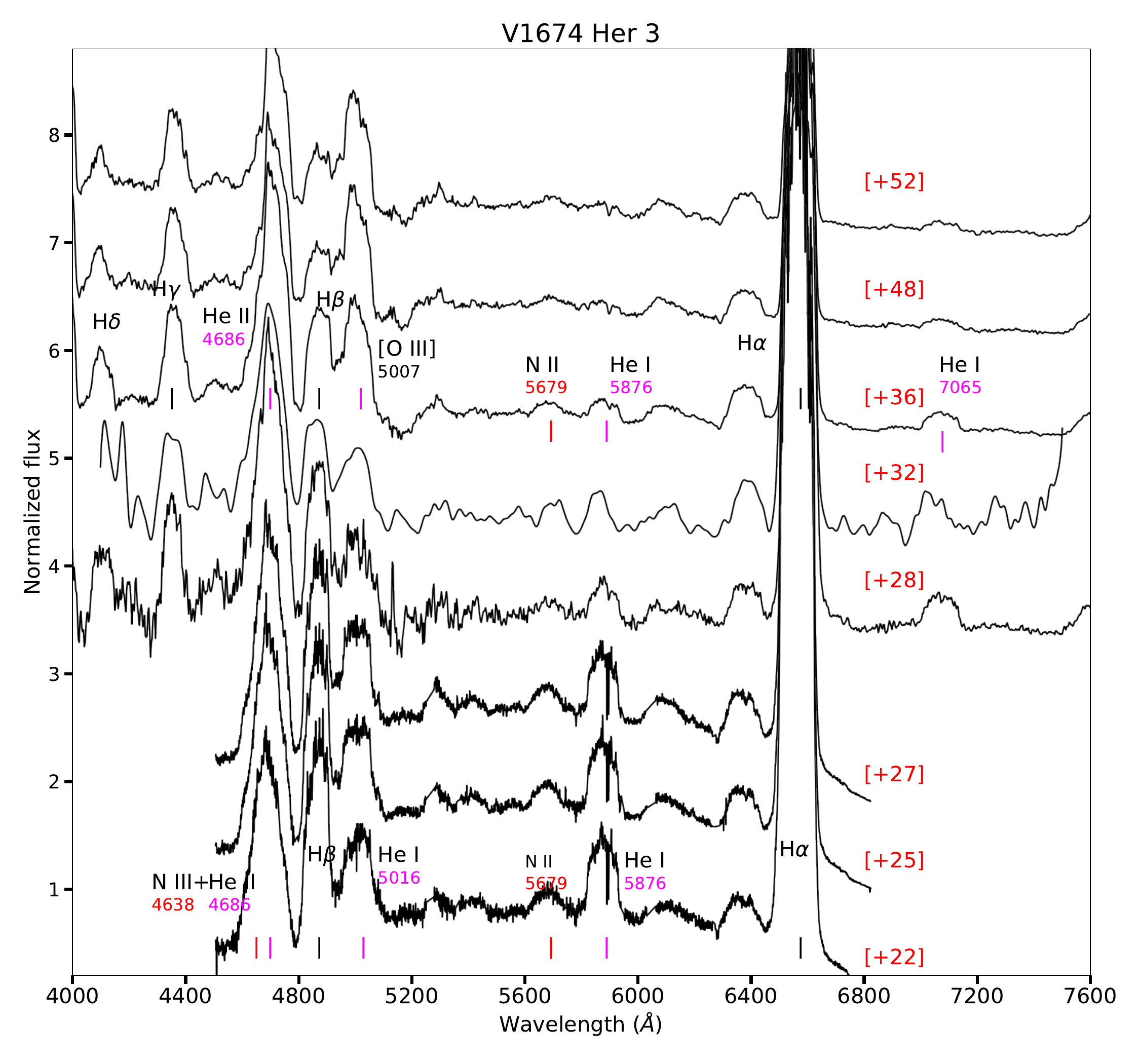}
\caption{\textbf{The spectral evolution of nova V1674~Her (part 3)}. The numbers between brackets are days since $t_0$. Line identifications are presented with tick marks under the lines for easier identification and they are color
coded based on the line species.}
\label{Fig:V1674_Her_spec_3}
\end{center}
\end{figure*}


\begin{figure*}
\begin{center}

\includegraphics[width=\textwidth]{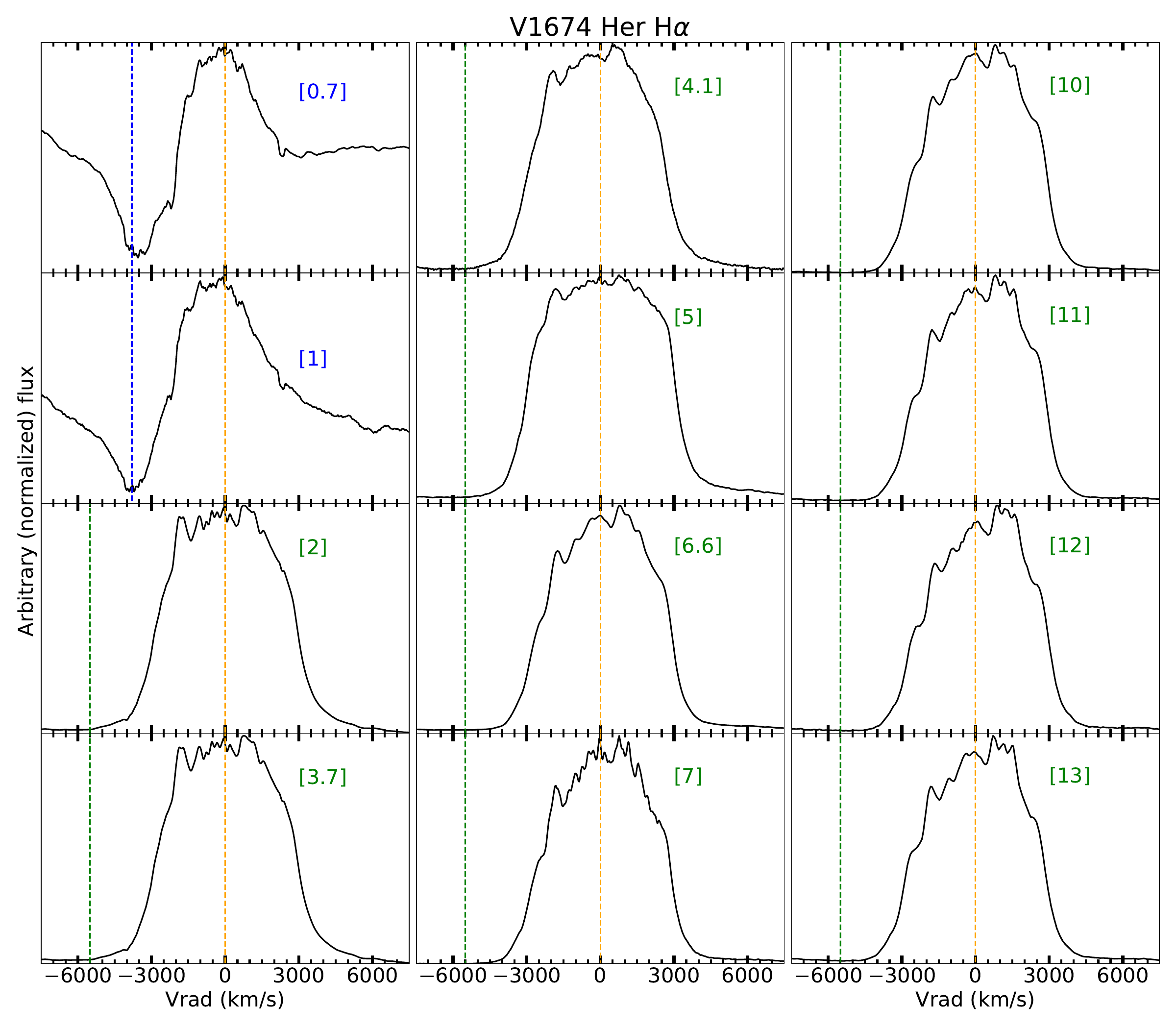}
\caption{The line profiles of H$\alpha$ for nova V1674~Her during the first $\approx$ 2 \textbf{weeks} of the eruption. The orange, blue, and green dashed lines represent $v_{\mathrm{rad}}$ = 0\,km\,s$^{-1}$, $v_{\mathrm{rad}}$ = $-3800$\,km\,s$^{-1}$, $v_{\mathrm{rad}}$ = $-5500$\,km\,s$^{-1}$, respectively. The numbers between brackets are days since $t_0$.}
\label{Fig:V1674_Her_Halpha_line_profile}
\end{center}
\end{figure*}


\begin{figure*}
\begin{center}

\includegraphics[width=0.75\textwidth]{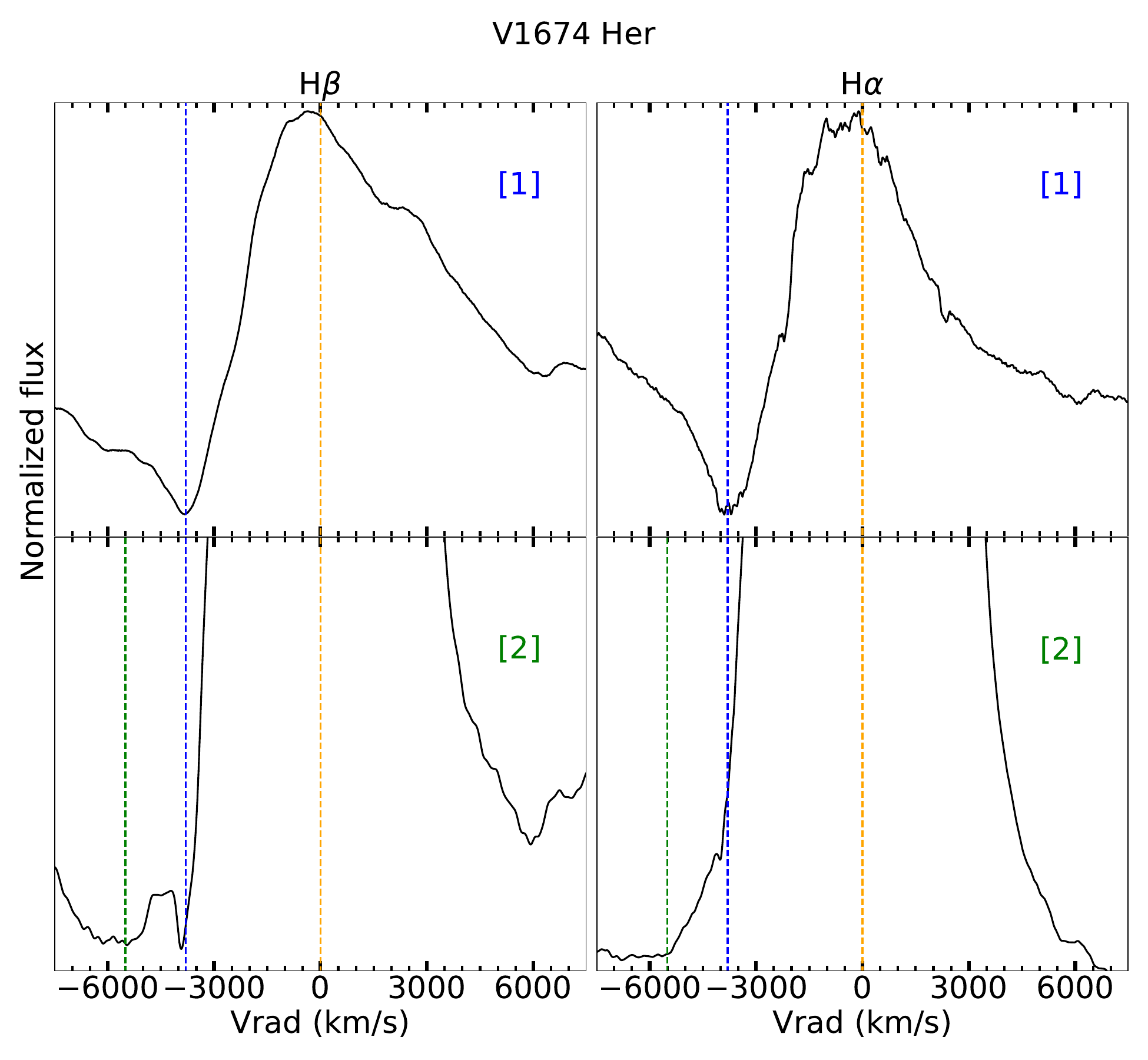}
\caption{A zoom-in view of the line profile evolution of the Balmer lines during the first 2 days of the eruption, with a focus on the absorption lines at the base of the spectra. The orange, blue, and green dashed lines represent $v_{\mathrm{rad}}$ = 0\,km\,s$^{-1}$, $v_{\mathrm{rad}}$ = $-3800$\,km\,s$^{-1}$, $v_{\mathrm{rad}}$ = $-5500$\,km\,s$^{-1}$, respectively. \textbf{The numbers between brackets are days since $t_0$.}}
\label{Fig:spec_zoom}
\end{center}
\end{figure*}


\begin{figure*}
\begin{center}

\includegraphics[width=0.75\textwidth]{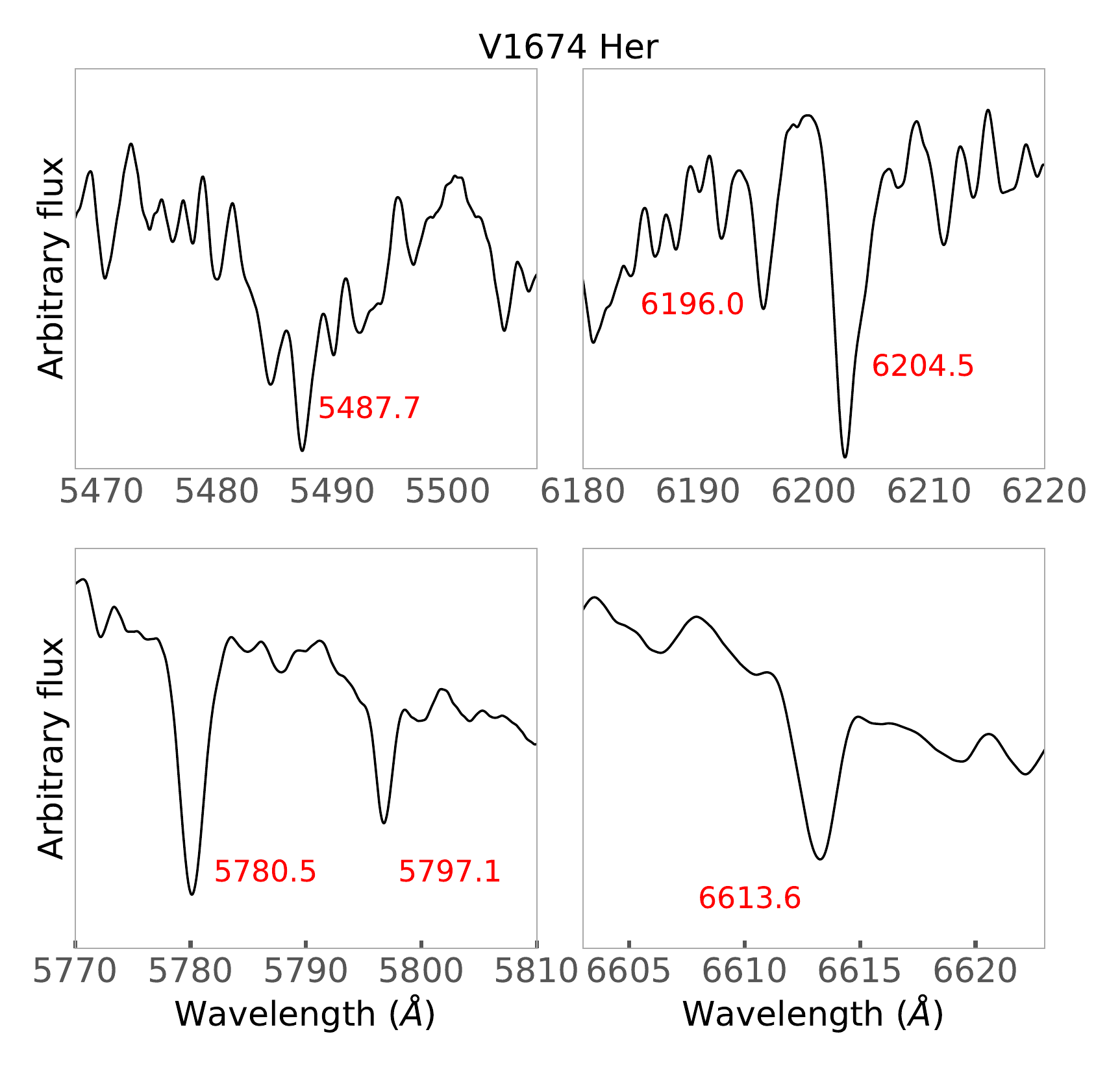}
\caption{A sample of the absorption lines from diffuse interstellar bands used to derive the reddening towards V1674~Her. The numbers in red are the wavelength of the diffuse interstellar bands asborption lines.}
\label{Fig:V1674_Her_DIBs}
\end{center}
\end{figure*}

\begin{figure}
    \centering
\vspace{-1.5cm}
    \includegraphics[width=0.9\linewidth]{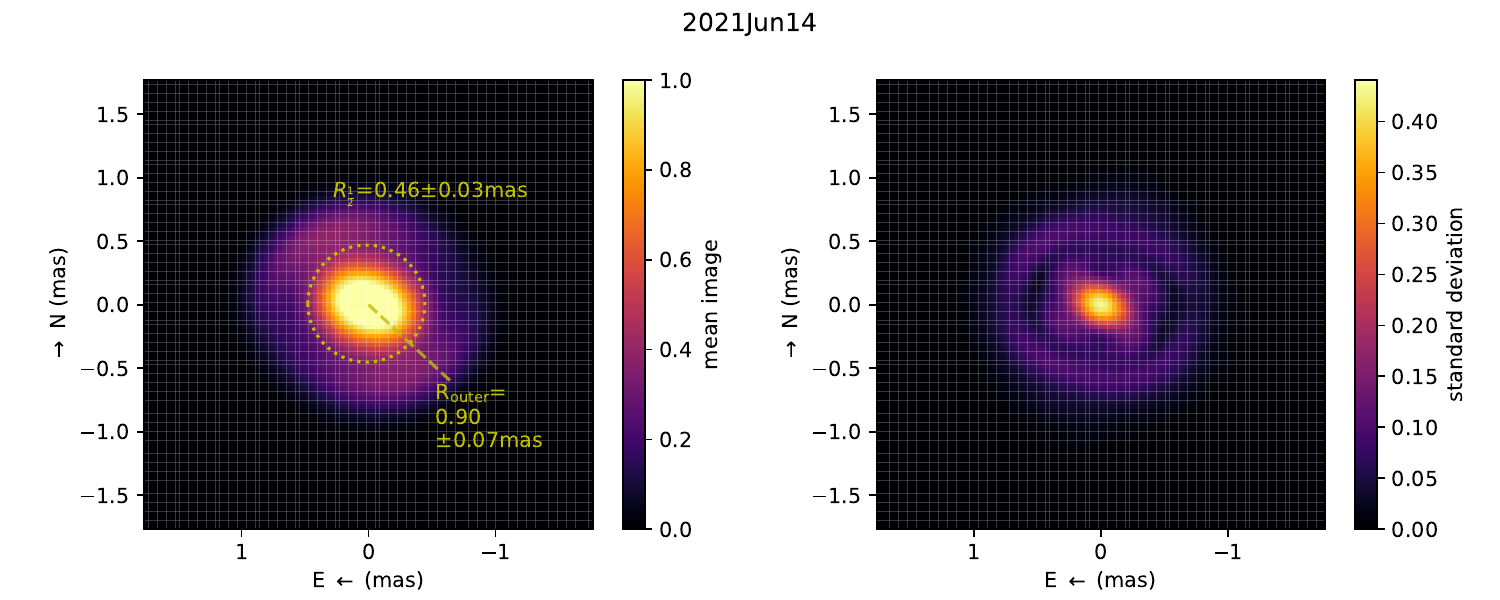}

    \includegraphics[width=0.9\linewidth]{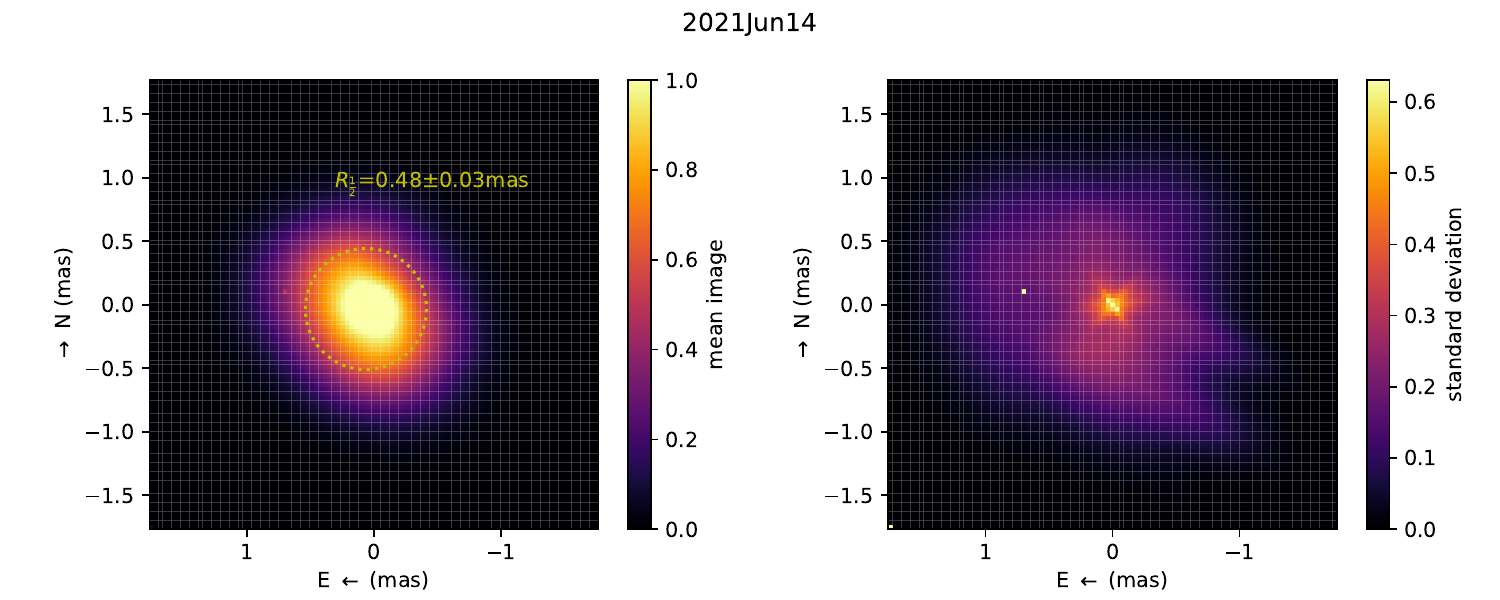}

    \includegraphics[width=0.7\linewidth]{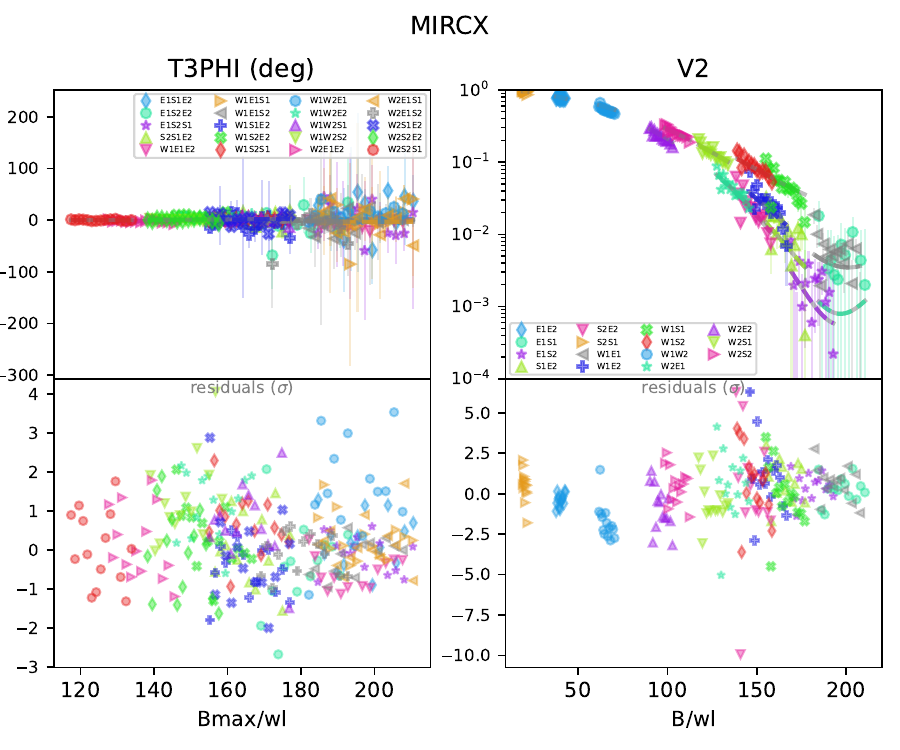}

    \caption{\footnotesize{Synthetic images of V1674 Her, based on \textsc{PMOIRED} models (for 2021 June 14 epoch; day 2.2). The first row corresponds to the smooth ring model, while the second row corresponds to the 3 elongated Gaussians model. For the first and second rows, the first column show the synthetic images, whereas the second column show the corresponding variance in the image, based on the noise in the data. For each model, the half-light radius is drawn. The bottom plots correspond to the fitness to the data, comparing closure phase (first colums) and squared visibilities (second column) data (points) and models (solid lines). The bottom panels are the residuals to the fit, normalized to the uncertainties in the data.}}
    \label{fig:PMOIRED_V1674_Her_1}
\end{figure}

\begin{figure}
    \centering
\vspace{-1.5cm}
    \includegraphics[width=0.9\linewidth]{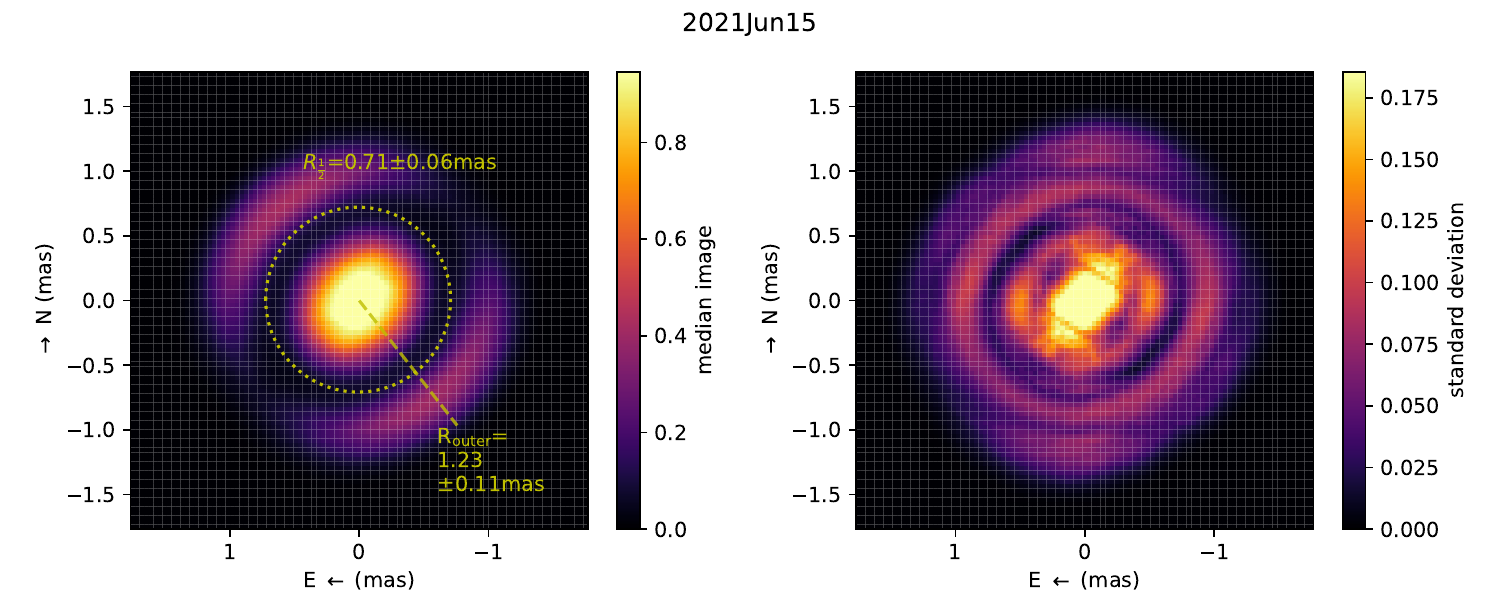}

    \includegraphics[width=0.9\linewidth]{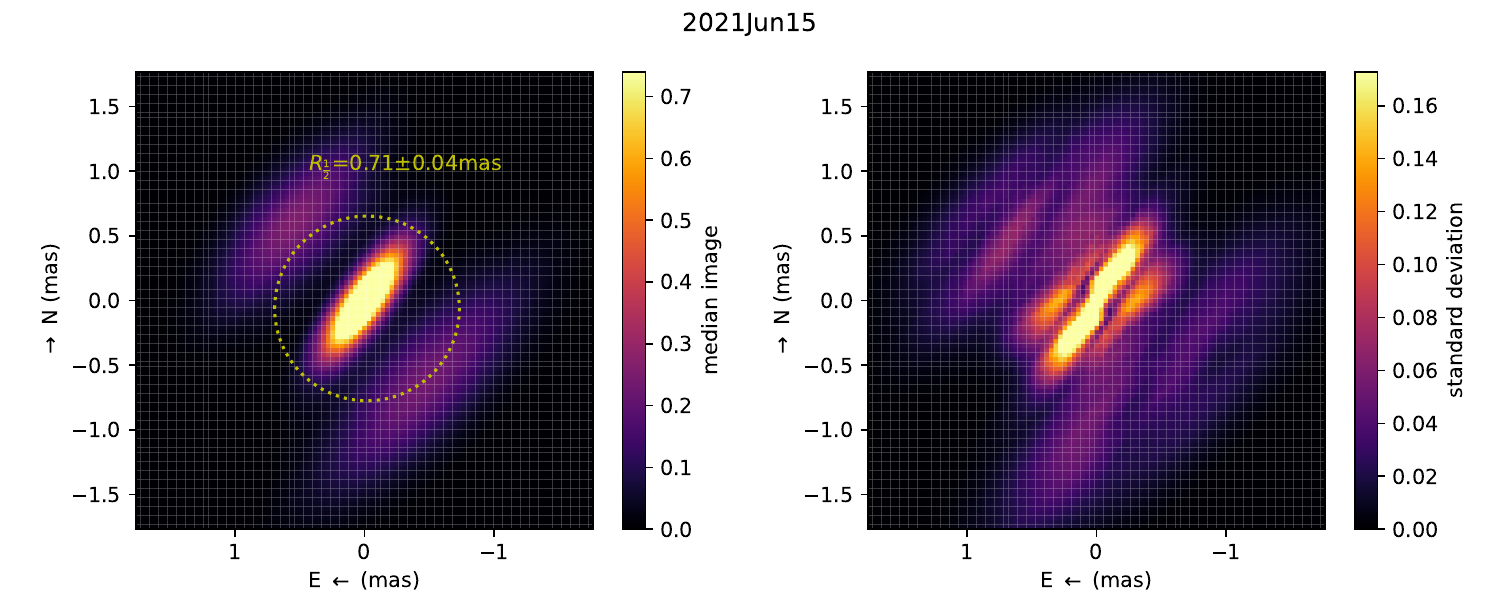}

    \includegraphics[width=0.7\linewidth]{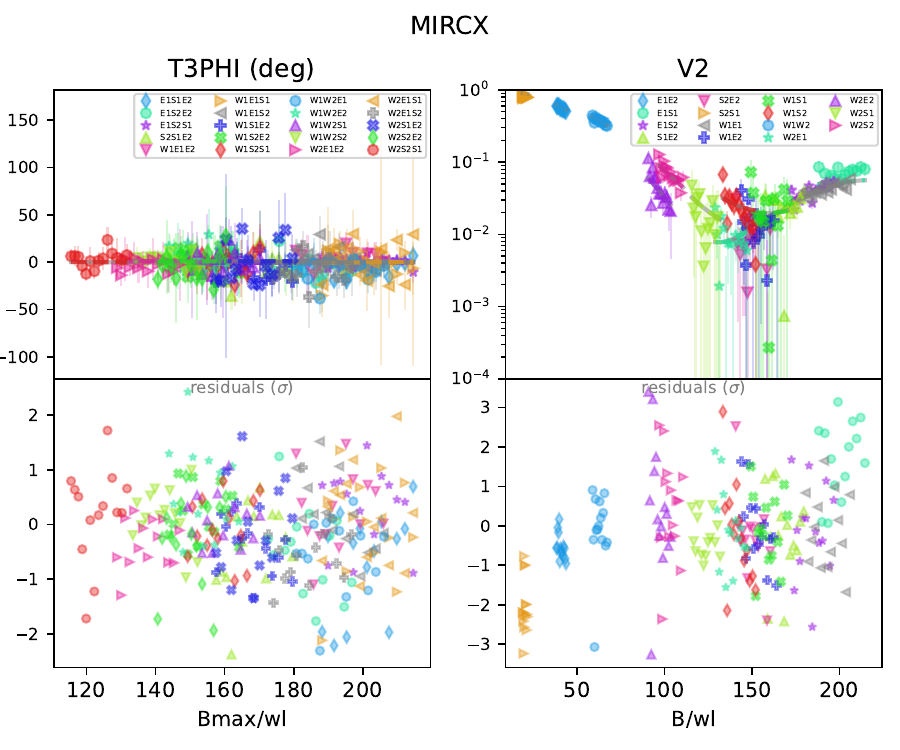}
    
    \caption{\footnotesize{Synthetic images of V1674 Her, based on \textsc{PMOIRED} models (for 2021 June 15 epoch; day 3.2). The first row corresponds to the smooth ring model, while the second row corresponds to the 3 elongated Gaussians model. For the first and second rows, the first column show the synthetic images, whereas the second column show the corresponding variance in the image, based on the noise in the data. For each model, the half-light radius is drawn. The bottom plots correspond to the fitness to the data, comparing closure phase (first colums) and squared visibilities (second column) data (points) and models (solid lines). The bottom panels are the residuals to the fit, normalized to the uncertainties in the data.}}
    \label{fig:PMOIRED_V1674_Her_2}
\end{figure}
\clearpage

\begin{table*}[t]

\centering
\def\arraystretch{1.0}
\begin{tabular}{lcccc}
\hline
\hline
\rule{0pt}{2ex} Date & $t-t_0$ & Source & Range & $R$\\
(UT date) & (days) & & $\AA$ & \\
\hline
2021-06-12 &   0.7 &   ARAS & 3900\,--\,7500 &       10,000 \\
2021-06-13 &   1.0 &   ARAS & 3900\,--\,7500 &       10,000 \\
2021-06-14 &   2.0 &   ARAS & 3900\,--\,7500 &       10,000 \\
2021-06-15 &   3.7 &   ARAS & 3900\,--\,7500 &       10,000 \\
2021-06-16 &   4.1 &   ARAS & 3900\,--\,7500 &       10,000 \\
2021-06-17 &   5.0 & Gemini & 4500\,--\,6800 &        4,400 \\
2021-06-18 &   6.6 &   ARAS & 3900\,--\,7500 &        1,000 \\
2021-06-19 &   7.0 & Gemini & 4500\,--\,6800 &        4,400 \\
2021-06-21 &  10.0 &   ARAS & 3900\,--\,7500 &        1,000 \\
2021-06-22 &  11.0 &   ARAS & 3900\,--\,7500 &       10,000 \\
2021-06-23 &  12.0 &   ARAS & 3900\,--\,7500 &       10,000 \\
2021-06-25 &  13.0 &   ARAS & 3900\,--\,7500 &        1,000 \\
2021-06-27 &  15.0 & Gemini & 4500\,--\,6800 &        4,400 \\
2021-06-30 &  18.0 &   ARAS & 3900\,--\,7500 &        1,000 \\
2021-07-01 &  19.0 & Gemini & 4500\,--\,6800 &        4,400 \\
2021-07-03 &  21.0 &  SNIFS & 3400\,--\,9000 &        1,000 \\
2021-07-04 &  22.0 & Gemini & 4500\,--\,6800 &        4,400 \\
2021-07-07 &  25.0 & Gemini & 4500\,--\,6800 &        4,400 \\
2021-07-09 &  27.0 & Gemini & 4500\,--\,6800 &        4,400 \\
2021-07-10 &  28.0 &  SNIFS & 3400\,--\,9000 &        1,000 \\
2021-07-14 &  32.0 &   ARAS & 3900\,--\,7500 &        1,000 \\
2021-07-18 &  36.0 &  SNIFS & 3400\,--\,9000 &        1,000 \\
2021-07-31 &  48.0 &  SNIFS & 3400\,--\,9000 &        1,000 \\
2021-08-03 &  52.0 &  SNIFS & 3400\,--\,9000 &        1,000 \\
\hline
\end{tabular}

\caption{Log of spectral observations of V1674~Her. The first column represents the date of the observations, the second column is the time relative to discovery epoch ($t_0$). The third columnd represents the instrument/source of the data. The last two columns are the spectral range and the resolving power $R$. }
\label{table:spec_log_V1674_Her}
\end{table*}

\clearpage 

\noindent {\bf {\large SI 2. Nova V1405~Cas}}

\noindent Nova V1405~Cas was discovered by Yuji Nakamura (Mie, Japan) on 2021-03-18.42 UT at an unfiltered visible magnitude of 9.6 mag. 
The transient coincides with a known variable star CzeV3217 (UCAC4\,756-077930, Gaia\,EDR3\,2015451512907540480)\cite{Wischnewski_2022} characterized by a quiescent magnitude of around 15.6 in the $V-$band. 
Spectroscopic follow up observations by \cite{ATel_14471,ATel_14472,ATel_14476} classified the transient as a classical nova. Although the progenitor system was previously classified as a contact eclipsing binary of the W Ursae Majoris type (a contact binary consisting of two hot stars, e.g., A- or F-type; \cite{Wischnewski_2022}), the occurrence of a nova eruption demonstrates that this classification was incorrect, and the progenitor system is in fact a cataclysmic variable.

\noindent \textbf{Light curve evolution.} In Supplementary figure~\ref{Fig:V1405_Cas_optica_LC} we present the complete optical light curve of nova V1405~Cas throughout the different stages of the eruption. 
The nova reached a $V-$band magnitude $\approx$ 7.5\,mag a few days after discovery. Thereafter, the brightness-increase halted for a couple of weeks, before rising again to reach a peak in $V-$band of 5.1\,mag around 53 days into the eruption. After this first peak, the nova stayed bright ($\gtrsim$ 7\,--\,9\,mag above quiescence) for more than 200 days, showing multiple flares/maxima, particularly two major ones around days 120 and 180. After day 220, the nova slowly declined in brightness. The light curve is characterized by $t_2 \approx 165\pm5$ days, making it a very slow nova.

Several slow novae are observed to stay near visible peak for an extended period of time (weeks to months; \cite{Strope_etal_2010}), many of which also exhibit a long-lasting pre-maximum halt and a gradual rise to visible peak spanning several weeks (e.g., V723~Cas \cite{Munari_etal_1996}, V463 Sct \cite{Hachisu_Kato_2004}, V5558 Sgr \cite{Tanaka_etal_2011}, and V1548 Aql \cite{Kato_Takamizawa_2001}). The origin of this extended pre-maximum halt and gradual rise to peak brightness has not been fully understood yet. \cite{Orio_Shaviv_1993} suggested that a local thermonuclear runaway on the surface of the white dwarf could be the reason behind the slow rise to peak for some novae, as the propagation time of the runaway along the meridian may be as long as days to weeks. \cite{Hillman_etal_2014} attempted to model the light curves of novae in different bands, with the aim of recreating some of the complex features observed in nova light curves. Their models show halts during the rise to peak, similar to the ones observed in many novae. They associate the halt to a phase during the early expansion of the nova envelope where convection ceases to play a significant role near the envelope surface, before mass loss begins and radiation takes over, leading to a reverse of the halt. \cite{Kato_Hachisu_1994,Hachisu_etal_2014} suggested that in the case of massive accreted envelopes on lower mass white dwarfs (0.5 to 0.8\,M$_{\odot}$), the nova resembles a super-giant star during the early days of the eruption where the envelope radius expands to $\sim$ 100\,R$_{\odot}$ while dropping in temperature to around 7000\,K. According to \cite{Hachisu_etal_2014} ''during this stage the envelope experiences minor changes in its radius and temperature, particularly relative to its high mass, leading to a saturation in the visual magnitude of the nova, which lasts an extended period of time before it declines -- this saturation is the pre-maximum halt.'' In a more recent work, \cite{Munari_etal_2017} suggested that the early rise to peak is caused by the expanding nova envelope, known as the fireball phase. This is followed by a later, often more pronounced, peak in the visible brightness caused by shock interaction, which powers both luminous $\gamma$-ray and optical emission (e.g., \cite{Li_etal_2017_nature,Aydi_etal_2020a}). The shift from one dominant source of emission to another resembles a halt in the light curve. Some studies have suggested that during this early rise to visible peak, the binary motion might play a role in driving the mass loss and eventually expelling the nova envelope (e.g., \cite{Livio_etal_1990}), during a short but important phase of common envelope interaction.

In Supplementary figure~\ref{Fig:V1405_Cas_colors}, we present the $(B-V)_0$, $(V-R)_0$, and $(R-I)_0$ color evolution of the nova during the first 100 days of the eruption. During the pre-maximum halt, the $(B-V)_0$ colors show slight variation around zero value. As the nova rises to peak after day 50, the colors show moderate changes, with $(B-V)_0$ reaching 0.4 mag. The main change in colors occurs as the nova declines from peak, especially in $(V-R)_0$ and $(R-I)_0$ colors. These changes are mostly driven by the emergence of emission lines after peak and are heavily affected by the strong H$\alpha$ emissions in the $R$-band.  

\noindent  \textbf{Spectral evolution.} In Supplementary figures~\ref{Fig:V1405_Cas_spec_1} and~\ref{Fig:V1405_Cas_spec_2} we present the spectral evolution of nova V1405~Cas throughout the eruption. A detailed description of the spectral evolution of the nova is presented in \cite{Aydi_etal_2023b,Taguchi_etal_2023}. The spectra are initially dominated by P Cygni profiles of H I, He I, He II, N III, during the early He/N phase as the nova rises to peak. Similarly to the optical range, the infrared spectra taken on days 5 and 10 after discovery (48 and 43 days before peak visible brightness; Figure~\ref{Fig:V1405_IR_spec}) show P Cygni lines of He I and H I of the Paschen and Brackett series. Near peak, the spectrum shows a shift to the first Fe II phase with the spectrum being dominated by Balmer and Fe II low-ionization lines of the (42), (48), and (49) multiples. While the light curves show multiple peaks/maxima, the spectra also show oscillatory changes, being dominated by Balmer and Fe II lines during the rise to each peak, while shifting to a He/N spectrum during the decline from peaks. \cite{Aydi_etal_2023b} discuss this oscillatory behavior and associate it with changes in the opacity of the ejecta. 

In Supplementary figures~\ref{Fig:V1405_Cas_Halpha_line_profile}, \ref{Fig:V1405_Cas_Halpha_line_profile_2}, \ref{Fig:V1405_Cas_Halpha_line_profile_3}, and~\ref{Fig:V1405_Cas_Halpha_line_profile_4} we present the line profile evolution of H$\alpha$ during the first 260 days of the eruption. For the first 55 days, during the slow rise to peak, H$\alpha$ and other Balmer lines were characterized by P Cygni profiles with absorption troughs at blueshifted velocities at around 1500\,km\,s$^{-1}$, which rapidly decelerated as the nova climbed gradually to peak, with the absorption troughs reaching a velocity of around 700\,km\,s$^{-1}$ at peak (around day 53--55). This can be seen clearly in the 2D dynamic spectrum presented in Figure~\ref{Fig:V1405_Cas_Dynamic_spectrum}, where the dark features, representing absorption components, show a clear shift towards slower blueshifted velocities as the nova rise to peak brightness. Similar deceleration has been observe in several other novae, particularly ones with slow evolution where multiple spectra can be obtained for the nova as it climbs to peak (see figure 11 in \cite{Aydi_etal_2020b}). These authors suggested that the deceleration is apparent, and they attributed it to a change in the opacity of the ejecta where the photosphere is expanding outward in radius-space while receding inward in velocity-space --- in a Hubble-like outflow during the early expansion of the nova atmosphere (see as well \cite{Hachisu_Kato_2004,Shore_2014}). In the infrared, the He I and H I lines also show similar velocities to the ones in the optical lines during the rise to peak, with velocities of around 1300\,--\,1500\,km\,s$^{-1}$ measured from the absorption troughs of the P Cygni profiles. After the first peak, a broad emission emerges with a full width of zero intensity (FWZI) extending to around 4200\,km\,s$^{-1}$ (i.e., wings extending to around 2100\,km\,s$^{-1}$ on each side), while the pre-maximum P Cygni profiles are still superimposed on top of it. The emergence of the higher velocity emission profiles coincides with the appearance of extended ejecta in the CHARA images.

After the first peak, the nova showed multiple flares in the light curve, some of them characterized with an amplitude of more than 2 magnitudes. The last flare peaked on day around 220 before the nova started to gradually dim. This flaring activity is common in novae (e.g., \cite{Strope_etal_2010,Munari_etal_1996,Kato_Takamizawa_2001,Tanaka_etal_2011,Pejcha_2009,Aydi_etal_2019_I}), with several explanations ranging from instabilities/pulsations in the envelope of the WD \cite{Schenker_1999,Pejcha_2009}, renewed mass ejection episodes leading to shock interaction\cite{Pejcha_2009,Aydi_etal_2019_I,Aydi_etal_2020a}, and instabilities in a large accretion disk that survived the eruption \cite{Goranskij_etal_2007}. Moreover, \cite{Kato_etal_2011}  suggested that for slow novae that show a relatively long-lasting multipeak phase followed by a wind-like phase, the emission can be explained in context of transition from evolution with no optically thick wind to evolution with optically thick winds. They found that when the companion star is deeply embedded in the extended nova envelope, the structure of the static envelope approaches that of the optically thick wind solution. Thus, the transition from static to wind solution is triggered by the effect of the companion.

While a definitive explanation is yet to be established for these complex light curves, such flares usually coincide with the appearance of new absorption features at increasing velocities, which led some previous studies to suggest that renewed mass ejection episodes being one of the explanation for the flares --- as these ejection interact with previously ejected material, they could lead to shock interaction which power emission in different wavelengths, including the visible (e.g., \cite{Cheung_etal_2016,Aydi_etal_2019_I,Aydi_etal_2020a,Aydi_etal_2020b}). Indeed the spectral line evolution of V1405~Cas show clear evidence for new absorption features developing during the flaring phase of the nova (Supplementary figures~\ref{Fig:V1405_Cas_Halpha_line_profile}, \ref{Fig:V1405_Cas_Halpha_line_profile_2}, \ref{Fig:V1405_Cas_Halpha_line_profile_3}, and~\ref{Fig:V1405_Cas_Halpha_line_profile_4}). This is particularly clear in the 2D dynamic spectrum (Supplementary figure~\ref{Fig:V1405_Cas_Dynamic_spectrum}), where new absorption features appear to emerge at higher blueshifted velocities.

\noindent  \textbf{Reddening and distance.} The  \emph{Gaia} EDR3 distance\cite{Bailer-Jones_etal_2021} to V1405~Cas is 1.73$\pm$0.08\,kpc, consistent with the near-naked-eye peak brightness of the nova. Similar to V1674~Her, we use absorption lines from DIBs to derive the reddening to V1405~Cas and place it in Galactic reddening maps to estimate the distance. The reddening value we derive from the DIBs is $E(B-V) \simeq$ 0.60$\pm$0.20\,mag, implying $A_V \simeq 1.86\pm0.62$\,mag (assuming $R_V = 3.1$), which is the reddening value derived in \cite{ATel_14476}. Using these values and the relations from \cite{Wang_etal_2019}, we transform $E(B-V) \simeq 0.60\pm0.2$ into $E(G-Ks) \simeq 1.32\pm0.30$\,mag, $E(B_p-G_r) \simeq 0.77\pm0.20$\,mag, and $E(H-K_s) \simeq 0.10\pm0.05$\,mag. Based on these reddening values, we place the system in the Galactic reddening maps of \cite{Chen_etal_2019} and we derive a distance to V1405~Cas $d \simeq 2.5\pm1.5$\,kpc, which is slightly larger than the \emph{Gaia} distance. We will adopt the distance derived from the \emph{Gaia} parallax as the distance to the nova.

\noindent  \textbf{CHARA imaging.} The CHARA images of V1405~Cas on days 53, 55, and 67 since $t_0$ show evidence for delayed ejection and substantial changes marked by $\gamma$-ray detection: the first CHARA epoch is challenging to interpret. While the image shows little to no ejecta around a resolved central structure, winds or ejected material could still be present around the system, particularly due to the presence of P Cygni profiles in the optical and IR spectral lines throughout the first 53 days of the eruption (see Section SI). However, since only $\sim$ 1\% of the emission is originating from any potential extended structure around the central source, these ejected material/winds are likely characterized by low densities (constituting only a few per cent of the nova envelope). Consequently, the bulk of the nova envelope is only puffed up but still bound to or surrounding the system, in a supergiant-like phase similar to that described by \cite{Hachisu_etal_2014,Shen_Quataert_2022}. Is this in  agreement with the size of the central emitting source? 
The diameter of the resolved central source in the first CHARA epoch is 0.99$\pm$0.02 mas. 
Using the Gaia-measured distance of 1.73\,kpc (see Section SI), the inferred source size is about 1.71\,AU (corresponding to a radius of $\sim$0.85\,AU). This indicates that the $H$-band photosphere has a size comparable to that of a red giant star. If we instead assume that the nova envelope was impulsively ejected at $t_0$ and allowed to expand for 53 days at velocities between 700 and 1500\,km\,s$^{-1}$ (as derived from the troughs of the P Cygni profiles in the optical and IR spectra), the radius of the line-emitting region at maximum light would be expected to fall between 22 and 46\,AU. While the line-emitting region does not necessarily represent the continuum-emitting region, at peak brightness it should provide a rough estimate of the photospheric size, which dominates the emission. In this scenario, most of the observed radiation should arise from a structure much larger than $\sim$1.7\,AU. The stark contrast between the expected photospheric radius ($\sim$22--46\,AU, assuming impulsive ejection) and the measured radius of the primary emitting source ($\sim$0.85\,AU) implies that the bulk of the nova envelope had not yet been expelled.

This also means that the P Cygni profiles in the optical/NIR spectra likely originate in low-density winds, which are minimally contributing to the emission in the $H$-band. This is supported by (1) the shallow absorption bands in the P Cygni profiles relative to the emission, unlike the spectra of other novae near peak where the absorption features are much more prominent relative to the emission (see figure 1 and 2 in \cite{Aydi_etal_2020b}); and (2) the dramatic deceleration observed in the P Cygni line profiles during the prolonged pre-maximum halt from $-1500$ to $-700$\,km\,s$^{-1}$. As discussed above, Similar deceleration has been observed in the spectra of other novae during the rise to peak\cite{Aydi_etal_2020b}, and is associated with a change in the optical depth as the photosphere recedes inward in velocity space as the nova envelope expand. The dramatic deceleration by more than 800\,km\,s$^{-1}$ indicates that the photosphere has receded inward rapidly in velocity space, likely due to the low density of the wind/outflow. 

The third CHARA epoch obtained two weeks from visible peak shows evidence for extended structures, while the emission from the central structure is now only responsible for less than half of the nova emission in IR. This coincides with the appearance of faster emission components ($v_r \approx 2100$\,km\,s$^{-1}$) in the optical spectra. Since the slower pre-maximum P Cygni profiles are still superimposed on top of these broader emission lines, this indicates the presence of a new physically distinct, faster ejection/outflow (see \cite{Aydi_etal_2020b}). During this phase, the nova also shows high energy emission from shock interaction detected by \textit{Swift} (Supplementary figure~\ref{Fig:V1405_Cas_BVRI_Swift}) and \textit{Fermi}-LAT (Main Figure~1). All this suggests that the bulk of the nova envelope engulfing the binary might have finally been expelled more than 55 days into the eruption. It also suggests that the appearance of faster spectral components in the optical spectra, the detection of high-energy X-ray and $\gamma$-ray emission, and the extended structure of the CHARA images, all happening around the same time (two months into the eruption), is not a matter of coincidence but a matter of common origin (new faster outflows and shock interaction). 

In summary, our results suggest that during the extended pre-maximum halt and rise to visible peak of nova V1405~Cas, the bulk of the nova envelope was in a quasi-static, puffed up state (extended to $\sim$ 0.9 AU), with only a small fraction ($<$5\%) of the accreted envelope being lost via winds (e.g., \cite{Shen_Quataert_2022}). During this phase, the binary motion could be playing a role in driving the expansion of the envelope. Our observations also suggest that the ejection of the bulk of the envelope only took place more than 53 days into the eruption, consistent with the (\textit{delayed}) detection of high-energy X-rays and $\gamma$-rays around the same time. This eventual ejection could have been driven by the combination of (1) a fast wind launched from the white dwarf surface (as marked by the emergence of faster components in the optical spectra) pushing through the puffed-up envelope and (2) the binary motion helping to expel the envelope. This has major implications for our understanding of how novae expel their envelopes, the role of the binary motion in driving the mass-loss during the nova, and how shocks form in novae.

\noindent \subsection{Fitting the CHARA data of V1405~Cas.} We use the same PMOIRED method discussed for nova V1674~Her to fit the CHARA images of nova V1405~Cas and derive the size of the ejecta. Observations for three epochs were obtained: UT 2021 May 10, 12 and 24. The interferometric data show a marginally resolved object, with the presence of a fully resolved component for the latest epoch. A simple model is used to fit the data: a compact component ($\lesssim$1~mas) and a very extended one, with the flux ratio between the two allowed to vary. We report the size of the compact component both as Gaussian full width at half maximum, and as uniform disk diameter (see Supplementary table~\ref{tab:parametersV1405Cas} and Supplementary figures~\ref{fig:PMOIRED_Nova_cas_1}, ~\ref{fig:PMOIRED_Nova_cas_2}, and~\ref{fig:PMOIRED_Nova_cas_3}).

\begin{table}[h!]
\def\arraystretch{1.0}
\begin{tabular}{lcccc}
\hline
\hline

\rule{0pt}{2ex} UT Date & Telescopes & Spectral Mode & Target & Calibrators\\
\hline
2021 May 24 & E1-W2-W1-S2-E2 & GRISM190 & V1405~Cas & HD 219080\\
2021 May 10 & E1-W2-W1-S2-E2 & GRISM190 & V1405~Cas & HD 211982\\
2021 May 12 & E1-W2-W1-S2-E2 & PRISM50 & V1405~Cas & HD 135969\\
&  &  &  &  HD 145965\\
&  &   & & HD 151259\\
\hline
\end{tabular}

\caption{Log of CHARA observations of nova V1405~Cas. The first column is the date of the observations. The second column represents the CHARA units used in the observation. The third column is the spectral mode. The last two columns are the science target and the calibrator sources.}
\footnotesize{Note: the calibrators on UT 2021 May 12 (declination -23$^\circ$) were located at a very large angular separation on sky from V1405~Cas (declination +61$^\circ$).}
\label{table:CHARA_log_V1405_Cas}
\end{table}
\vspace{1cm}

\begin{table}[h!]
    \centering

    \begin{tabular}{cccc}
\hline
\hline
         & May 10, 2021 & May 12, 2021& May 24, 2021\\
    \hline
         compact FWHM (mas) & $0.62\pm0.01$  & $0.61\pm0.01$   & $0.32\pm0.02$ \\
         compact UD (mas)  & $0.99\pm0.02$ & $0.96\pm0.02$ & $0.54\pm0.04$  \\
         resolved flux (\% of total)  & $<1$ & $<1$ &$52\pm1$ \\    
        resolved FWHM (mas) &  --- &--- & $\gtrsim$ 5 \\        
\hline 
\end{tabular}
        \caption{Fitted parameters to CHARA observations of V1405~Cas. The model has 2 components: a compact source, and a fully (spatially) resolved one. For May 10 and 12, only the compact source is detected, and the extended source only manifests itself on the latest epoch. Unfortunately, the smallest baseline of CHARA fully resolved this component, leading to a lower limit of its FWHM of 5~mas.}
    \label{tab:parametersV1405Cas}
\end{table}

\clearpage


\begin{figure*}
\begin{center}
\vspace{-1.cm}
\includegraphics[width=\textwidth]{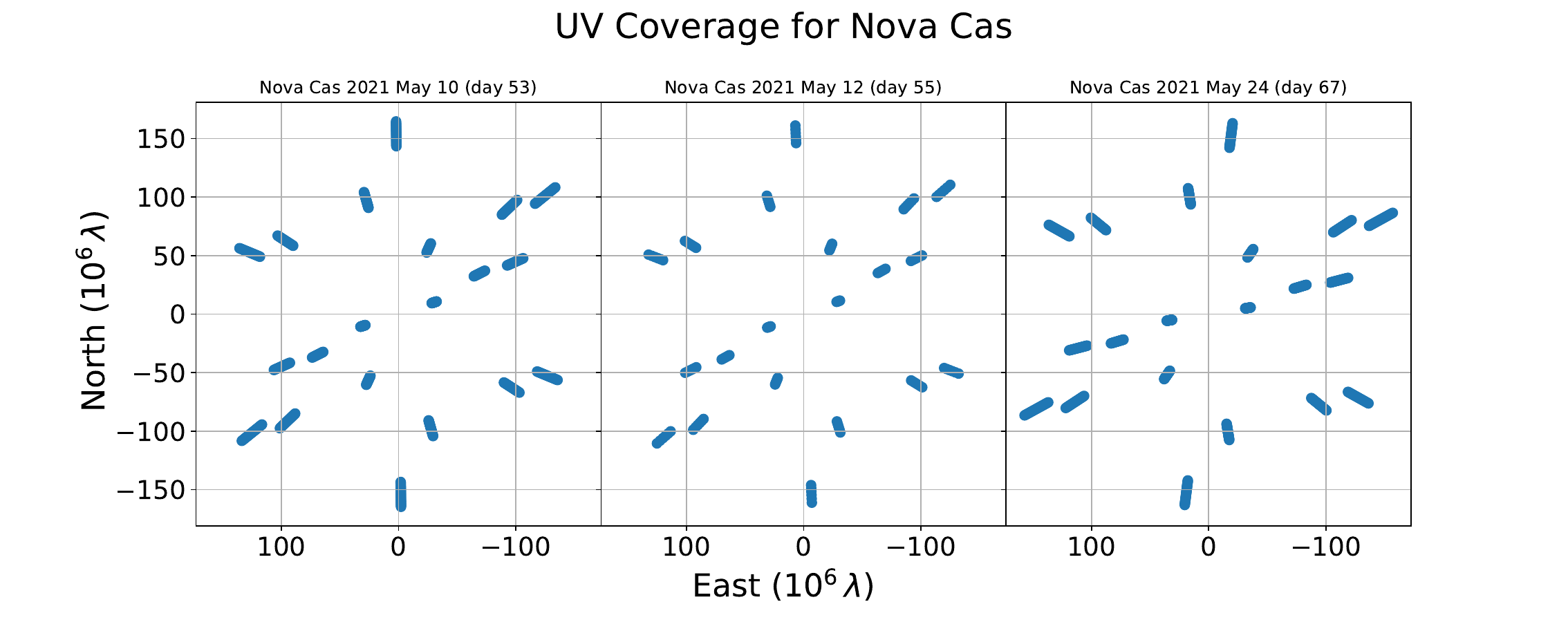}
\vspace{-0.5cm}
\caption{The CHARA (u,v) coverage of interferometric baselines projected on the  plane of the sky in right ascension (RA) and declination (Dec.), during the observations of nova V1405~Cas. \textit{Left} panel is the 2021 May 10 epoch (day 53), \textit{Middle} panel is the 2021 May 12 epoch (day 55), and the \textit{right} panel is the 2021 May 24 epoch (day 67).}
\label{Fig:UV_V1405_Cas}
\end{center}
\end{figure*}

\begin{figure*}
\begin{center}
\vspace{-1.5cm}
\includegraphics[width=0.7\textwidth]{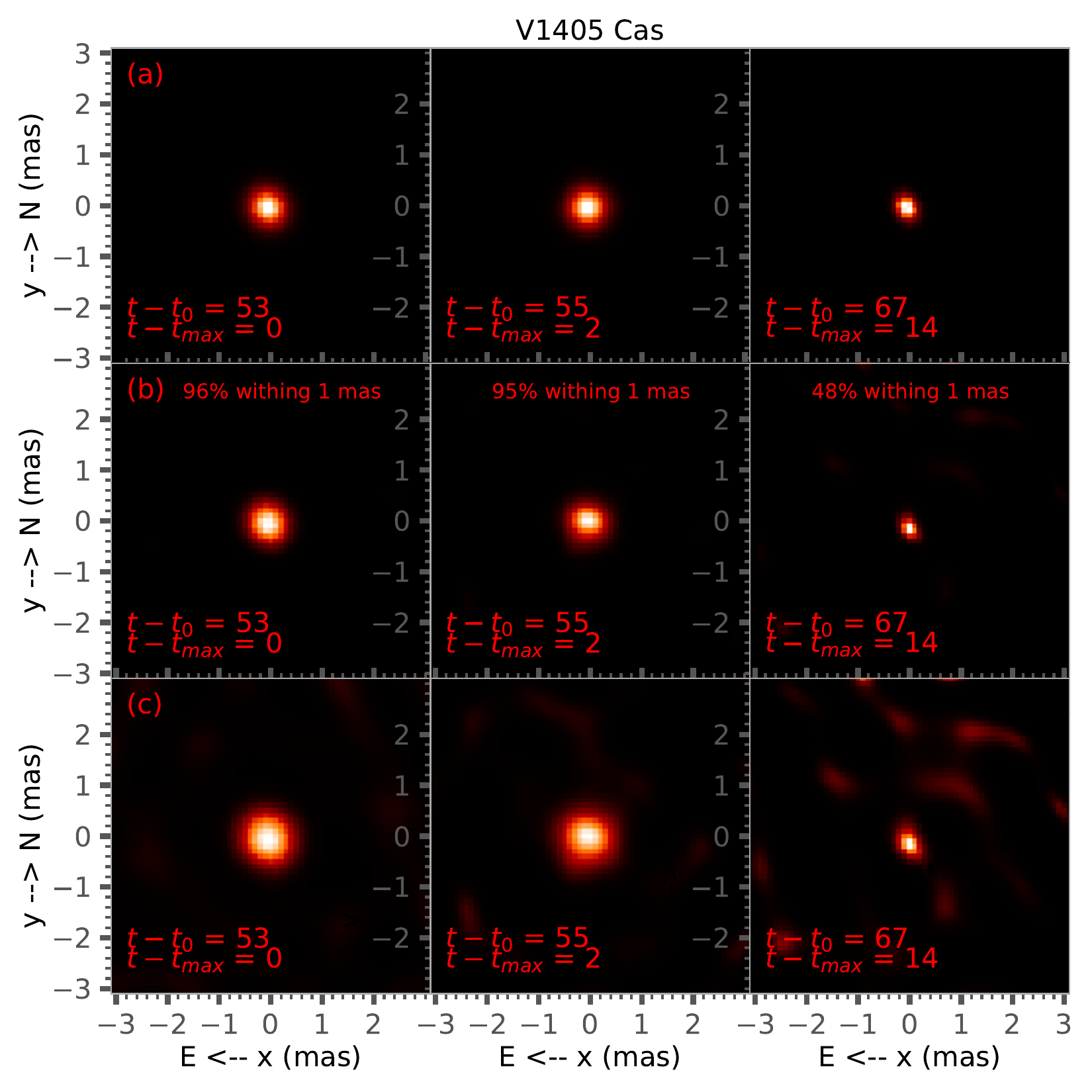}
\vspace{-0.5cm}
\caption{\textbf{The prior and \textsc{BSMEM} images of nova V1405~Cas.} \textit{Panels (a)}: the prior images used based on an elliptical Gaussian fit to the visibility data. \textit{Panels (b)}: the \textsc{BSMEM} images using a linear scale. \textit{Panels (c)}: the \textsc{BSMEM} images using square-root intensity, to highlight low-surface brightness emission within the field-of-view.}
\label{Fig:methods_images_2}
\end{center}
\end{figure*}

\begin{figure*}[h!]
\begin{center}
\includegraphics[width=\textwidth]{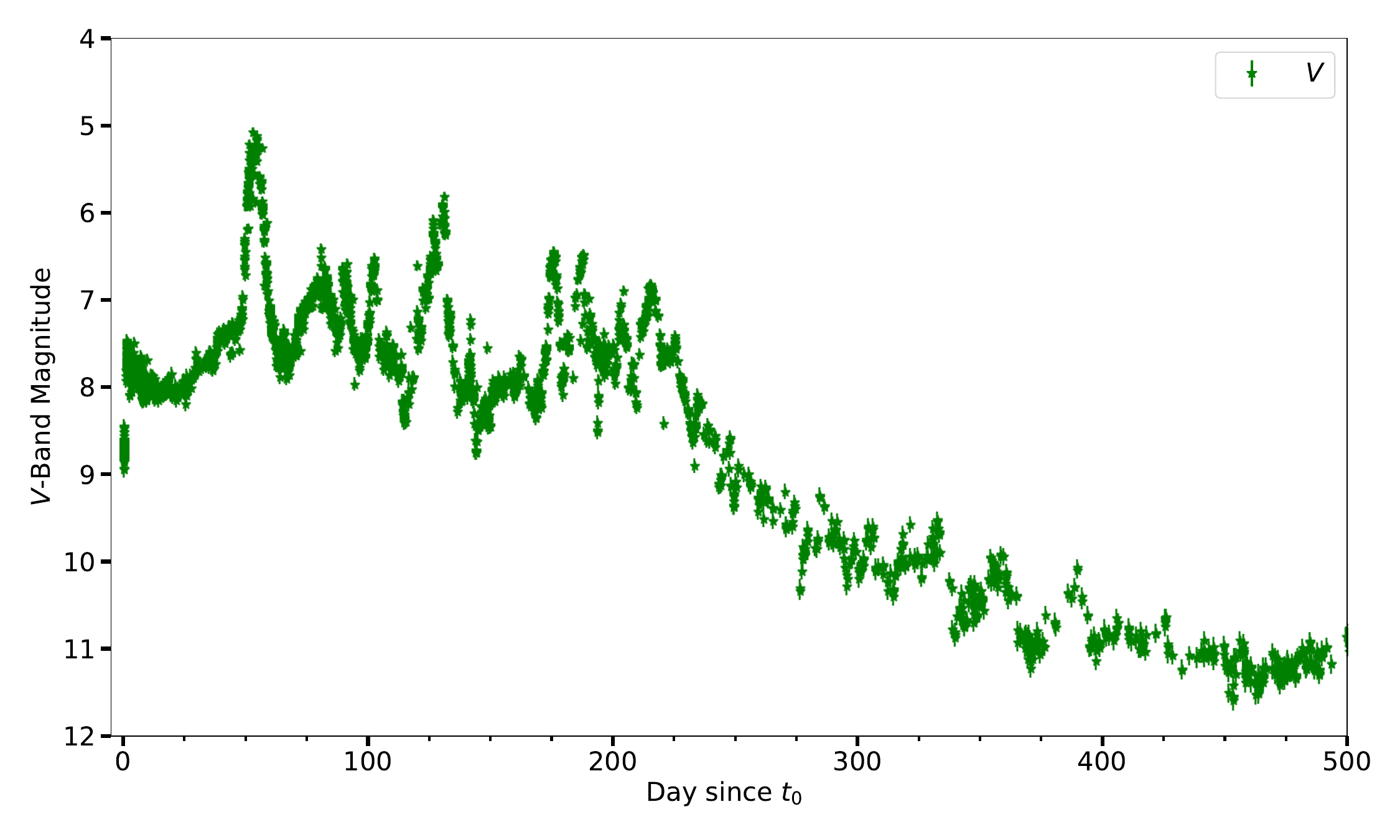}
\caption{The AAVSO optical light curve of nova V1405~Cas over 500 days of the eruption. The green symbols are $V$-band measurements. The error bars represent 1-$\sigma$ uncertainties.}
\label{Fig:V1405_Cas_optica_LC}
\end{center}
\end{figure*}


\begin{figure*}
\begin{center}
\includegraphics[width=\textwidth]{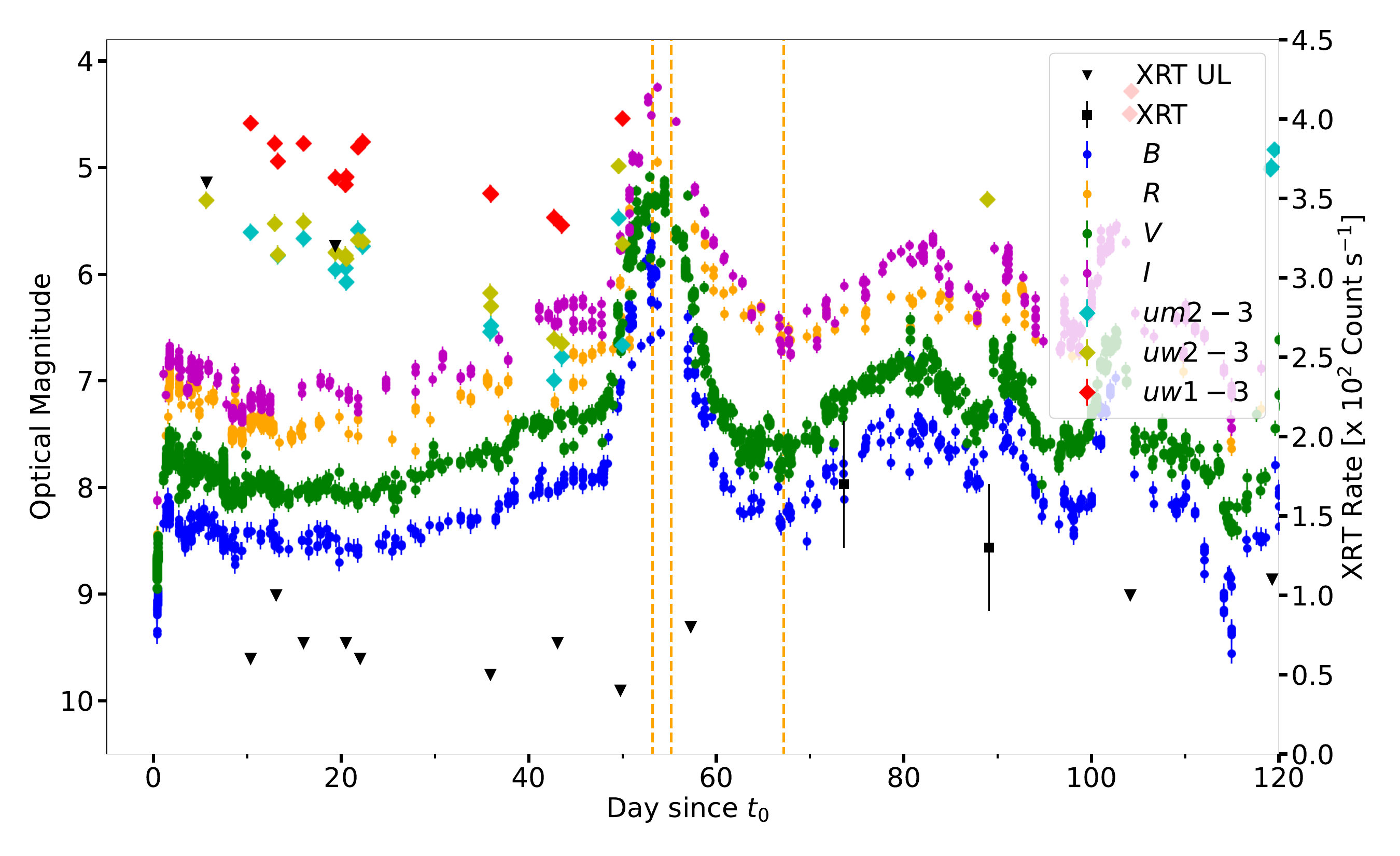}
\caption{Multi-wavelength light curves of nova V1405 Cas over the first 120 days of the eruption showing: optical $B$, $V$, $R$, and $I$ data (blue, green, orange, and magenta symbols, respectively), \textit{Swift}-UVOT ultraviolet data (red symbols for $uw1$ filter, yellow symbols for $uw2$ filter, and cyan symbols for $um2$ filters), and \textit{Swift}-XRT X-ray data (black symbols). The orange dashed lines represent the dates of the CHARA epochs. An offset of 3 magnitudes is applied to the ultraviolet measurements for visualization purposes. The error bars represent 1-$\sigma$ uncertainties..}
\label{Fig:V1405_Cas_BVRI_Swift}
\end{center}
\end{figure*}


\begin{figure*}
\begin{center}
\hspace{0.1cm}\includegraphics[width=0.8\textwidth]{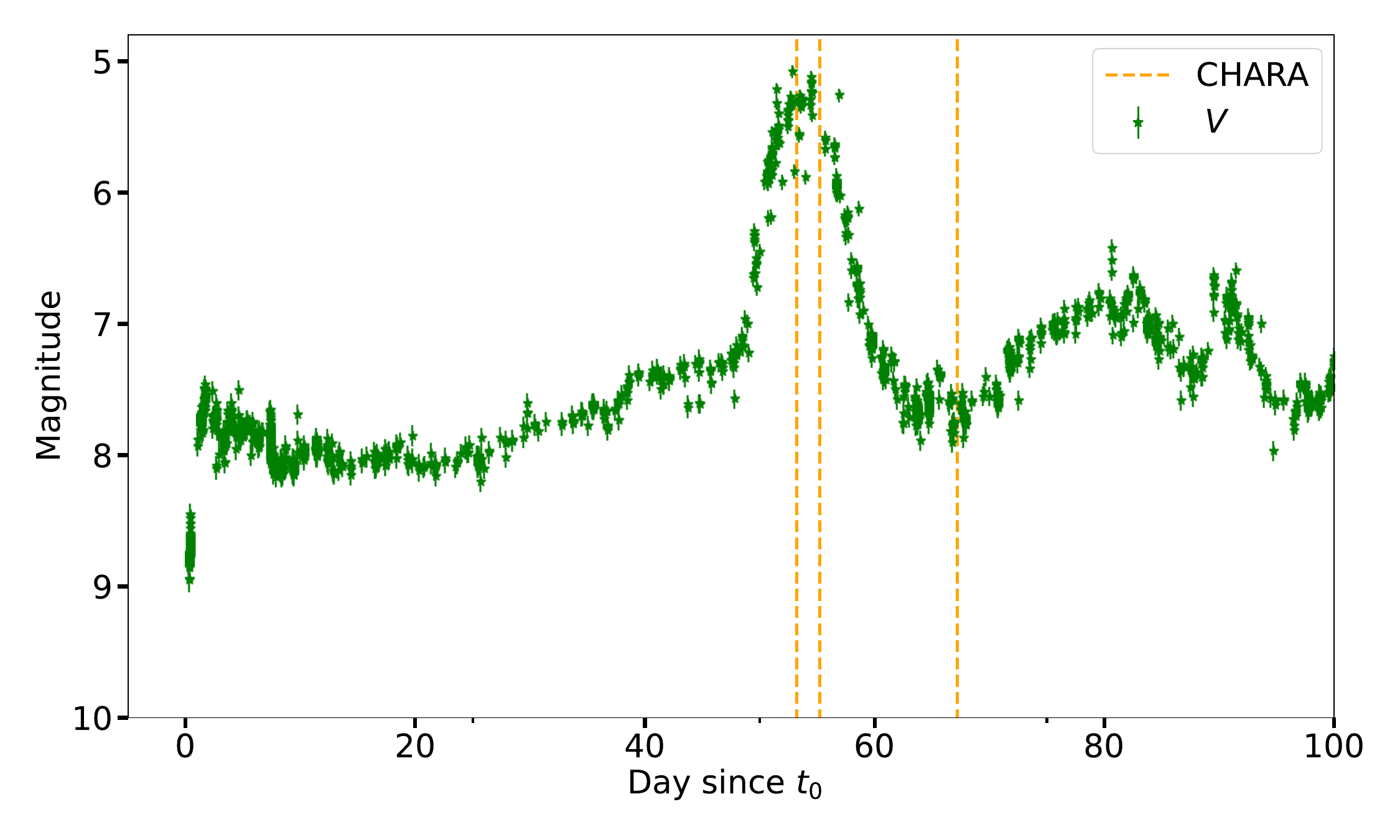}
\vspace*{-0.22cm}

\includegraphics[width=0.81\textwidth]{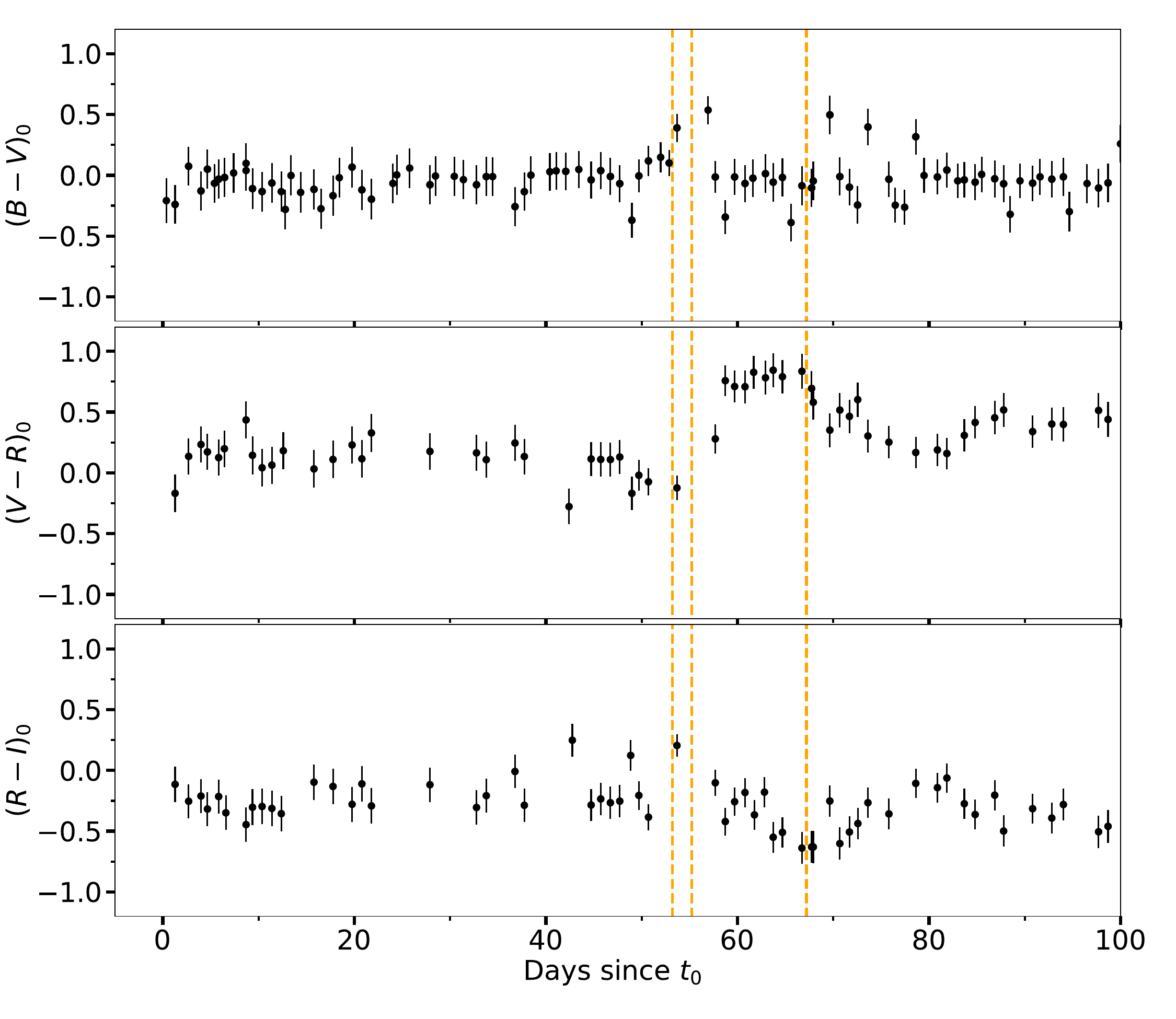}
\caption{\textit{Top panel}: the optical $V$-band light curve of nova V1405~Cas over the first 120 days of the eruption. \textit{Bottom panel}: from top to bottom the $(B-V)_0$, $(V-I)_0$, and $(R-I)_0$ color evolution of nova V1405 Cas during the first 120 days of the eruption. The orange dashed lines represent the dates of the CHARA epochs. The error bars represent 1-$\sigma$ uncertainties.}
\label{Fig:V1405_Cas_colors}
\end{center}
\end{figure*}


\begin{figure*}
\begin{center}
\includegraphics[width=\textwidth]{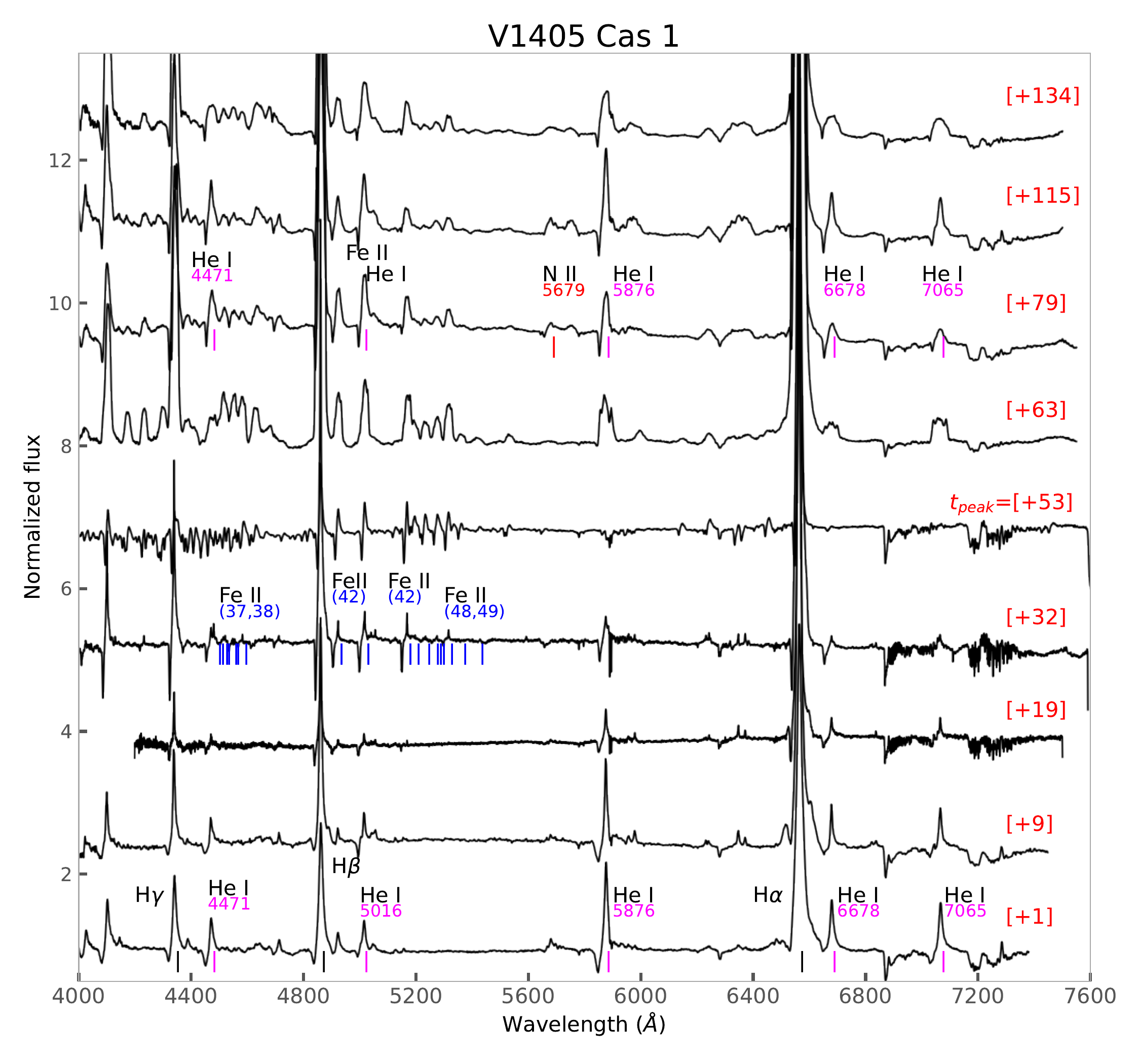}
\caption{\textbf{The spectral evolution of nova V1405~Cas (part 1)}. The numbers between brackets are days since $t_0$. Line identifications are presented with tick marks under the lines for easier identification and they are color
coded based on the line species.}
\label{Fig:V1405_Cas_spec_1}
\end{center}
\end{figure*}

\begin{figure*}
\begin{center}
\includegraphics[width=\textwidth]{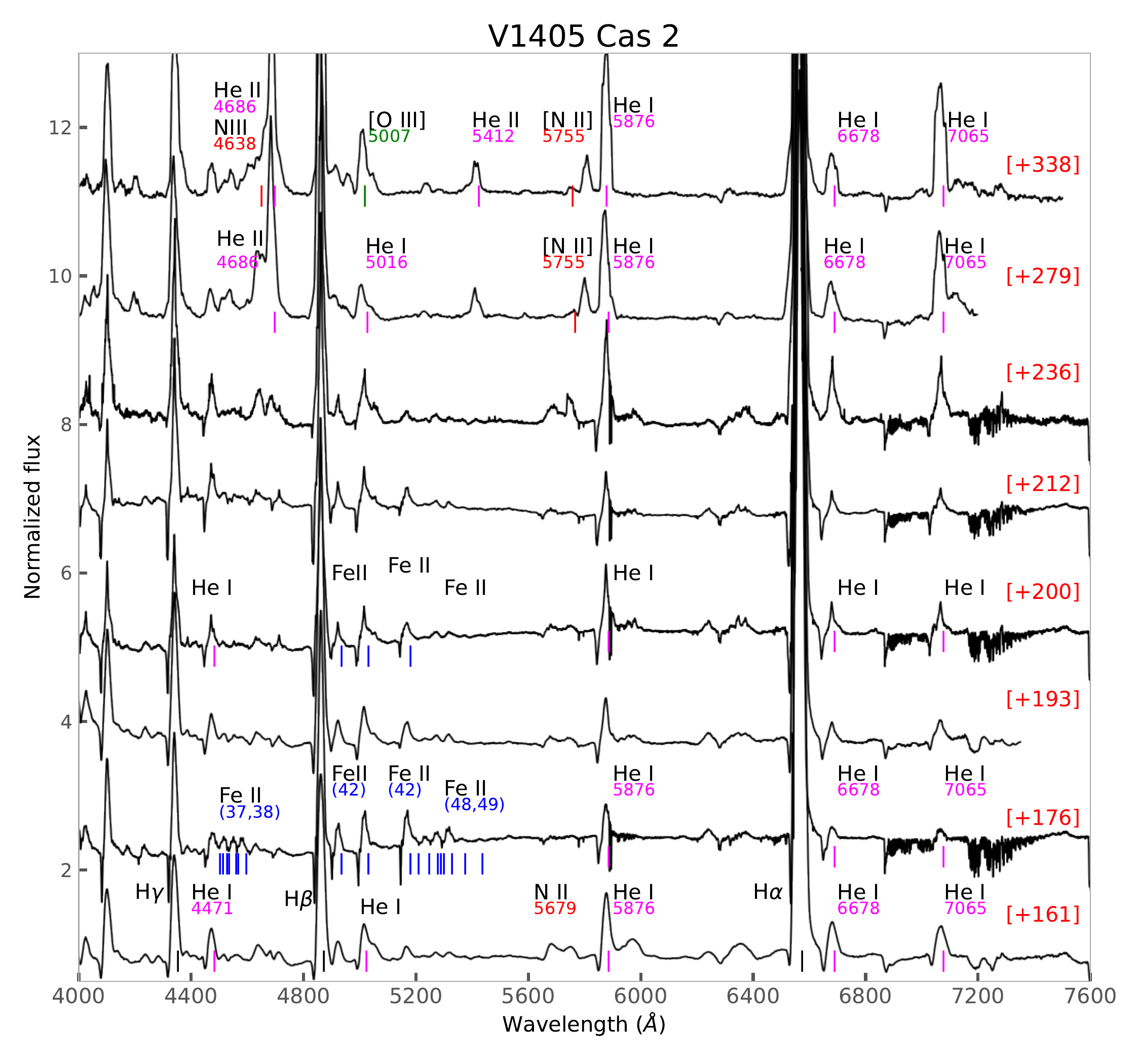}
\caption{\textbf{The spectral evolution of nova V1405~Cas (part 2)}. The numbers between brackets are days since $t_0$. Line identifications are presented with tick marks under the lines for easier identification and they are color
coded based on the line species.}
\label{Fig:V1405_Cas_spec_2}
\end{center}
\end{figure*}



\begin{figure*}
\begin{center}
\includegraphics[width=\textwidth]{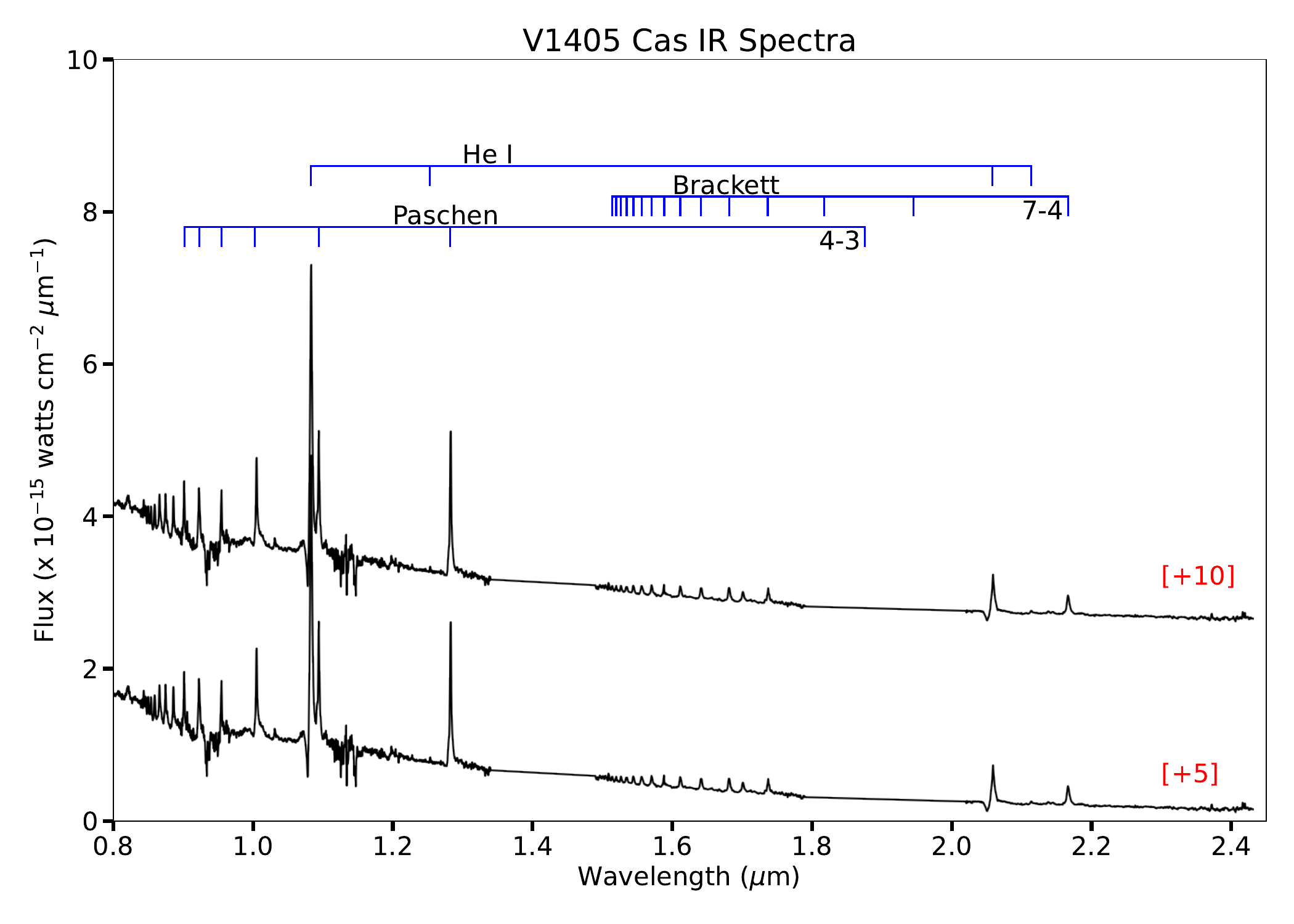}
\caption{\textbf{Infrared spectra of V1405~Cas taken on days 5 and 10 after discovery}, showing P Cygni lines of He I and H I of the Paschen and Bracket series. The numbers between brackets are days since $t_0$.}
\label{Fig:V1405_IR_spec}
\end{center}
\end{figure*}


\clearpage
\begin{figure*}
\begin{center}
\includegraphics[width=\textwidth]{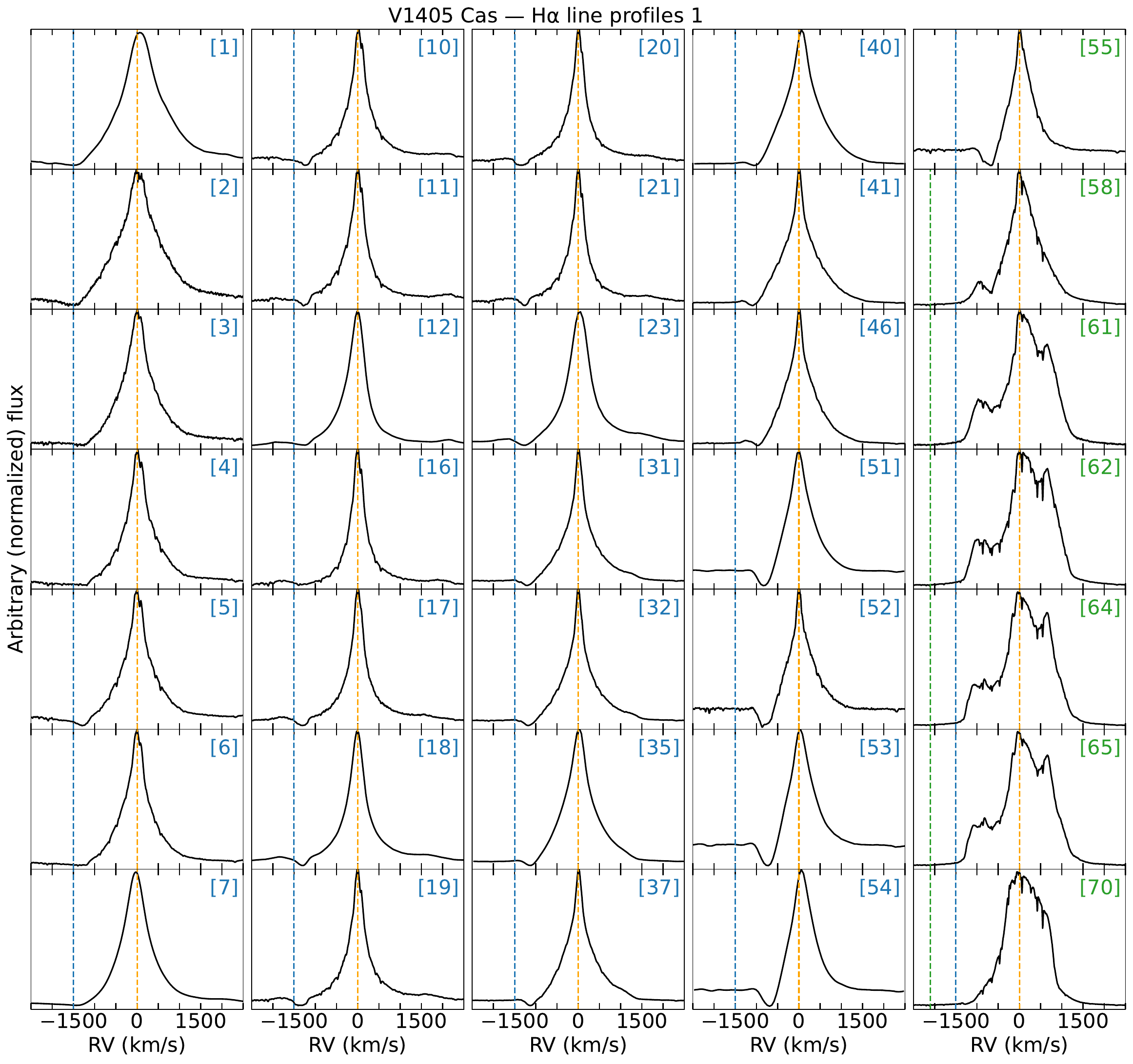}
\caption{The evolution of H$\alpha$ line profiles for V1405~Cas. The orange, blue, and green dashed lines represent $v_{\mathrm{rad}}$ = 0\,km\,s$^{-1}$, $v_{\mathrm{rad}}$ = $-1500$\,km\,s$^{-1}$, and $v_{\mathrm{rad}}$ = $-2100$\,km\,s$^{-1}$, respectively. The numbers between brackets are days since $t_0$.}
\label{Fig:V1405_Cas_Halpha_line_profile}
\end{center}
\end{figure*}

\begin{figure*}
\begin{center}
\includegraphics[width=\textwidth]{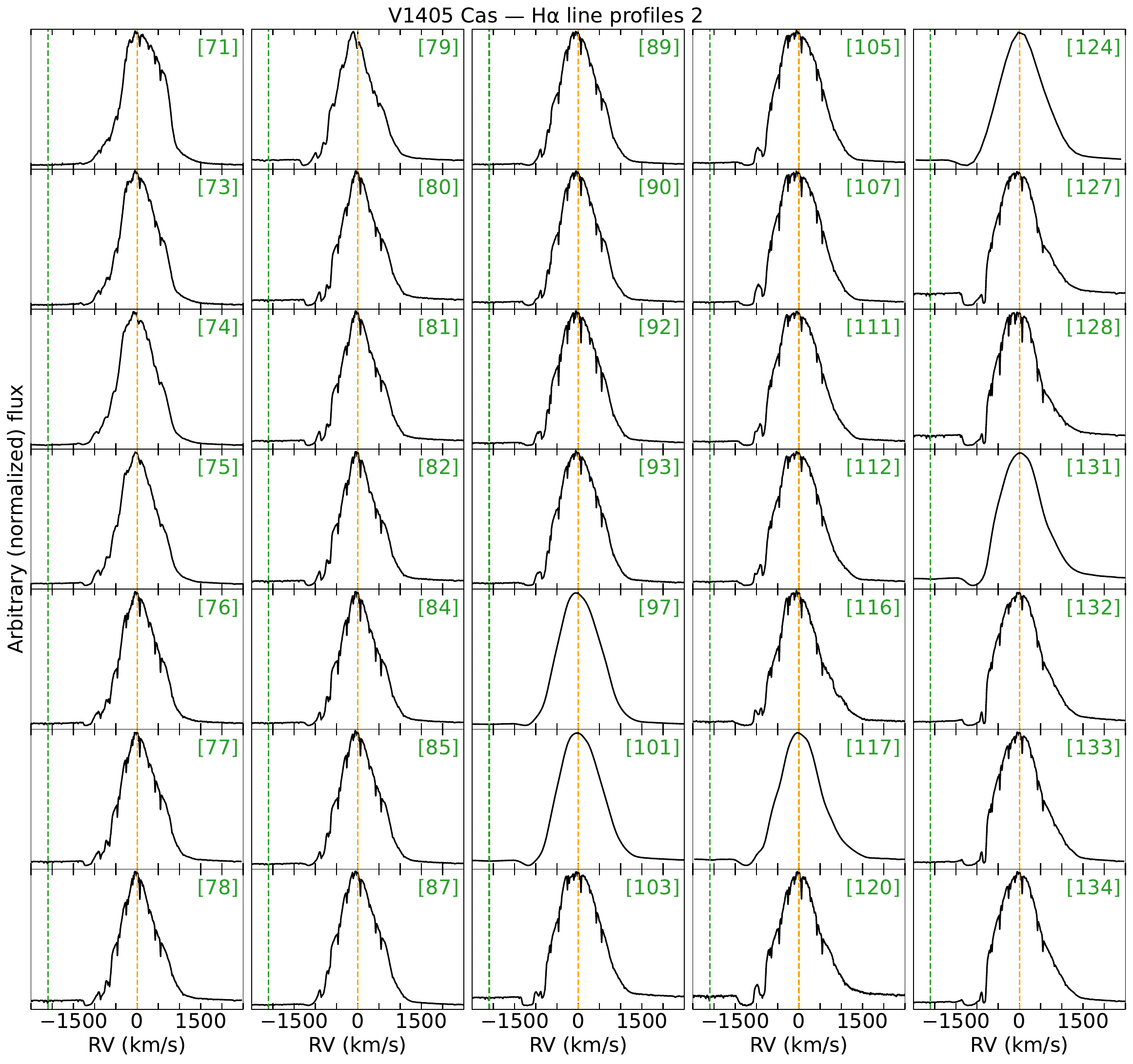}
\caption{The evolution of H$\alpha$ line profiles for V1405~Cas. The orange and green dashed lines represent $v_{\mathrm{rad}}$ = 0\,km\,s$^{-1}$ and $v_{\mathrm{rad}}$ = $-2100$\,km\,s$^{-1}$, respectively. The numbers between brackets are days since $t_0$.}
\label{Fig:V1405_Cas_Halpha_line_profile_2}
\end{center}
\end{figure*}

\begin{figure*}
\begin{center}
\includegraphics[width=\textwidth]{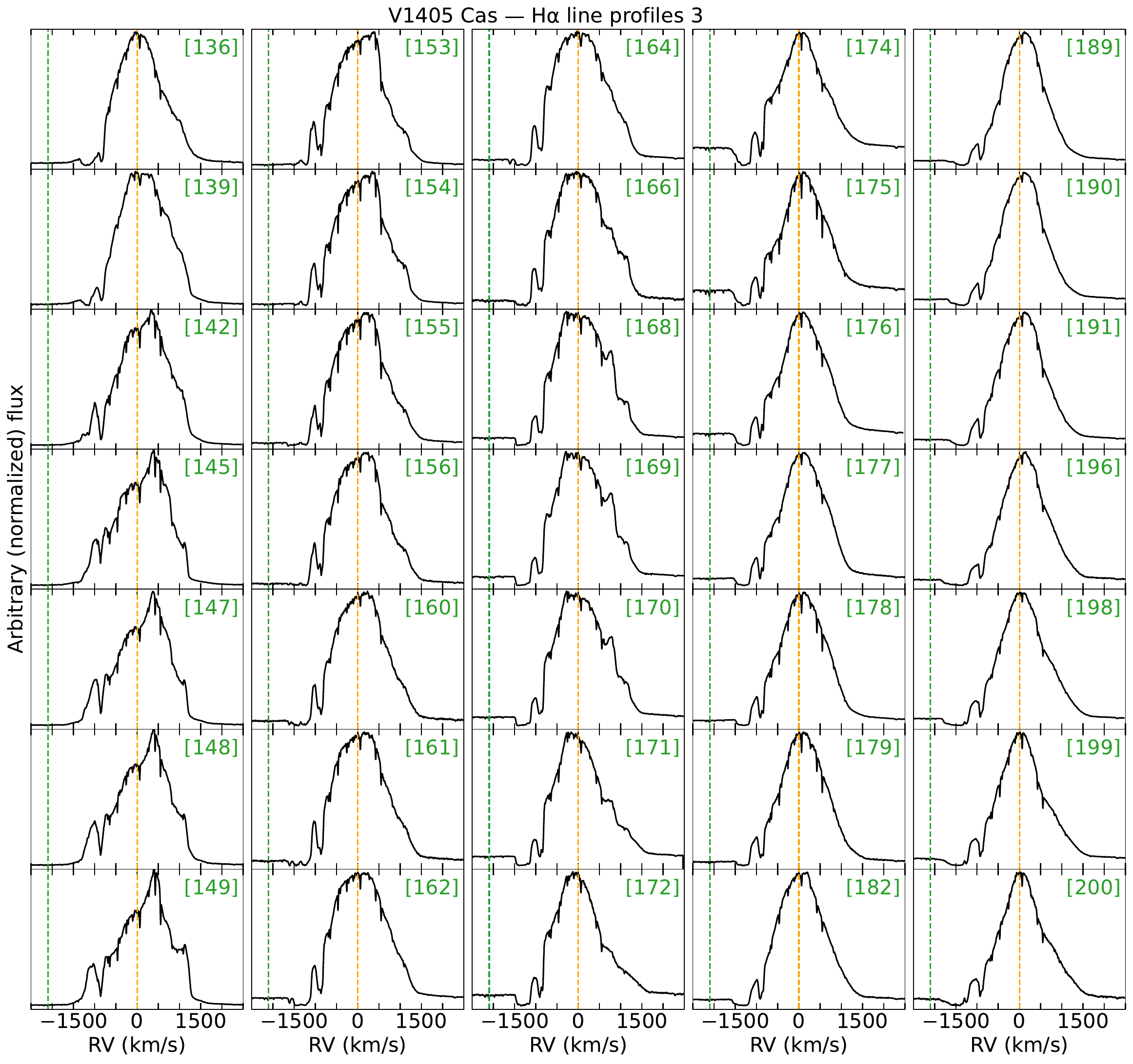}
\caption{The evolution of H$\alpha$ line profiles for V1405~Cas. The orange and green dashed lines represent $v_{\mathrm{rad}}$ = 0\,km\,s$^{-1}$ and $v_{\mathrm{rad}}$ = $-2100$\,km\,s$^{-1}$, respectively. The numbers between brackets are days since $t_0$.}
\label{Fig:V1405_Cas_Halpha_line_profile_3}
\end{center}
\end{figure*}

\begin{figure*}
\begin{center}
\includegraphics[width=\textwidth]{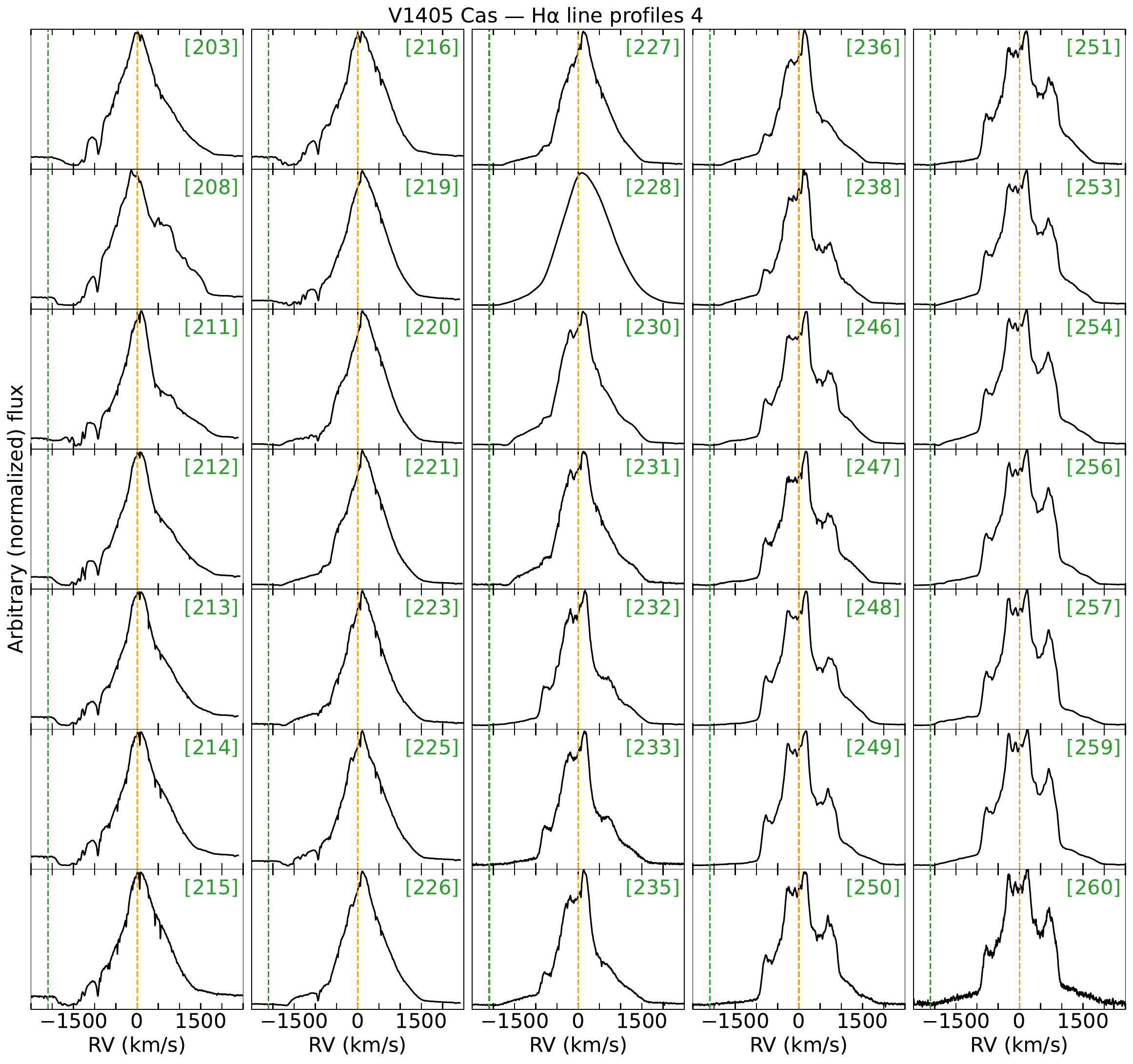}
\caption{The evolution of H$\alpha$ line profiles for V1405~Cas. The orange and green dashed lines represent $v_{\mathrm{rad}}$ = 0\,km\,s$^{-1}$ and $v_{\mathrm{rad}}$ = $-2100$\,km\,s$^{-1}$, respectively. The numbers between brackets are days since $t_0$.}
\label{Fig:V1405_Cas_Halpha_line_profile_4}
\end{center}
\end{figure*}

\begin{figure*}
\begin{center}
\includegraphics[width=0.62\textwidth]{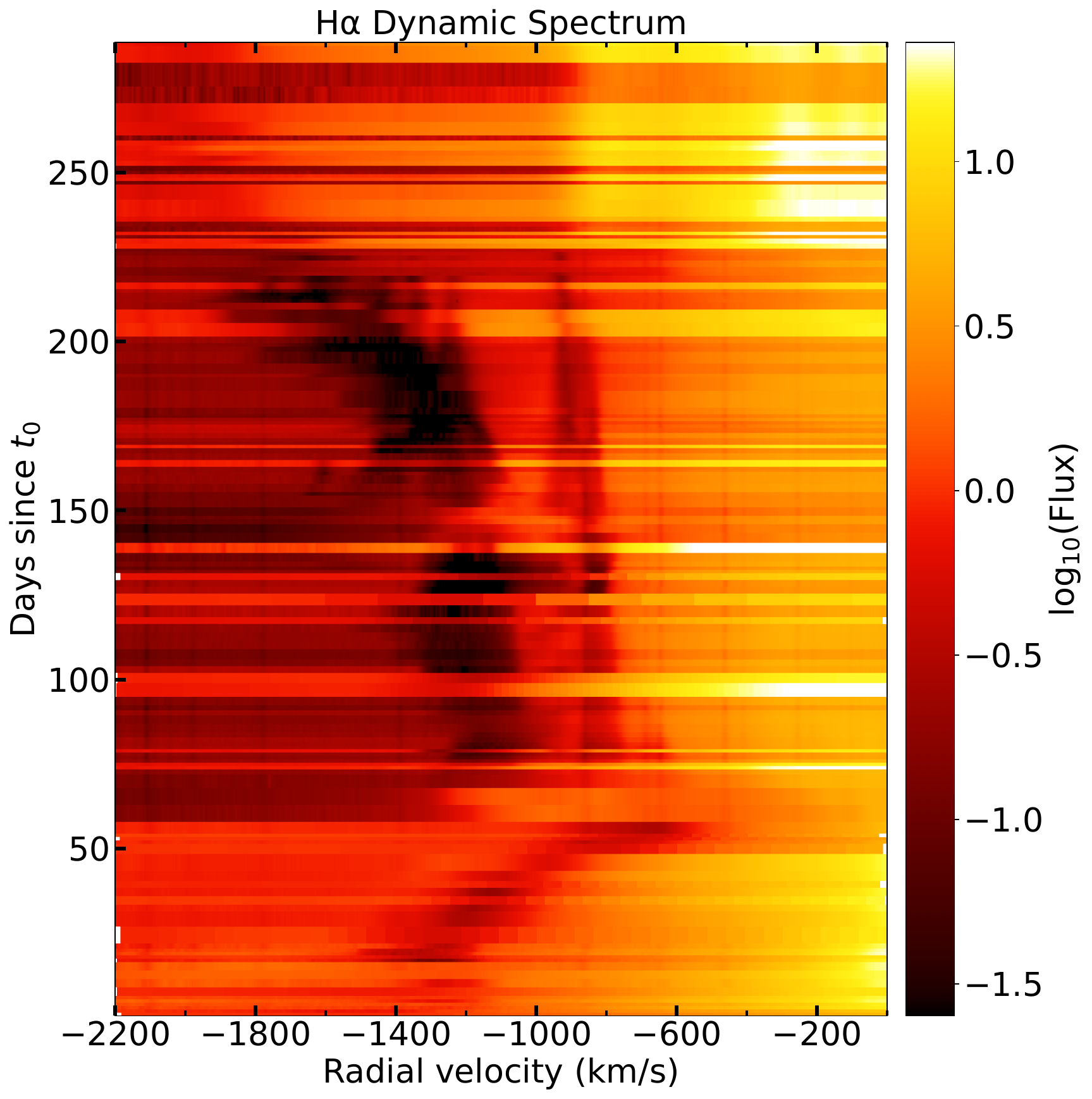}\includegraphics[angle=90,width=0.415\textwidth]{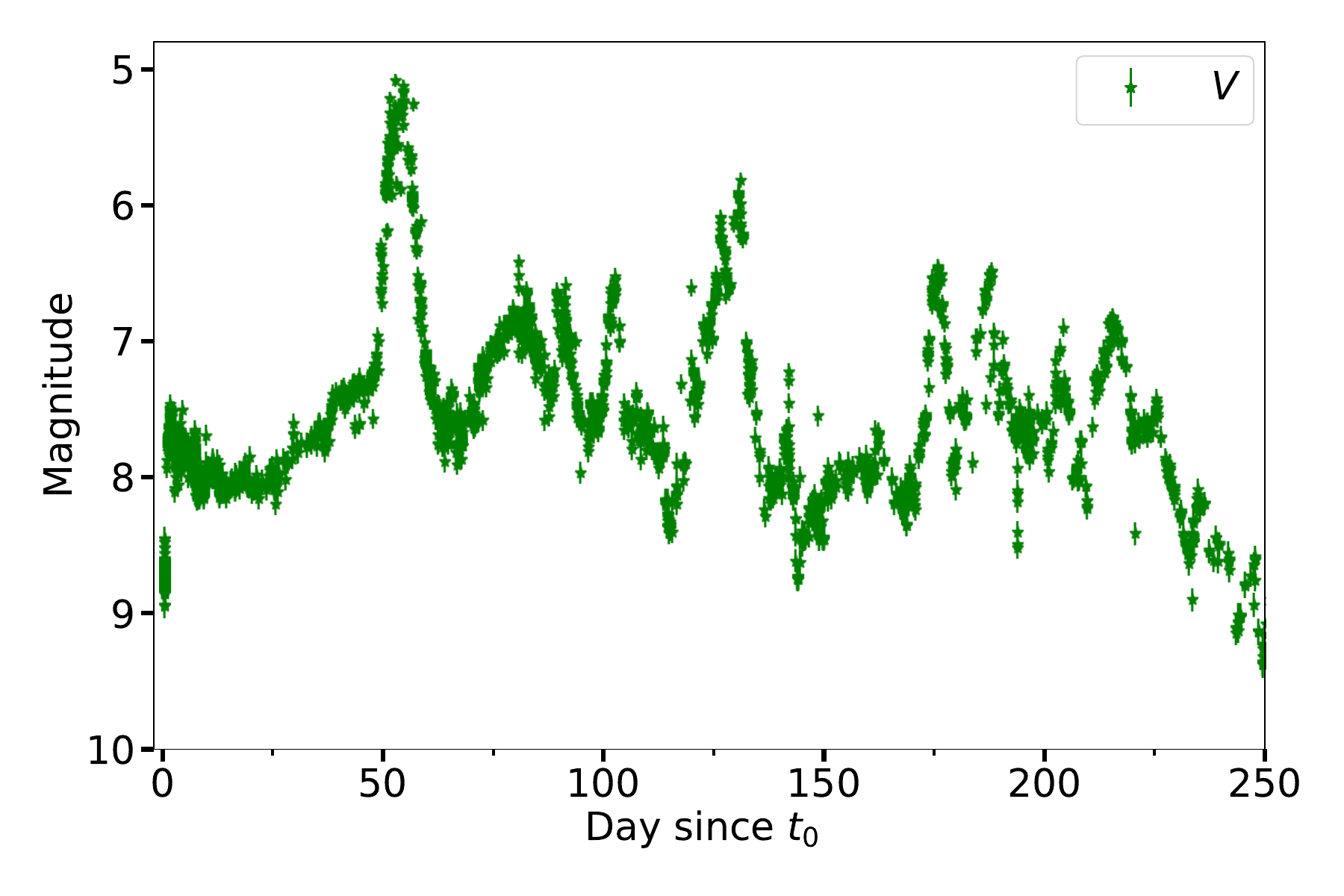}
\caption{\textit{Left}: 2D dynamic spectrum of nova V1405~Cas illustrating the evolution of the H$\alpha$ line profiles during the first 250~days of the eruption. Dark features correspond to absorption components in the spectra. \textit{Right:} the optical $V$-band light curve of V1405~Cas during the first 250 days of the eruption. The error bars represent 1-$\sigma$ uncertainties.}
\label{Fig:V1405_Cas_Dynamic_spectrum}
\end{center}
\end{figure*}

\begin{figure}
    \centering
    \hspace{-1.5cm}\includegraphics[width=0.85\linewidth]{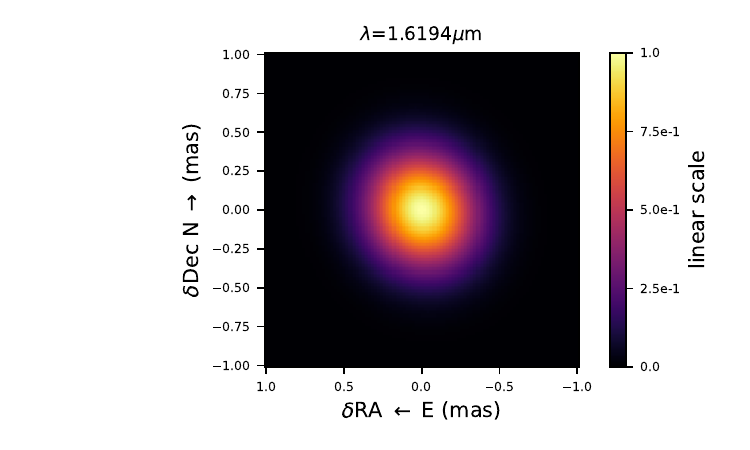}
    
    \includegraphics[width=0.6\linewidth]{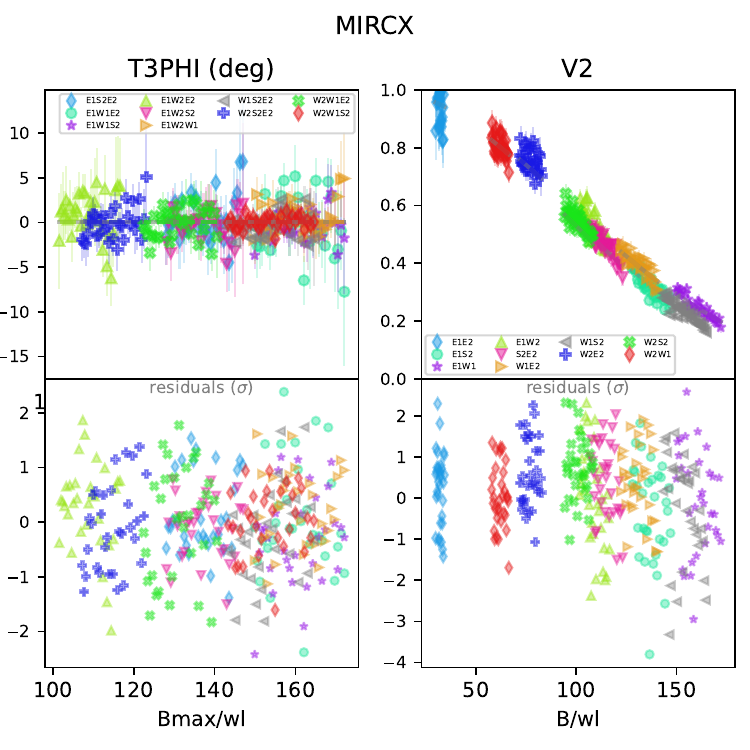}

    \caption{Synthetic images of V1405 Cas, based on \textsc{PMOIRED} models. 
    top row: synthetic image for 2021 May 10 (day 53). Bottom row: closure phase and squared visibility data and model, as well as residuals normalized to uncertainties.}
    \label{fig:PMOIRED_Nova_cas_1}
\end{figure}

\begin{figure}
    \centering

    \includegraphics[width=0.85\linewidth]{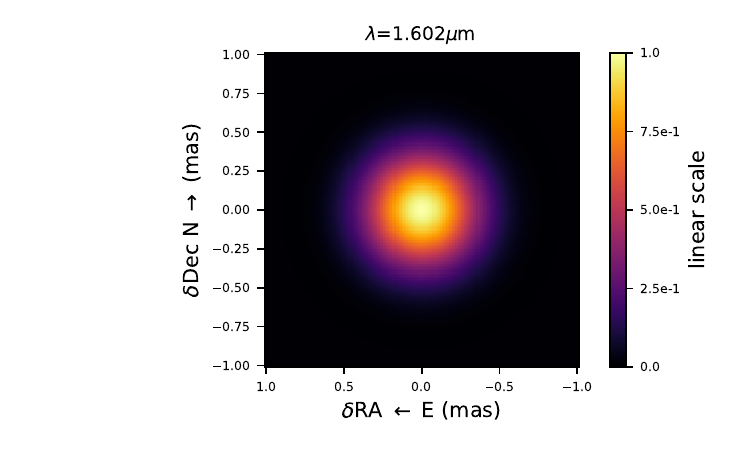}

    \includegraphics[width=0.6\linewidth]{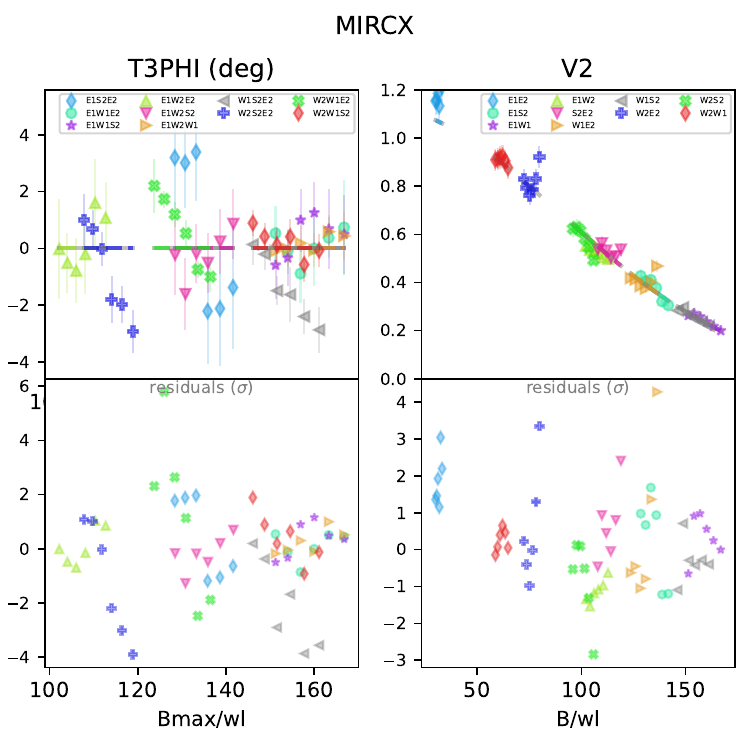}

    \caption{Synthetic images of V1405 Cas, based on \textsc{PMOIRED} models. 
    top row: synthetic image for 2021 May 12 (day 55). Bottom row: closure phase and squared visibility data and model, as well as residuals normalized to uncertainties.}
    \label{fig:PMOIRED_Nova_cas_2}
\end{figure}

\begin{figure}
    \centering

    \includegraphics[width=0.85\linewidth]{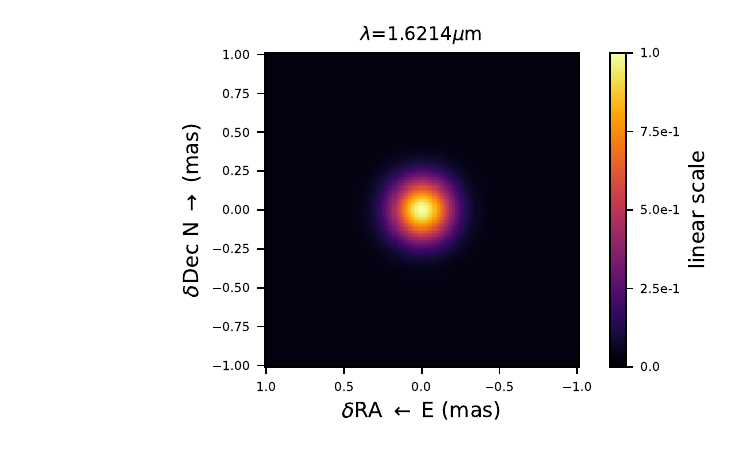}
   
    \includegraphics[width=0.6\linewidth]{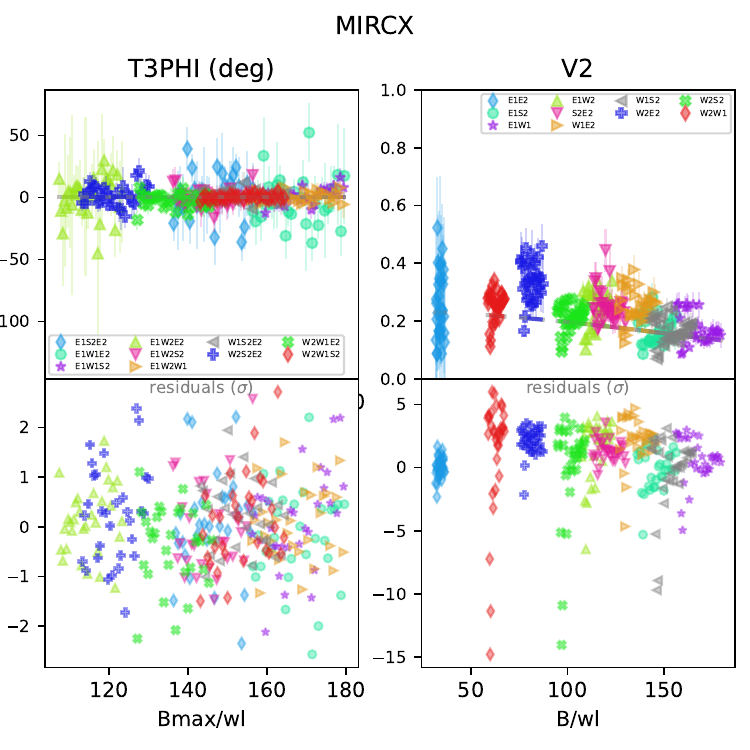}
   
    \caption{Synthetic images of V1405 Cas, based on \textsc{PMOIRED} models. 
    top row: synthetic image for 2021 May 24 (day 67). Bottom row: closure phase and squared visibility data and model, as well as residuals normalized to uncertainties.}
    \label{fig:PMOIRED_Nova_cas_3}
\end{figure}

\clearpage

\begin{table*}
\vspace{-1.5cm}
\caption{\footnotesize{Log of spectral observations of V1405~Cas. The first column represents the date of the observations, the second column is the time relative to discovery epoch ($t_0$). The third columnd represents the instrument/source of the data. The last two columns are the spectral range and the resolving power $R$.} }
\label{table:spec_log_V1405_Cas}
\centering
\def\arraystretch{1.0}
\begin{tabular}{lcccc}

\hline
\hline
\rule{0pt}{2ex} Date & $t-t_0$ & Source & Range & $R$\\
(UT date) & (days) & & $\AA$ & \\
\hline
2021-03-19 &   1  & ARAS & 3900\,--\,7500 & 10,000 \\
2021-03-20 &   2  & ARAS & 3900\,--\,7500 & 10,000 \\
2021-03-22 &   4  & ARAS & 3900\,--\,7500 & 10,000 \\
2021-03-23 &   5  & VNIRIS & 8000\,--\,24000 & 1000\\ 
2021-03-25 &   7  & ARAS & 3900\,--\,7500 & 10,000 \\
2021-03-27 &   9  & ARAS & 3900\,--\,7500 & 10,000 \\
2021-03-28 &   10  & VNIRIS & 8000\,--\,24000 & 1000\\ 
2021-03-29 &  11  & ARAS & 3900\,--\,7500 & 10,000 \\
2021-04-03 &  16  & ARAS & 3900\,--\,7500 & 10,000 \\
2021-04-07 &  19  & ARAS & 3900\,--\,7500 & 10,000 \\
2021-04-16 &  29  & ARAS & 3900\,--\,7500 & 10,000 \\
2021-04-18 &  31  & ARAS & 3900\,--\,7500 & 10,000 \\
2021-04-19 &  32  & ARAS & 3900\,--\,7500 & 10,000 \\
2021-04-24 &  37  & ARAS & 3900\,--\,7500 & 10,000 \\
2021-04-25 &  38  & ARAS & 3900\,--\,7500 & 10,000 \\
2021-04-28 &  41  & ARAS & 3900\,--\,7500 & 10,000 \\
2021-05-01 &  44  & ARAS & 3900\,--\,7500 & 10,000 \\
2021-05-03 &  46  & ARAS & 3900\,--\,7500 & 10,000 \\
2021-05-05 &  48  & ARAS & 3900\,--\,7500 & 10,000 \\
2021-05-07 &  50  & ARAS & 3900\,--\,7500 & 10,000 \\
2021-05-09 &  52  & ARAS & 3900\,--\,7500 & 10,000 \\
2021-05-10 &  53  & ARAS & 3900\,--\,7500 & 10,000 \\
2021-05-18 &  61  & ARAS & 3900\,--\,7500 & 10,000 \\
2021-05-25 &  68  & ARAS & 3900\,--\,7500 & 10,000 \\
2021-05-29 &  72  & ARAS & 3900\,--\,7500 & 10,000 \\
2021-06-02 &  76  & ARAS & 3900\,--\,7500 & 10,000 \\
2021-06-08 &  82  & ARAS & 3900\,--\,7500 & 10,000 \\
2021-06-13 &  87  & ARAS & 3900\,--\,7500 & 10,000 \\
2021-06-16 &  90  & ARAS & 3900\,--\,7500 & 10,000 \\
2021-06-19 &  93  & ARAS & 3900\,--\,7500 & 10,000 \\
2021-06-26 & 100  & ARAS & 3900\,--\,7500 & 10,000 \\
2021-07-01 & 105  & ARAS & 3900\,--\,7500 & 10,000 \\
2021-07-04 & 108  & ARAS & 3900\,--\,7500 & 10,000 \\
2021-07-11 & 115  & ARAS & 3900\,--\,7500 & 1,000  \\
2021-07-30 & 134  & ARAS & 3900\,--\,7500 & 1,000  \\
2021-08-26 & 161  & ARAS & 3900\,--\,7500 & 1,000  \\
2021-09-10 & 176  & ARAS & 3900\,--\,7500 & 10,000 \\
2021-09-27 & 193  & ARAS & 3900\,--\,7500 & 1,000  \\
2021-10-04 & 200  & ARAS & 3900\,--\,7500 & 10,000 \\
2021-10-16 & 212  & ARAS & 3900\,--\,7500 & 10,000 \\
2021-11-09 & 236  & ARAS & 3900\,--\,7500 & 10,000 \\
2021-12-22 & 279  & ARAS & 3900\,--\,7500 & 1,000  \\
2022-02-19 & 338  & ARAS & 3900\,--\,7500 & 1,000  \\
\hline
\end{tabular}
\end{table*}
\clearpage


\end{document}